\documentclass[aps,prd,english,preprintnumbers,nofootinbib,floatfix,twocolumn,10pt]{revtex4-1}

\usepackage{amsfonts,amsmath,amssymb}
\usepackage{graphicx,epsfig}
\usepackage{float}
\usepackage{subfig}
\usepackage{subfloat}
\usepackage[utf8]{inputenc}
\usepackage{hyperref}
\usepackage{babel}
\usepackage[justification=justified]{caption}

\usepackage[font=footnotesize,labelfont=rm,justification=raggedright]{caption}

\newcommand\be{\begin{equation}}
\newcommand\ee{\end{equation}}
\newcommand\bea{\begin{eqnarray}}
\newcommand\eea{\end{eqnarray}}

\begin{document}

\def\rhoo{\rho_{_0}\!} 
\def\rhooo{\rho_{_{0,0}}\!} 

\begin{flushright}
\phantom{
{\tt arXiv:2006.$\_\_\_\_$}
}
\end{flushright}

{\flushleft\vskip-1.4cm\vbox{\includegraphics[width=1.15in]{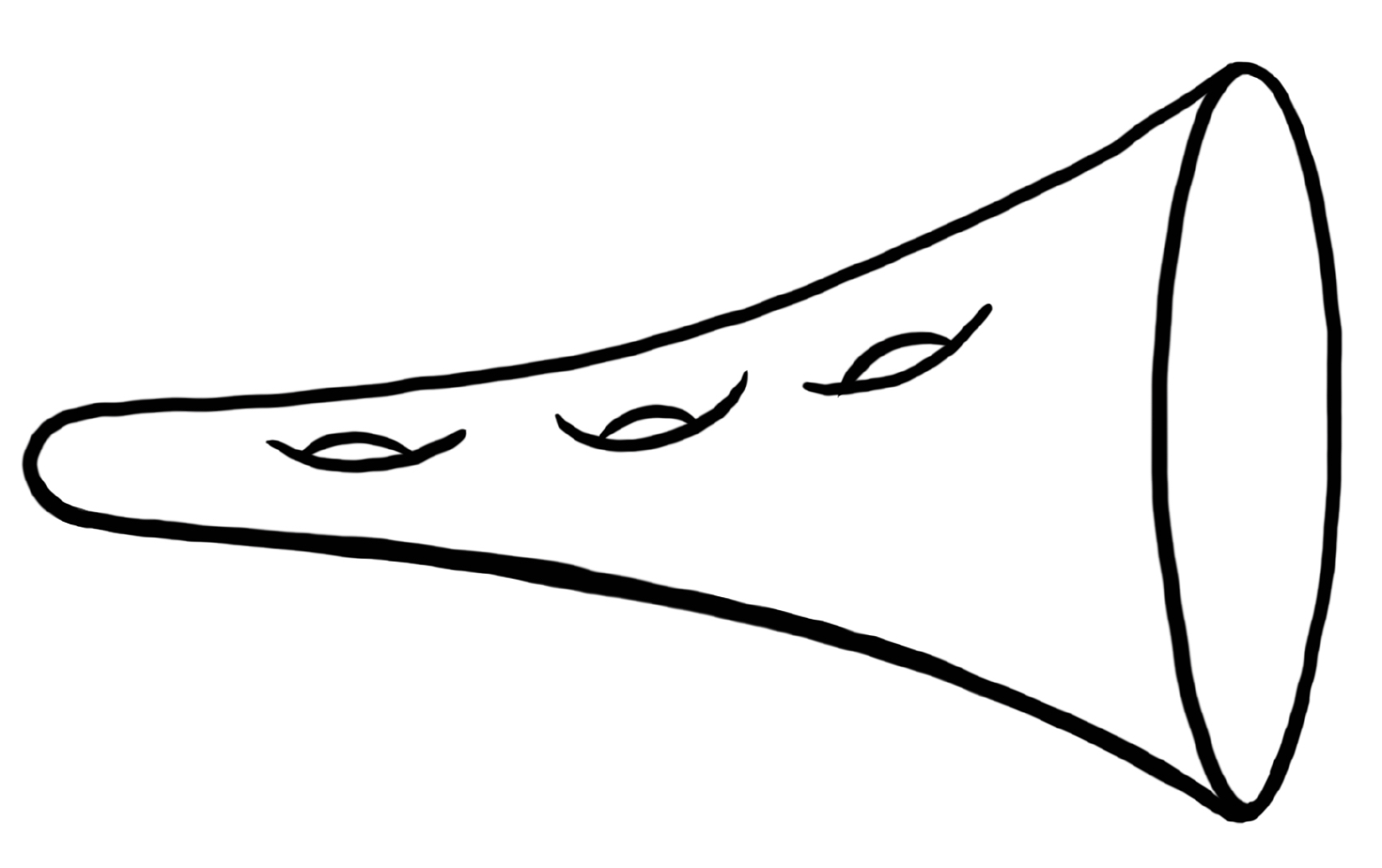}}}

\title{Explorations of Non--Perturbative JT Gravity and Supergravity}
\author{Clifford V. Johnson}
\email{johnson1@usc.edu}
\affiliation{Department of Physics and Astronomy\\ University of
Southern California \\
 Los Angeles, CA 90089-0484, U.S.A.}


\begin{abstract}
Some recently proposed definitions of  Jackiw--Teitelboim gravity and supergravities  in terms of  combinations of minimal string models are explored, with a focus on  physics beyond  the perturbative expansion in spacetime topology. While this formally involves solving  infinite order non--linear differential equations, it is shown that the physics can be  extracted to arbitrarily high accuracy  in a simple controlled truncation scheme, using a combination of analytical and numerical methods. 
The  non--perturbative spectral densities are explicitly computed and exhibited. 
The full spectral form factors, involving crucial non--perturbative contributions  from   wormhole geometries,  are also computed and displayed, showing the non--perturbative details of the characteristic ``slope'', ``dip", ``ramp'' and ``plateau'' features.  It is emphasized that results of this kind can most likely be readily extracted for other types of JT gravity using the same methods. 
\end{abstract}


\maketitle

\section{Introduction}
\label{sec:introduction}

There are many reasons to study
Jackiw--Teitelboim (JT) gravity~\cite{Jackiw:1984je,Teitelboim:1983ux}. One of  them is the fact that it is a theory of a two dimensional quantum gravity, where the spacetime is allowed to split and join, changing its topology (characterized by Euler characteristic $\chi{=}2{-}2g{-}b{-}c$ where $g$ counts handles, $b$ boundaries, and $c$ crosscaps). In the full theory  the partition function $Z(\beta)$ is  a sum over the contributions from all topologies 
 as well as a non--perturbative part that is not captured by the perturbative expansion in topology:
\begin{equation}
Z(\beta) = \sum_{\rm \chi} Z_{\chi}(\beta) + \mbox{\rm non--perturb.}
\end{equation} 
Here, $Z_\chi(\beta)$ stands for the  contribution to the partition function from surfaces of Euler characteristic $\chi$. It comes with a factor $e^{\chi  S_0}$, as $S_0$ is a coupling that multiplies the Einstein--Hilbert action in the model.   (Although $\chi{=}1$ for the (leading) disc order quantities, the subscript 0 will be widely used at leading order henceforth. So the disc level partition function is $Z_0$, spectral density is $\rho_0$, {\it etc}.)

The focus of this paper will be on characterizing the full partition function of the theory, including the  {\it full non--perturbative} physics, by making explicit aspects of the double--scaled matrix model definitions suggested in refs.~\cite{Johnson:2019eik,Johnson:2020heh}, which should be considered companion papers to this one. The beautiful work of refs.~\cite{Saad:2019lba,Stanford:2019vob} in defining double--scaled matrix models of (various kinds of) JT gravity is intrinsically perturbative in spirit, since they use recursion relations connecting different topologies, and the work in refs.~\cite{Johnson:2019eik,Johnson:2020heh} is intended as a complementary construction (using minimal strings) that allows more direct access to non--perturbative quantities. The output of this paper will be the first explicit computation of the full spectral densities (and hence the partition functions, by Laplace transform), and   explorations of  several important phenomena that depend crucially on being able to compute non--perturbative physics. 
 \begin{figure}[h]
\centering
\includegraphics[width=0.45\textwidth]{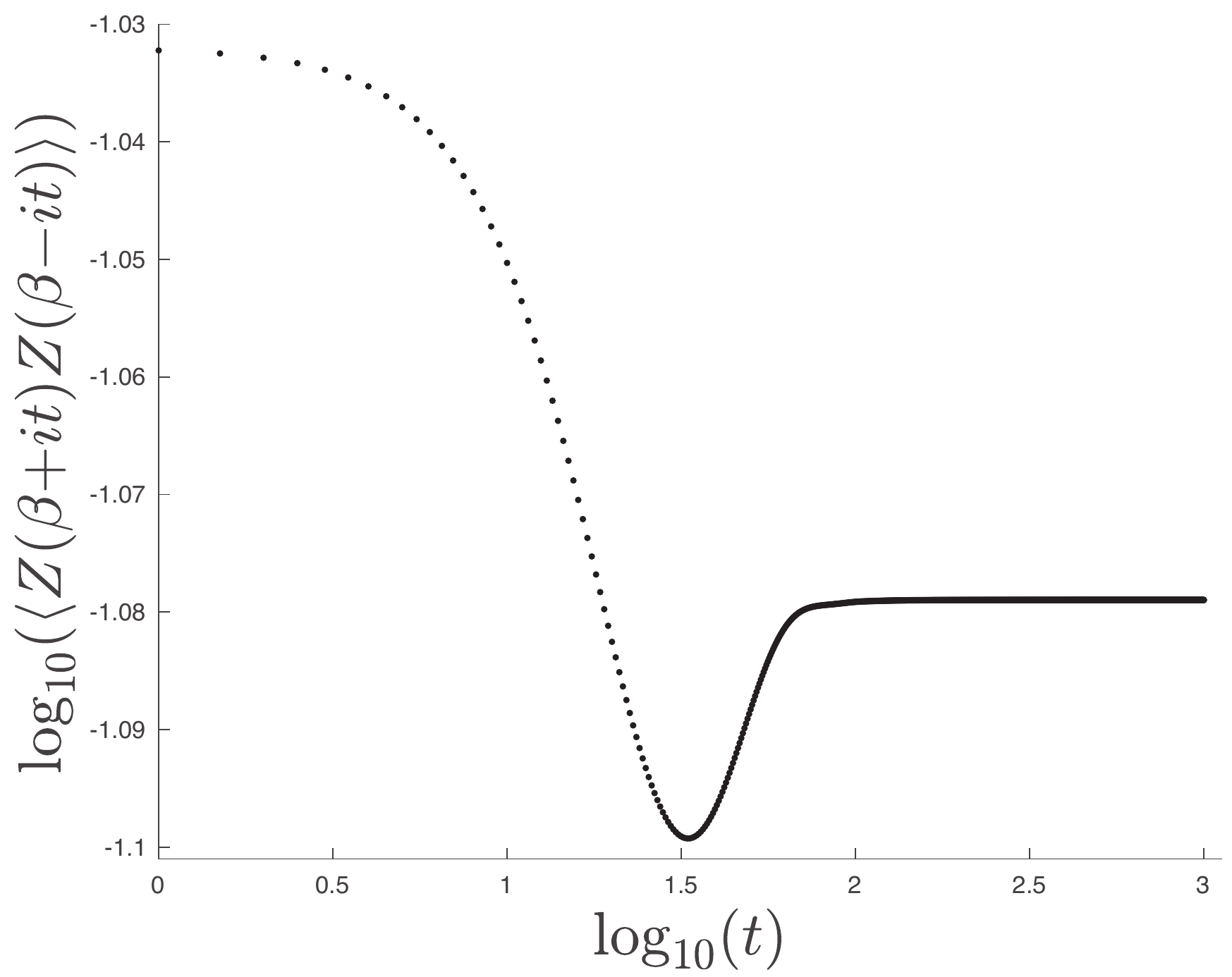}
\caption{\label{fig:combined-sff-SJT1-alt} The full spectral form factor, showing the classic (saxophone) shape made up of a slope, dip, ramp, and plateau. This is computed using the methods of this paper for the (2,2) model of JT supergravity. Here, $\beta{=}35$, $\hbar{=}1/5$.  (See text.)}
\end{figure}

An example of the latter is the  2--point ``spectral form factor'' shown in figure~\ref{fig:combined-sff-SJT1-alt}, a quantity that helps in  diagnosing  universal aspects of quantum chaotic behaviour~\cite{Guhr:1997ve,Liu:2018hlr}. It was computed using the methods of this paper. This is the first time this quantity (and others like it to be presented later) has been computed fully in JT gravity or supergravity for generic values of $\beta$ and $S_0$, and so some time will be spent unpacking the techniques and results\footnote{An interesting recent paper~\cite{Okuyama:2020ncd} presented an expression for the spectral form factor of JT gravity, but  in a very special ultra--low temperature scaling limit that allowed a closed form to be written. Also, key aspects of parts of the spectral form factor in special limits were discussed using matrix model techniques in ref.~\cite{Blommaert:2019wfy}. In this paper, no special limits on the parameters are taken, and while no closed forms are presented, answers can be systematically extracted for a  range of $\beta$ and $S_0$.}. The  late time ``plateau'' feature of the curve, and the transition to it from the ``ramp'' behaviour, are intrinsically non--perturbative features of wide interest. There are important non--perturbative effects that show up in the slope part too, in some cases, as will be demonstrated. They can sometimes be dramatic, as will be seen in the supergravity examples presented.

\begin{figure}[h]
\centering
\includegraphics[width=0.45\textwidth]{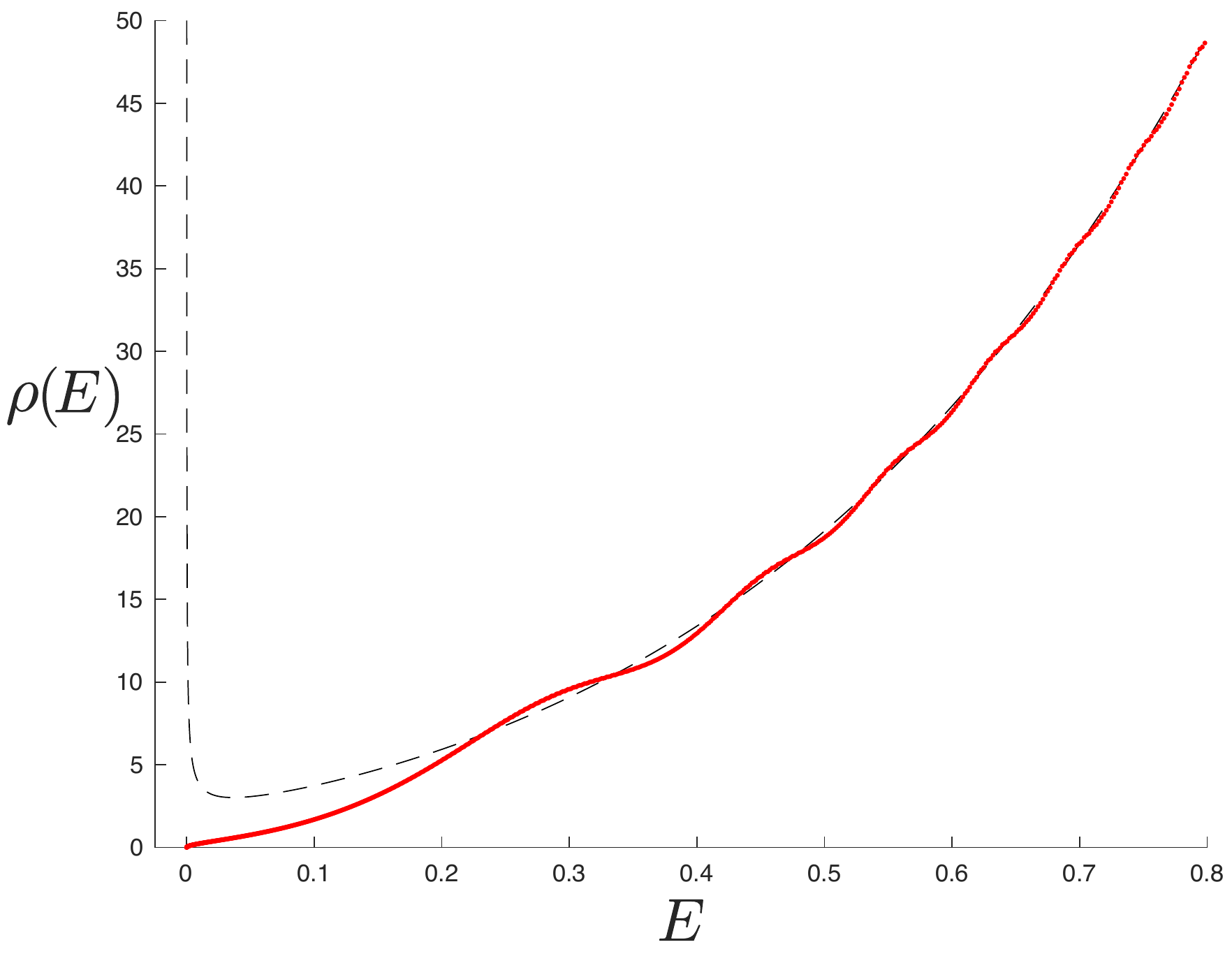}
\caption{\label{fig:non-pert-sd1} The full spectral density, computed using the methods of this paper, for the (2,2) model of JT supergravity. The dashed blue line is the disc level result of equation~(\ref{eq:spectral_SJT}). Here $\hbar{=}1$.}
\end{figure}

Another  example of this paper's results is given in figure~\ref{fig:non-pert-sd1}. It is the full spectral density (the thicker line, actually made out of data dots) of the model $\rho^{\rm SJT}(E)$ with the classical result (see equation~(\ref{eq:spectral_SJT})) plotted as a dashed  line for comparison.  {\it By Laplace transform, this function defines the full non--perturbative partition function for the supergravity theory, and is computed here explicitly for the first time.} 
This JT supergravity model is in fact the $(\boldsymbol{\alpha},\boldsymbol{\beta}){=}(2,2)$ matrix model in the Altland--Zirnbauer\cite{Altland:1997zz} classification scheme, or the case $\Gamma{=}\frac12$ in the notation of ref.~\cite{Johnson:2020heh}. The  result for the companion $(0,2)$ case ($\Gamma{=}{-}\frac12$)  will be displayed later (see figure~\ref{fig:non-pert-sd2} on page~\pageref{fig:non-pert-sd2}). As can be seen in  figure~\ref{fig:non-pert-sd1}, for the (2,2) case non--perturbative effects entirely erase the characteristic classical peak in the spectrum at low energy, which dramatically alters the ``slope'' part of the spectral form factor as compared to the analogous result for the (0,2) case where a peak persists in the full spectrum.

While ordinary JT gravity is important and interesting (and results {\it will} be presented for it), a good deal of attention will be given to these two particular models of JT supergravity. They are of particular interest because the non--perturbative physics is more dramatic, in a sense. It was observed in ref.~\cite{Stanford:2019vob} (and confirmed to be manifest in the minimal model construction of ref.~\cite{Johnson:2020heh}) that beyond the first one or two leading  orders of perturbation theory (depending upon the quantity being computed) {\it the entire topological perturbative series vanishes}. Therefore the non--perturbative effects uncovered in these models (as will be done here) are  placed more in stark relief than other JT gravity systems.

Having shown examples of the key results, the job of the rest of the paper is to  explain how to get them, and then to interpret them.  The results  follow from the non--perturbative construction, proposed in refs.~\cite{Johnson:2019eik,Johnson:2020heh}, of JT gravity and supergravity in terms of  minimal string models (of a special type). The basic idea, building on suggestions in  refs.~\cite{Saad:2019lba,Okuyama:2019xbv}, is to reinterpret the JT system as an infinite set of minimal models (non--linearly) coupled together in a particular way, or equivalently (as explained in ref.~\cite{Johnson:2020heh}) by turning on an infinite set of operators in the minimal string model obtained by taking the $k{\to}\infty$ limit\footnote{Other recent work exploring connections between the formalism of Liouville theory and minimal strings on the one hand, and JT gravity on the other, includes refs.~\cite{Okuyama:2020ncd,Betzios:2020nry,Mertens:2020hbs}.}. Since the full information about the $k$th minimal string model in question (see section~\ref{sec:minimal-model-deconstruction} for a quick review of the essentials) involves solving an order $2k{+}1$ highly non--linear ordinary differential equation (ODE), this way of  defining JT gravity or supergravity involves solving an infinite order differential equation.  This might seem rather daunting, or even formal, but from a pragmatic point of view it is rather straightforward to implement an approximation scheme that allows computation of an answer to a specific concrete question, to whatever accuracy is desired.  The point is that the contribution to the  model  of successively higher orders of derivatives in the ODE grows smaller with increasing $k$, and so there is a point at which truncating the ODE  and solving a finite order equation will give access  to the full spectrum all the way up to a given desired energy, to some required accuracy. In other words, this is hardly any different from computing Feynman diagrams up to some sufficiently high order  for some field theory problem (except that here the formalism is computing non--perturbative physics, and moreover the series is convergent, not asymptotic.)

{\bf An outline of the paper is as follows:} Section~\ref{sec:JT-gravity-summary} is a brief summary of some of the (now standard) key ideas in the study of JT gravity that will be used in this paper. It is entirely optional for those who know the subject well, but serves to set context, notation, and (perhaps) some motivation.   The deconstruction in terms of minimal models will be lightly explained in  Section~\ref{sec:minimal-model-deconstruction}. Refs.~\cite{Johnson:2019eik,Johnson:2020heh}  should be consulted for further details, and  the non--perturbative explorations of key toy models presented there. The main task of this paper is to show how to extract non--perturbative results for the full JT (super)gravities. In particular, this section will explain how (using the supergravity examples) the truncation scheme of the previous paragraph works. Section~\ref{sec:spectral-density} will  solve the full quantum mechanical system to yield the non--perturbative spectral density (and hence the partition function), for the supergravity  cases. Then Section~\ref{sec:spectral-form-factor} turns to the non--perturbative  spectral form factor for the supergravities, explaining how it is computed and then displaying several results. 

Section~\ref{sec:JT-gravity-regular} then discusses the analogous construction and results for a non--perturbative definition of ordinary JT  gravity obtained (as first presented in ref.~\cite{Johnson:2019eik}) by embedding it into a larger framework that it matches perturbatively (at high energy) but which supplies it with non--perturbatively well--behaved low energy physics. 

Since  most of the results of this paper come from numerically unpacking the highly non--linear system of equations (and also using computer algebra to help unpack them), some  Appendices  are included with some (it is hoped) helpful technical notes and suggestions about the  methods employed, for the reader interested in  computing these or other results using this formalism. Appendix~\ref{app:airy-model} presents a numerical study of the spectral form factor of the  Airy model (the double--scaled Gaussian Hermitian matrix model) and compares the results to the known exact expressions, showing how the effects of the truncation to a numerical system are extremely well controlled. This serves as a demonstration of the trustworthiness of  the numerical results obtained for the JT gravity and supergravity models in the main body of the paper. Appendix~\ref{app:notes-and-tips-1} describes aspects of solving high order differential equations numerically, and Appendix~\ref{app:notes-and-tips-2} describes how to solve for the energies and eigenfunctions needed to build the spectrum and spectral form factor. Appendix~\ref{app:gelfand-dikii} lists some important quantities needed in the body of the paper (the Gel'fand--Dikii differential polynomials) and a recursion relation for getting the higher order expressions.

There are some brief closing remarks in the final section,~\ref{sec:closing-remarks}, with thoughts about  the potential application of these methods to other systems.

\section{JT Gravity Lightning Tour}
\label{sec:JT-gravity-summary}
\begin{figure}[h]
\centering
\includegraphics[width=0.5\textwidth]{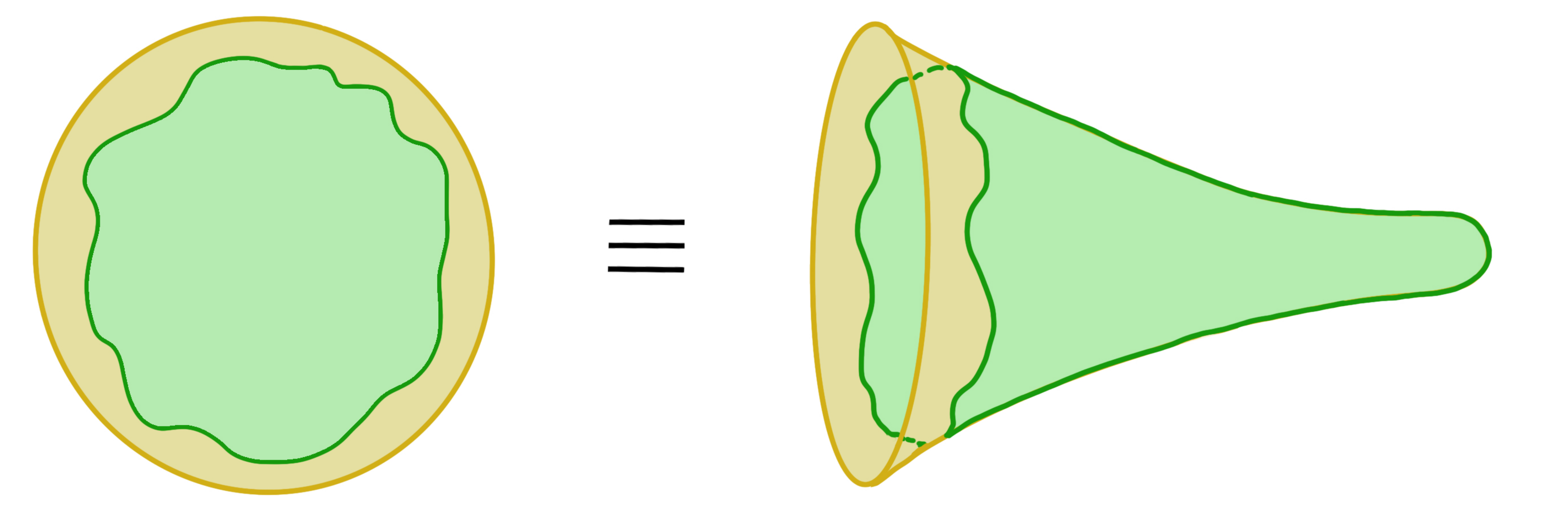}
\caption{\label{fig:disc-diagram} The ``nearly AdS$_2$'' geometry, presented in two equivalent ways.}
\end{figure}

Although it is a 2D theory of quantum gravity, by virtue of a coupling to a scalar, the dynamics of  JT gravity is all on the 1D spacetime boundary. (A good review of much of this is ref.~\cite{Sarosi:2017ykf}.) The boundary can change its shape while keeping its total  length fixed to be the inverse temperature $\beta{=}1/T$, the period of Euclidean time. Meanwhile, the bulk spacetime has constant negative curvature (the Ricci scalar $R{=}{-}2$). So the theory is locally AdS$_2$, and the leading spacetime (disc topology, {\it i.e.}, no handles or crosscaps, one boundary) is often called ``nearly--AdS$_2$''~\cite{Jensen:2016pah,Maldacena:2016hyu,Maldacena:2016upp,Engelsoy:2016xyb}, in the sense that, {\it e.g.} in Poincar\'e coordinates, the boundary is not a fixed circle an infinite distance away, but instead a finite loop of length~$\beta$ that is allowed to change its shape. See figure~\ref{fig:disc-diagram}.

At this order the dynamics of the loop is controlled by a Schwarzian action\cite{Maldacena:2016hyu}, and the result is:
\begin{equation}
\label{eq:disc-partition-function-JT}
Z^{\rm JT}_{0}(\beta)=\frac{e^{S_0}e^{\frac{\pi^2}{\beta}}}{4\pi^{1/2}\beta^{3/2}} = \int_0^\infty \rho_0^{\rm JT}(E) e^{-\beta E}dE\ ,
\end{equation}
related to the disc order spectral density $\rho_0^{\rm JT}(E)$ by a Laplace transform.
There is a JT supergravity generalization of this result~\cite{Stanford:2017thb,Stanford:2019vob}:
\begin{equation}
\label{eq:disc-partition-function-SJT}
Z^{\rm SJT}_{0}(\beta)=\sqrt{2}\frac{e^{S_0}e^{\frac{\pi^2}{\beta}}}{\pi^{1/2}\beta^{1/2}}= \int_0^\infty \rho_0^{\rm SJT}(E) e^{-\beta E}dE\ ,
\end{equation}
defining a disc order spectral density $\rho_0^{\rm SJT}(E)$. In each case, the densities are given by:
\begin{eqnarray}
\rho_0^{\rm JT}(E)&=&e^{S_0}\frac{\sinh(2\pi\sqrt{E})}{4\pi^2}\ ,\quad\mbox{\rm and}\label{eq:spectral_JT}\\
\rho_0^{\rm SJT}(E)&=& \sqrt{2}e^{S_0}\frac{\cosh(2\pi\sqrt{E})}{\pi\sqrt{E}}\ .\label{eq:spectral_SJT}
\end{eqnarray}

(Henceforth the redefinition $\sqrt{2}\rho_0^{\rm SJT}{\to}\rho_0^{\rm SJT}$ will be done, to adapt JT conventions of ref.~\cite{Stanford:2019vob} to the matrix model normalization to be used here.)
 The coupling $e^{-S_0}$ will be denoted $\hbar$ in what follows, and indeed it will be the $\hbar$ of a key quantum--mechanical system to appear shortly. One interpretation of $S_0$ is that it is simply the leading ($T{=}0$, disc topology) contribution to the entropy. For the ordinary JT case:
\begin{equation}
S=\left(1-\beta\frac{\partial}{\partial\beta}\right)\ln Z_{0}(\beta) = S_0+\frac{2\pi^2}{\beta}-\frac{3}{2}\ln\beta+\cdots\ ,
\end{equation}
This leads to a second reason (beyond the one mentioned in the introduction) to study JT gravity. It is a model of the low--temperature  (near--extremal) dynamics of certain higher dimensional black holes and branes (see e.g. refs.~\cite{Achucarro:1993fd,Fabbri:2000xh,Nayak:2018qej,Kolekar:2018sba,Ghosh:2019rcj}). For example, the metric of a charged black hole in $d{=}4$ is well known to  become AdS$_2{\times}S^2$ at $T{=}0$, and the area $A$ of the two--sphere, $S^2$ sets the $T{=}0$ entropy: $A{=}4S_0$. Turning on a small temperature replaces AdS$_2$ by ``nearly-AdS$_2$'', and the horizon area and hence the entropy gets corrections. The JT gravity model captures the dynamics of these corrections. (The dynamical scalar represents the deviation of the area away from extremality.)  The 2D dynamics can be thought of as containing black holes in its own right as well, worth studying in their own terms. These are, at leading order, the disc geometries already described. 

A third reason for studying JT gravity is that it is a low energy holographic dual, in a certain sense~\cite{Almheiri:2014cka,Jensen:2016pah,Maldacena:2016upp,Engelsoy:2016xyb} of a class of  1D quantum systems that exhibit quantum chaos, such as the    Sachdev--Ye--Kitaev (SYK) model~\cite{Sachdev:1992fk,Kitaev:talks,Maldacena:2016hyu}. A key diagnostic of the quantum chaotic behaviour of the system is the 2--point ``spectral form factor'' $\langle Z(\beta{-}it)Z(\beta{+}it)\rangle$, which exhibits certain key universal features~\cite{Guhr:1997ve,Liu:2018hlr,Papadodimas:2015xma,Maldacena:2001kr}. Starting out at  $\langle Z(\beta)^2\rangle$, it decays down a ``slope'' to a ``dip'' at during the first epoch of  time $t$, rises along a ``ramp'' at intermediate times, before levelling off to a ``plateau'' at late times at a value given by $\langle Z(2\beta)\rangle$. (See all these features in figure~\ref{fig:combined-sff-SJT1-alt}, but recall that it is not an SYK spectral form factor, but a gravity one; See below).

The timescales over which these features manifest are important, especially the time to when the plateau sets in, as it gives a measure of how long correlations take to wash away. No single SYK dual cleanly exhibits the universal behaviour individually. There are wild  oscillations in the spectral form factor at intermediate and late times~\footnote{In the phraseology of the moment,  these later eras are ``difficult times'' for an SYK  model.}. Instead, these features emerge as the time--averaged behaviour, as can be seen by averaging over an ensemble of models~\cite{Cotler:2016fpe}. An important idea in quantum chaos is the notion that random matrix ensembles should capture the universal features seen in the averaged behaviour of a chaotic system~ (for a review see ref.~\cite{Guhr:1997ve}). This led to the suggestion of refs.~\cite{Garcia-Garcia:2016mno,Cotler:2016fpe} that a random matrix description of averaged SYK could be available. On the other hand, random matrix models are known to describe, in a ``double--scaling'' limit~\cite{Brezin:1990rb,Douglas:1990ve,Gross:1990vs,Gross:1990aw}, the sum over  surfaces of a 2D quantum gravity,  so this is another way of seeing that there ought to be a dual gravitational description of SYK--like models. This was shown to be more than a coincidence of ideas in ref.~\cite{Saad:2019lba}, where JT gravity was demonstrated to be explicitly consistent with ---order by order in the topological expansion--- the properties of a double--scaled matrix model. Ref.~\cite{Stanford:2019vob} furnished several more examples and a classification of the possibilities in terms of the ten standard random matrix ensembles.

\begin{figure}[h]
\centering
\includegraphics[width=0.5\textwidth]{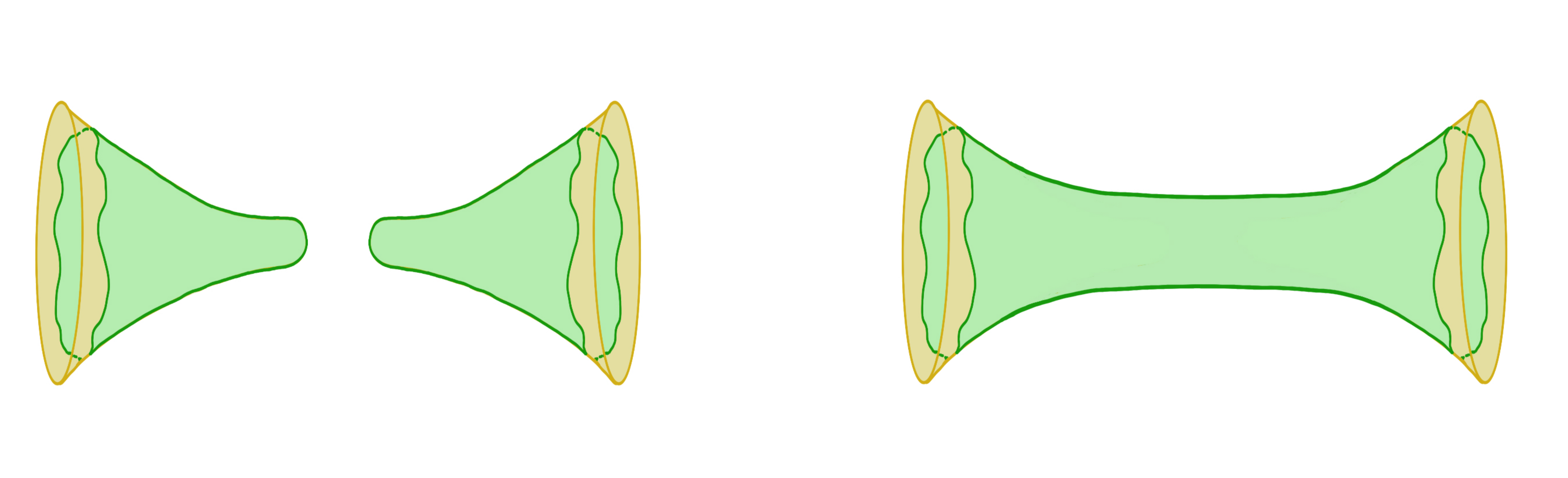}
\caption{\label{fig:spectral_form_factor_diagrams} Black holes {\it vs.} wormholes.}
\end{figure}

So the  JT gravity  dual (or supergravity dual, for the appropriate generalization of SYK~\cite{Fu:2016vas,Stanford:2017thb,Li:2017hdt,Kanazawa:2017dpd,Sun:2019yqp,Forste:2017kwy,Fu:2016vas,Li:2017hdt,Kanazawa:2017dpd,Forste:2017kwy,Murugan:2017eto,Sun:2019yqp,Stanford:2019vob}) performs the ensemble average directly. The early time behaviour is controlled by the disconnected diagram constructed of two discs (a pair of AdS$_2$ black holes), plus corrections, while the later ramp and plateau features come from the  cylinder diagram (an AdS$_2$ wormhole)~\cite{Saad:2018bqo} plus corrections. See figure~\ref{fig:spectral_form_factor_diagrams}. 
These amplitudes do not fluctuate chaotically in time, but have smooth behaviour to be expected from geometric objects in a theory of gravity. This can be seen already in the leading computation for the cylinder diagram~\cite{Saad:2019lba,Ginsparg:1993is}, which yields a simple~$t$ dependence: $\langle Z(\beta{-}it)Z(\beta{+}it)\rangle{\sim}\beta^{-1}\sqrt{\beta^2+t^2}$  gives a rise for the initial part of the ramp behaviour in a regime that would already be beset by fluctuations in any given SYK model.  For~$t{\gg}\beta$, assuming the transition to the plateau has not yet occurred, this yields a linear rise.  In this paper it will be observed that non--perturbative effects can, depending upon the value of $\hbar$, take over rapidly to generate the ramp, and so the linear part is hardly visible at moderate~$\beta$.

As already mentioned,  the plateau in the spectral form  factor (and the transition to it from the ramp) is a result of perturbative and, especially, non--perturbative corrections to the leading cylinder contribution. The purpose of this paper is to focus on unpacking the non--perturbative definitions of refs.~\cite{Johnson:2019eik,Johnson:2020heh} in order to  explicitly uncover such effects. Figure~\ref{fig:combined-sff-SJT1-alt}, already shown above, is a sample of the work reported on in this paper. It is the full spectral form factor for  a particular model of JT supergravity. It will be discussed more fully in section~\ref{sec:spectral-form-factor}. Now, on to the computations.

\section{Constructing JT (Super)gravity from Minimal Strings}
\label{sec:minimal-model-deconstruction}

The key ingredients are certain double scaled matrix models that have been used in the past to study certain kinds of ``minimal'' string theories. (See {\it e.g.} refs.~\cite{Ginsparg:1993is,Seiberg:2004at} for reviews.) The details of the string theory constructions do not matter here. The most important fact to know is that some of the models (a subset of the ``one--cut'' matrix models) can be described in terms of an  associated 1D quantum mechanics problem~\cite{Banks:1990df,Douglas:1990dd}, with Hamiltonian: 
\be
\label{eq:schrodinger}
{\cal H}=-\hbar^2\frac{\partial^2}{\partial x^2}+u(x)\ , 
\ee
 where the potential $u(x)$ satisfies a non--linear ordinary differential equation (ODE) called a ``string equation''. The key task is to build the correct $u(x)$ for the problem in hand. Once it is known, the full spectral density can be extracted by simply solving the spectrum of ${\cal H}$ and evaluating the fully non--perturbative $\rho(E)$, using an expression given in the next section.   It is useful to  note that in the limit where just the disc--level physics is kept,  the spectral density at this order can be written as a simple integral involving the leading part of the potential, $u_0(x){=}\lim_{\hbar\to0}u(x)$:
 \be
 \label{eq:integral-density-classical}
 \rho_0(E) = \frac{1}{\pi\hbar}\int_0^E \frac{f(u_0)}{\sqrt{E-u_0}}du_0\ ,
 \ee
 where $f(u_0){=}{-}\partial x/\partial u_0(x)$. (This is in a slightly different normalization from that used in ref.~\cite{Johnson:2020heh}.)
 
Turning back to the ingredients,  
the minimal models will be labelled by an integer index, $k$. As mentioned before, the models will be combined together to yield the JT (super)gravity. There is a parameter, $t_k$, that will be used to turn on the $k$th model in the mix. The model is turned on if~$t_k$ is non--zero.  The minimal models in question can be obtained~\cite{Morris:1991cq,Dalley:1992qg,Dalley:1992br,Dalley:1992vr,Dalley:1992yi}\footnote{These minimal models were later identified  by ref.~\cite{Klebanov:2003wg} as the $(4k,2)$ superconformal minimal models coupled to gravity with a type~0A projection.} by taking the double--scaling limit of models of a complex $N{\times}N$ matrix $M$, with a potential $V(M^\dagger M)$ (see also footnote~\ref{fn:matrix-model-remark}). 
The string equation that needs to be solved is:
 \be
\label{eq:string-equation-2}
u{\cal R}^2-\frac{\hbar^2}{2}{\cal R}{\cal R}^{\prime\prime}+\frac{\hbar^2}{4}({\cal R}^\prime)^2=\hbar^2\Gamma^2\ ,
\ee
where the constant $\Gamma$ will be discussed shortly and 
\be
\label{eq:flow-object}
 {\cal R} \equiv \sum_{k=1}^\infty t_k {\tilde R}_k[u] + x\ .
\ee
Here, ${\tilde R}_k[u]$ is proportional to the  $k$th order polynomial in $u(x)$ and its $x$--derivatives defined by Gel'fand and Dikii~\cite{Gelfand:1975rn}. They have a purely polynomial in $u(x)$ piece, which is $u(x)^k$,  a purely derivative linear piece, $u(x)$ $x$--differentiated $2k{-}2$ times, and then non--linear mixed terms involving $u(x)$ and its $x$--derivatives.  Here, they are normalized so that the coefficient of $u^k$ is unity. The first three are:
\bea
\label{eq:gelfand-dikii}
{\tilde R}_1[u]&=&u\ ,\qquad 
{\tilde R}_2[u]=u^2-\frac{1}{3}u^{''}\ , \quad {\rm and }\nonumber\\
{\tilde R}_3[u]&=&u^3-\frac12(u^{'})^2-uu^{''}+\frac{1}{10}u^{''''} \ , 
\eea
where a prime denotes an $x$--derivative times a factor of~$\hbar$.  It will transpire that ${\tilde R}_4, {\tilde R}_5$, ${\tilde R}_6$ and ${\tilde R}_7$ will be used  in this paper too, but since they are rather lengthy, some are listed in Appendix~\ref{app:gelfand-dikii}, along with methods for generating others if needed.

The boundary condition that ensures good non--perturbative behaviour is, for each model, 
\bea
\label{eq:boundary-conditions}
u(x)&\to&0\quad{\rm as}\quad x\to+\infty\ ,\nonumber\\
u(x)&\to&(-x)^{\frac1k}\quad {\rm as}\quad x\to-\infty\ . 
\eea
 Note  the presence of $\hbar{=}e^{-S_0}$ in the string equation (and the various quantities that make it up). It is  very useful for separating the classical parts  from the rest, by  sending $\hbar{\to}0$, or equivalently, dropping derivatives. For the study of non--perturbative physics, solutions $u(x)$ of the equation will be extracted for $\hbar{=}1$. Several results presented in  figures to come will be for  this value, because it allows the non--perturbative effects to be writ large in the results (for spectra, {\it etc.}), and therefore seen easily. When instructive to do so, comparison to results with $\hbar$ dialled down  will be discussed. It is interesting that it is in fact {\it more difficult} to solve the string equations for smaller $\hbar$. This is  because when derivatives have smaller coefficients, they are allowed to fluctuate more, contributing to the  sensitivity  when solving these highly non--linear equations, as will be discussed later. (This increased difficulty to get smaller $\hbar$ results has the character of  a sort of strong/weak coupling duality, in fact.)

 Turn now to the constant $\Gamma$ in the string equation~(\ref{eq:string-equation-2}). With it present, the matrix model is in the $(2\Gamma{+}1,2)$ class in  the $(\boldsymbol{\alpha},\boldsymbol{\beta})$ Altland--Zirnbauer classification of matrix ensembles~\footnote{\label{fn:matrix-model-remark}Double scaling means that in the matrix model\cite{Morris:1991cq,Dalley:1992qg,Dalley:1992br,Dalley:1992vr,Dalley:1992yi} of the complex matrix $M$, the size $N$ is taken to infinity while couplings in the potential $V(M^\dagger M)$ are tuned to certain critical~\cite{Kazakov:1989bc} values. Diagonalizing $M$ turns this into a problem involving its eigenvalues $\lambda_i$ ($i{=}1\cdots N$) at a cost of introducing a van der Monde determinant $J{=}\prod_{i<j} (\lambda_i^2-\lambda_j^2)^2$ for the Jacobian. The effective Dyson gas problem for the $\lambda_i^2$  can be thought of as existing on the positive real line, with a wall at zero. The constant~$\Gamma$ in equation~(\ref{eq:string-equation-2}) can be thought of as arising from the coefficient of a logarithmic term in the potential of the model (see {\it e.g.} ref.~\cite{Kostov:1990nf}), and as such, results in an extra factor  $\lambda_i^{2\Gamma}$ to the effective integration measure over the $i$th eigenvalue, giving $\prod_i \lambda_i^{2\Gamma+1}d\lambda_i$. With the  factor~$J$ included, the model is seen to be in the $(2\Gamma{+}1,2)$ class in the $(\boldsymbol{\alpha},\boldsymbol{\beta})$ Altland--Zirnbauer classification of matrix ensembles, defined for $\boldsymbol{\alpha}{=}0,1,2$. Actually $\Gamma$ can be more general integers or half--integers that just these values.}.   The two choices $\Gamma{=}{\pm}\frac12$ will mostly be considered in this paper, and  the two JT supergravity models discussed here are will be labelled~$(2,2)$ and~$(0,2)$.

A particular (say, the $m$th) minimal model can be studied by setting all the $t_k$ to zero except for $k{=}m$. As discussed in the previous two papers~\cite{Johnson:2019eik,Johnson:2020heh}, the $m{=}1$ case in particular was important as it models the   low energy tail of the eigenvalue distribution very well. To get the full behaviour, all of the $t_k$ must be turned on in a particular combination. For example, in the case of JT supergravity, the combination (derived in ref.~\cite{Johnson:2020heh}) is:
\begin{equation}
\label{eq:teekay_SJT}
t_k=\frac{\pi^{2k}}{(k!)^2}\ .
\end{equation}
So all the infinite models are turned on and  the string equation becomes a highly complicated object. But the purpose of this paper is to show that physics can be readily extracted nonetheless. 

Here is the reason  why. First, look at the disc level. The string equation is expression~(\ref{eq:string-equation-2}) with the three parts with $\hbar^2$ in front of them removed, and  the solution for $u_0(x)$ comes in two branches. Either $u_0(x){=}0$ or 
\be
\label{eq:leading-potential}
{\cal R} = \sum_{k=1}^\infty t_k u_0^k+x = 0\ , 
\ee
corresponding to the asymptotics given in equation~(\ref{eq:boundary-conditions}). This equation has $u_0(x){=}0$  at $x{=}0$, and so the two branches join at $x{=}0$.
 For the second branch, the $x{<}0$ regime, the combination~(\ref{eq:teekay_SJT}) of $t_k$'s amounts to $f(u_0){=}\pi I_1(2\pi\sqrt{u_0})/\sqrt{u_0}$ in equation~(\ref{eq:integral-density-classical}), yielding the part of the spectral density expanded in positive powers of $E$. The simple $E^{-\frac12}$ part comes from the $u_0{=}0$ behaviour. Integrating $f(u_0)$ with respect to $u_0$, or simply by looking at equation~(\ref{eq:leading-potential}), the explicit potential that gives JT supergravity on the disc is given by the equation
 \be
 \label{eq:class-pot-SJT}
 x= 1-I_0(2\pi\sqrt{u_0})\ ,
 \ee 
 where $I_0(s)$ is the zeroth modified Bessel function of $s$. This is a remarkably simple form. 
 
\begin{figure}[h]
\centering
\includegraphics[width=0.5\textwidth]{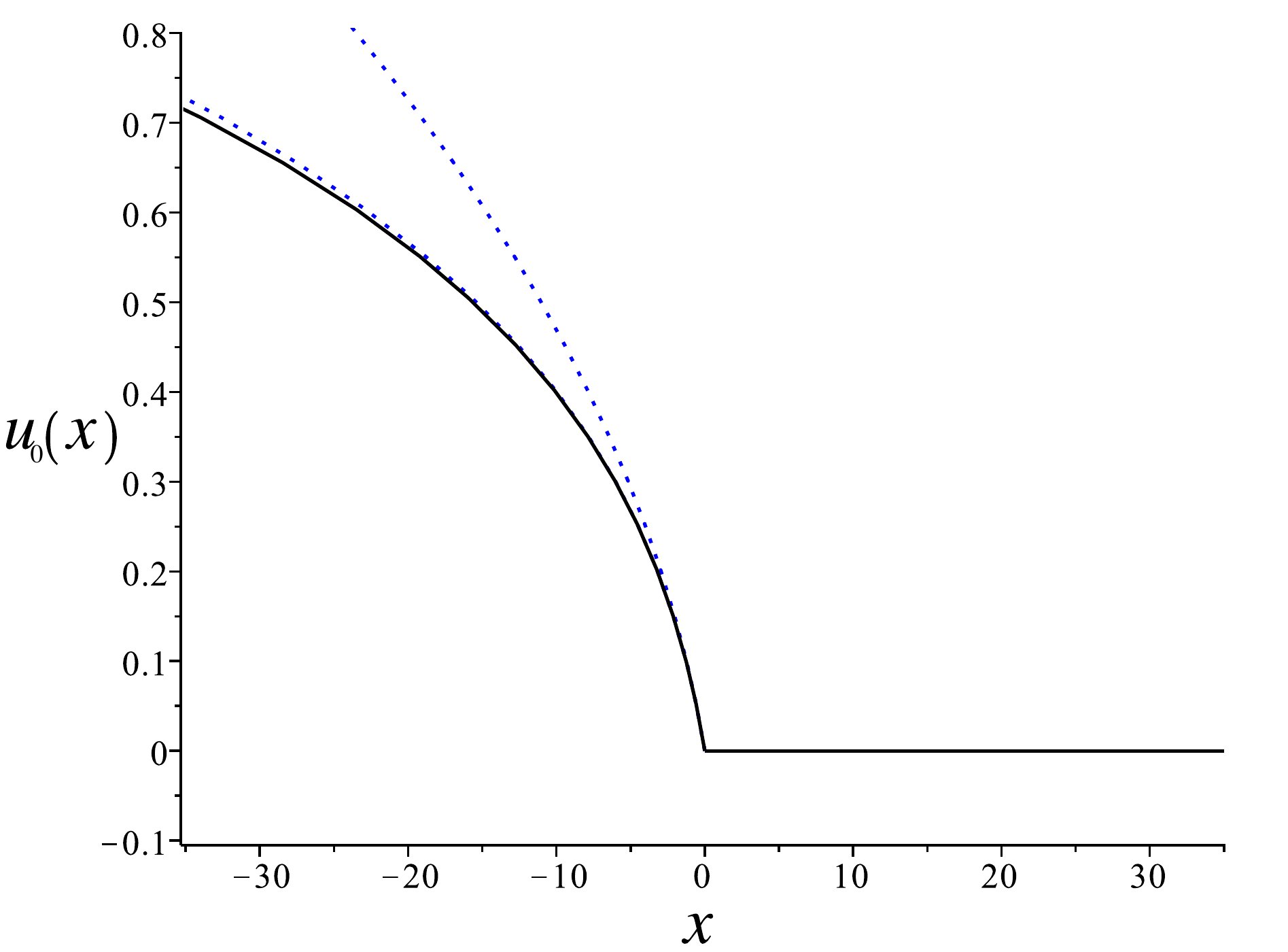}
\caption{\label{fig:truncation-examples} The complete classical potential for JT supergravity (solid line). The uppermost dotted line is a truncation up to $t_2$, the lower dotted is a truncation up to $t_4$.}
\end{figure}

 The issue of tractability becomes the simple issue of how well this potential can be approximated by truncating to a finite number of $t_k$s. The answer boils down to what maximum energy scale $E$ one wants to know the spectrum up to, and to what accuracy. As an example, the full classical potential~(\ref{eq:class-pot-SJT}) is plotted in figure~\ref{fig:truncation-examples} alongside two truncations. 

 The first truncation contains just $t_1$ and $t_2$:
 \be
 \label{eq:truncation-1}
 x= -{\pi}^{2}u_0-\frac14\,{\pi}^{4}u_0^{2}\ ,
 \ee
 and it is clear that it is a good approximation for energies up to approximately $E{\sim}0.1$ after which it begins to deviate considerably. The next example truncation  adds~$t_3$ and~$t_4$:
 \be
 \label{eq:truncation-2}
 x=-{\pi}^{2}u_0-\frac14\,{\pi}^{4}u_0^{2}-\frac{1}{36}{\pi}^{6}u_0^{3}-{\frac {1}{576}} {\pi}
^{8}u_0^{4}\ ,
 \ee
and for energies up to order $E{\sim}0.5$ it serves rather well. Further improvements come by adding higher orders. 

The next issue to appreciate is how much the solution changes when all the non--perturbative corrections are included. For all~$k$ a shallow well can develop in the central region (slightly to the right of $x{=}0$). Crucially, moving away from that region, the deviation of the solution from the disc level behaviour rapidly  dwindles, as it matches on to the asymptotic behaviour. The same is true for the coupled solution. Moreover as can be seen from equation~(\ref{eq:teekay_SJT}), the form of the $t_k$ as $k$ grows is such that good approximations at the disc level can be found by adding only a small number of minimal models, for a given needed accuracy. For larger $E$s, the solution becomes hard to distinguish from the classical result, and in that case the exact classical potential can be used, to a good approximation.

\begin{figure}[h]
\centering
\includegraphics[width=0.45\textwidth]{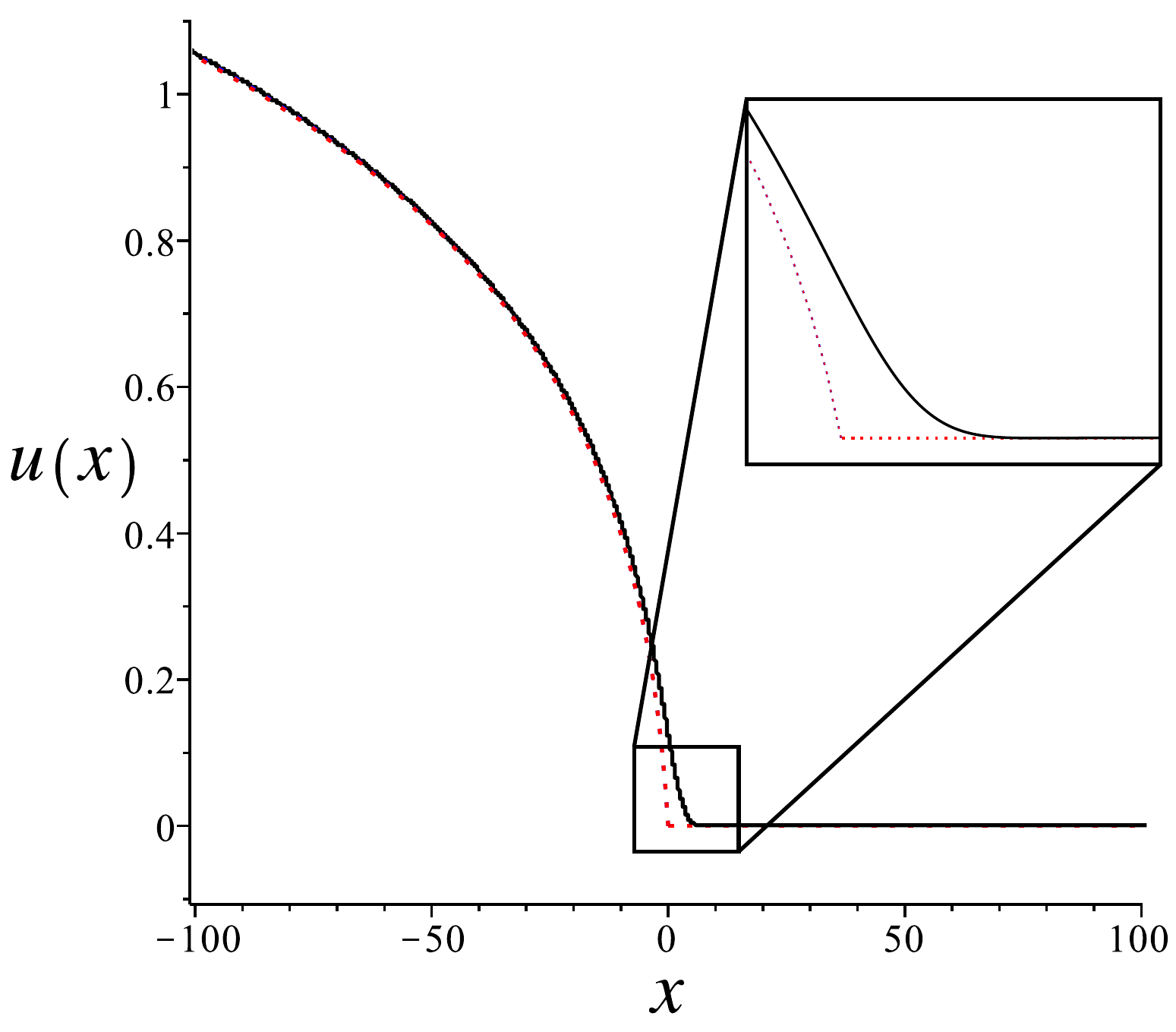}
\caption{\label{fig:truncation-examples-A} The $\Gamma{=}{+}\frac12$ solution (solid line) of the string equation for truncation up to $t_6$. The inset shows a closeup of the smooth transitional region in the interior. For comparison, the full classical solution line is shown too (dotted).}
\end{figure}

The highest truncation levels chosen for the purposes of this paper was to keep all the minimal models up to $k{=}6$ (this section) and $k{=}7$ (for section~\ref{sec:JT-gravity-regular}), although very good results were obtained for lower order truncations too. Since $R_7[k]$ has the twelfth order derivative of $u(x)$ in it,  the string equation~(\ref{eq:string-equation-2}) is a $14$th order differential equation. (It is 12th order for the $k{=}6$ truncation.) In general, it is easier to take a derivative of the string equation, whereupon an overall factor of ${\cal R}$ can be divided out, reducing some of the non--linearity somewhat, at the expense of an increase in the order. For the boundary conditions in question, a 15th (or 13th) order differential equation  is not too hard to solve numerically, with care. Some suggestions and notes are given in appendix~\ref{app:notes-and-tips-1}, for those who wish to carry out their own computations using this framework. Part of the full non--perturbative potential $u(x)$ for the truncation to $k{=}6$  is displayed in figure~\ref{fig:truncation-examples-A}. 

This is for the case $\Gamma{=}{+}\frac12$, {\it i.e.,} the (2,2) JT supergravity. (It was solved between $x{=}{-}200$ and $x{=}{+}200$.) 
Notice that it approaches the classical solution and  agrees with it rather well up (and beyond, it turns out) to $E{\sim}1.3$, and so the full solution, out to beyond the $x{\simeq}{-}100$ shown, can be used to capture the spectrum of (2,2) JT supergravity with good accuracy (see more in appendix~\ref{app:notes-and-tips-2} on how to do this). It is (relatively) easy to do better, if desired, but little visible change was noticed in going to higher orders, in exchange for accessing only a slightly larger maximum energy. 
 
\begin{figure}[h]
\centering
\includegraphics[width=0.45\textwidth]{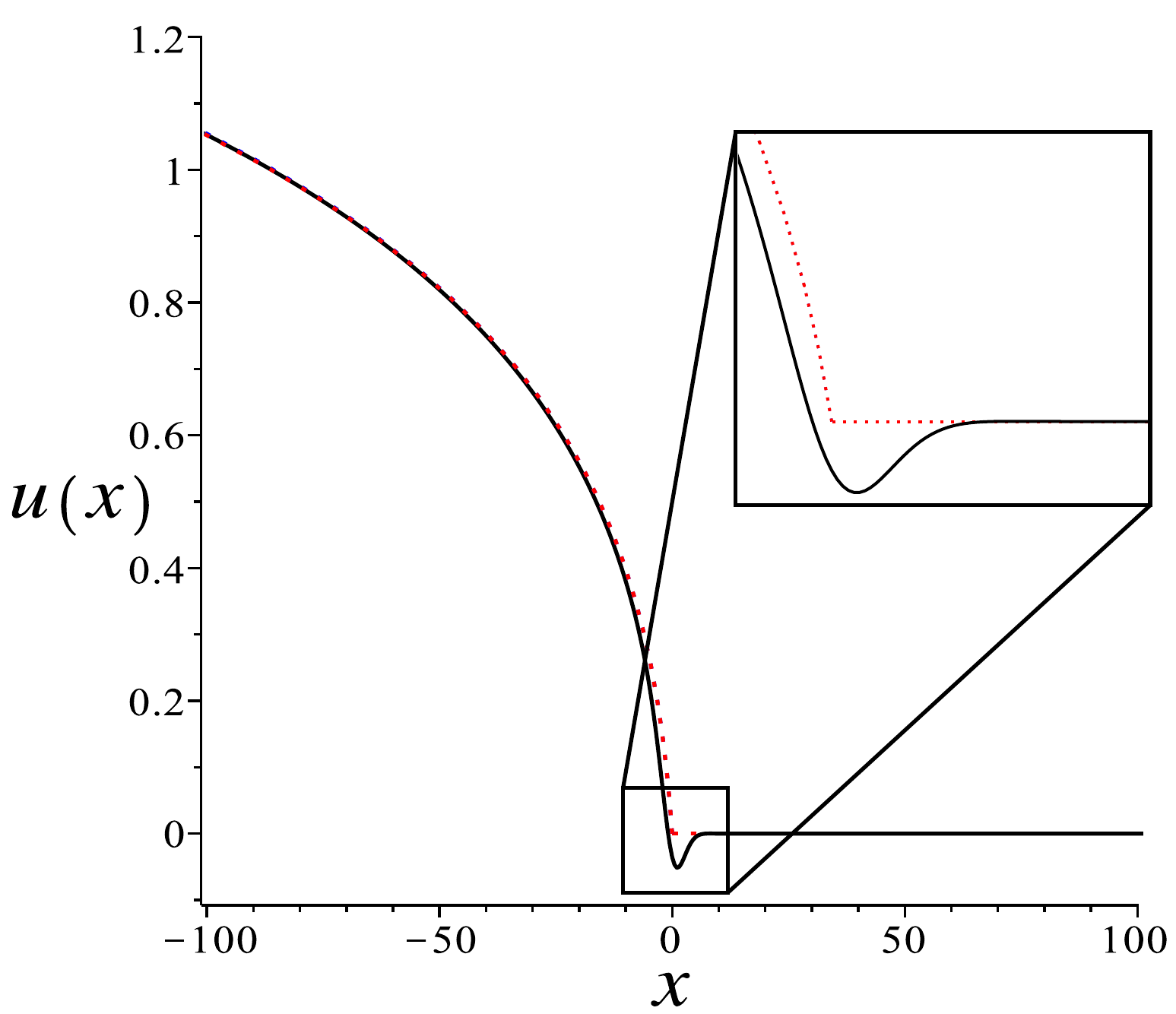}
\caption{\label{fig:truncation-examples-B} The $\Gamma{=}{-}\frac12$ solution (solid line) of the string equation for truncation up to $t_6$. The inset shows the smooth well that developed in the interior. For comparison, the full classical solution  is shown too (dotted).}
\end{figure}

Figure~\ref{fig:truncation-examples-B} shows  the solution for  $\Gamma{=}{-}\frac12$, which will be used to study the properties of the (0,2) JT supergravity. 

This solution has a well in the interior (as is quite typical of these solutions), and so slightly more rapid changes take place there. Rather than use numerical methods to solve for this directly (which are inevitably more sensitive to error in this case), a handy solution--generating technique derived in ref.~\cite{Carlisle:2005mk} was used\footnote{It is inherited from  the rather rich underlying Korteweg--de Vries (KdV) hierarchy integrable structure that underpins this entire formalism.}, that allowed it to be generated from the $\Gamma{=}{+}\frac12$ solution already found. See Appendix~\ref{app:notes-and-tips-1} for more on this.

\section{The Spectral Density}
\label{sec:spectral-density}

\subsection{General Remarks}
The next step is to solve the full spectral problem for the Hamiltonian ${\cal H}$, given the potential $u(x)$ found in the previous section by solving the (truncated) string equation. 
The relation between the spectrum of ${\cal H}$ and the JT partition function is as follows. From the minimal string perspective, the JT (super)gravity partition function is simply~\cite{Saad:2019lba,Okuyama:2019xbv} the expectation value of a ``macroscopic loop'' of length~$\beta$. The technology for working this out was derived long ago in ref.~\cite{Banks:1990df}. (Ref.~\cite{Ginsparg:1993is} unpacks the formalism in a useful review.) It is the trace of the exponentiated ${\cal H}$, with a projection ${\cal P}$ inserted:
\be
\label{eq:partition-function-from-H}
Z(\beta)={\rm Tr}(e^{-\beta{\cal H}}{\cal P})
\ee
where the   operator ${\cal P}{\equiv}\int_{-\infty}^\mu dx\, |x\rangle \langle x| $, and the upper limit $\mu$ will be discussed shortly.  Inserting a complete set of states:
\be
\int d\psi |\psi\rangle\langle\psi | = 1\ ,
\ee
yields
\bea
Z(\beta)&=& \int_{-\infty}^\mu  dx \langle x| e^{-\beta {\cal H}} |x\rangle  \nonumber\\
             &=& \int_{-\infty}^\mu  dx \int d\psi  \langle x| e^{-\beta {\cal H}}|\psi\rangle\langle\psi | x\rangle  \nonumber\\
             &=& \int_{-\infty}^\mu  dx \int d\psi_E  \langle x |\psi_E\rangle\langle\psi_E | x\rangle e^{-\beta E} \nonumber\\
             &=&   \int dE  e^{-\beta E}  \rho(E) \ ,
             \label{eq:partition-function-integration}
\eea
where 
\be
\label{eq:spectral-density-integration}
\rho(E) = \int_{-\infty}^\mu \psi(x,E)\psi^*\!(x,E) dx\ ,
\ee
is the spectral density. 

To understand the upper limit $\mu$, return to the leading order expression for the density given in equation~(\ref{eq:integral-density-classical}), and change variables to $x$. The solid black line in figure~\ref{fig:truncation-examples} is a reminder of the behaviour of $u_0(x)$. The part of the integral where $u_0(x){=}0$ begins at $x{=}0$ and extends to positive $x$ to some value denoted $\mu$, while (at a given $E$) the lowest value for $x$ is set by the turning point where $E{=}u(x)$, at a position denoted ${-}|x_0|$  and so:
\be
\label{eq:integral-density-classical-2}
\rhoo(E)=\frac{1}{\pi \hbar}\int_{{-}|x_0|}^\mu\frac{dx}{\sqrt{E-u_0(x)}}\ .
\ee
In the full quantum expression~(\ref{eq:spectral-density-integration}), there are contributions to the spectral density from the whole integration range down to $x{=}{-}\infty$. Intuitively, this is because wavefunctions penetrate to the left beyond the classical turning point $E{=}u(x)$.  (Often, in the classical expression~(\ref{eq:integral-density-classical-2}), the lower limit is also written as $x{=}{-}\infty$ with the understanding that only the real part of the integral contributes to $\rhoo$)\footnote{Of course, the  integrand of~(\ref{eq:spectral-density-integration}) is  not written in terms of~$E$ and the full $u(x)$, but in terms of the wavefunctions. The precise connection between the two expressions is {\it via} the  diagonal of the resolvent, $({\cal H} {-} E)^{-1}$,  of ${\cal H}$, denoted~${\widehat R}(x)$. It can be  built out of $\psi(x,E)\psi^*\!(x,E)$, using standard Green function methods. Gel'fand and Dikii~\cite{Gelfand:1975rn} derived an equation for ${\widehat R}(x)$ which is $4(u(x){-}E){\widehat R}^2{-}2 {\widehat R}{\widehat R}^{\prime\prime}{+}({\widehat R}^\prime)^2 {=}1$. The leading order solution  comes again from dropping derivatives, and is ${\widehat R(x)}{=}{\pm}(2\sqrt{u(x){-}E})^{-1}$. In the normalization of this paper $\rhoo{=}(2/\pi\hbar){\rm Im}\int {\widehat R}(x) dx$, yielding the form~(\ref{eq:integral-density-classical-2}).}. The role of $\mu$ becomes clear from focussing on just the contribution from the $u_0(x){=}0$ part in equation~\ref{eq:integral-density-classical-2}. It adds $\mu/(\pi\hbar\sqrt{E})$ to the classical spectral density. Physically, $\mu$ controls how much the classical spectral density is ``piled up'' against the natural wall at $E{=}0$ that stops the energies in the complex matrix models from going negative (see footnote~\ref{fn:matrix-model-remark}). Vanishing~$\mu$ corresponds to {\it just} touching the wall.  In the normalization of equation~(\ref{eq:spectral_SJT})  (without the $\sqrt{2}$), the value of $\mu$  is $1$, and this will be used for much of the rest of this paper. (Ref.~\cite{Johnson:2020heh} uses the value $\mu{=}2$.) However, it is useful for later to write a more general classical spectral density with a different $\mu$:
\be
\label{eq:spectral_SJT_mu}
\rhoo^{\rm SJT}(E,\mu)= \frac{1}{\pi\hbar}\left[\frac{\cosh(2\pi\sqrt{E})}{\sqrt{E}}+\frac{\mu-1}{\sqrt{E}}\right] \ .
\ee

\subsection{Computation}

Now  to the matter of  computing the full spectral density~(\ref{eq:spectral-density-integration}). Just as in refs.~\cite{Johnson:2019eik,Johnson:2020heh},  a matrix  Numerov method~\cite{doi:10.1119/1.4748813} was used to solve for the spectrum of ${\cal H}$. Conceptually it is a simple problem in finding eigenvalues and eigenfunctions for a Schr\"odinger operator in 1D. The wavefunctions are free and oscillatory to the far right ($x{>}0$), begin to feel the presence of the potential as they move further into the interior (it starts as a $1/x^2$ dependence), and then once they hit the potential there is an exponential decay to the left ($x{<}0$). As a guide to extracting them with good accuracy, some extra suggestions and notes for interested readers are given in appendix~\ref{app:notes-and-tips-2}.  The same normalization method as the one used in ref.~\cite{Johnson:2019eik} was used for the resulting eigenfunctions. The key point explained there is that in the far $x{>}0$ region, wavefunctions are known to asymptote to a simple form involving the Bessel function of order $\Gamma$, where the normalization can be analytically chosen to yield the correct contribution to the disc level spectral density.  

\begin{figure}[h]
\centering
\includegraphics[width=0.48\textwidth]{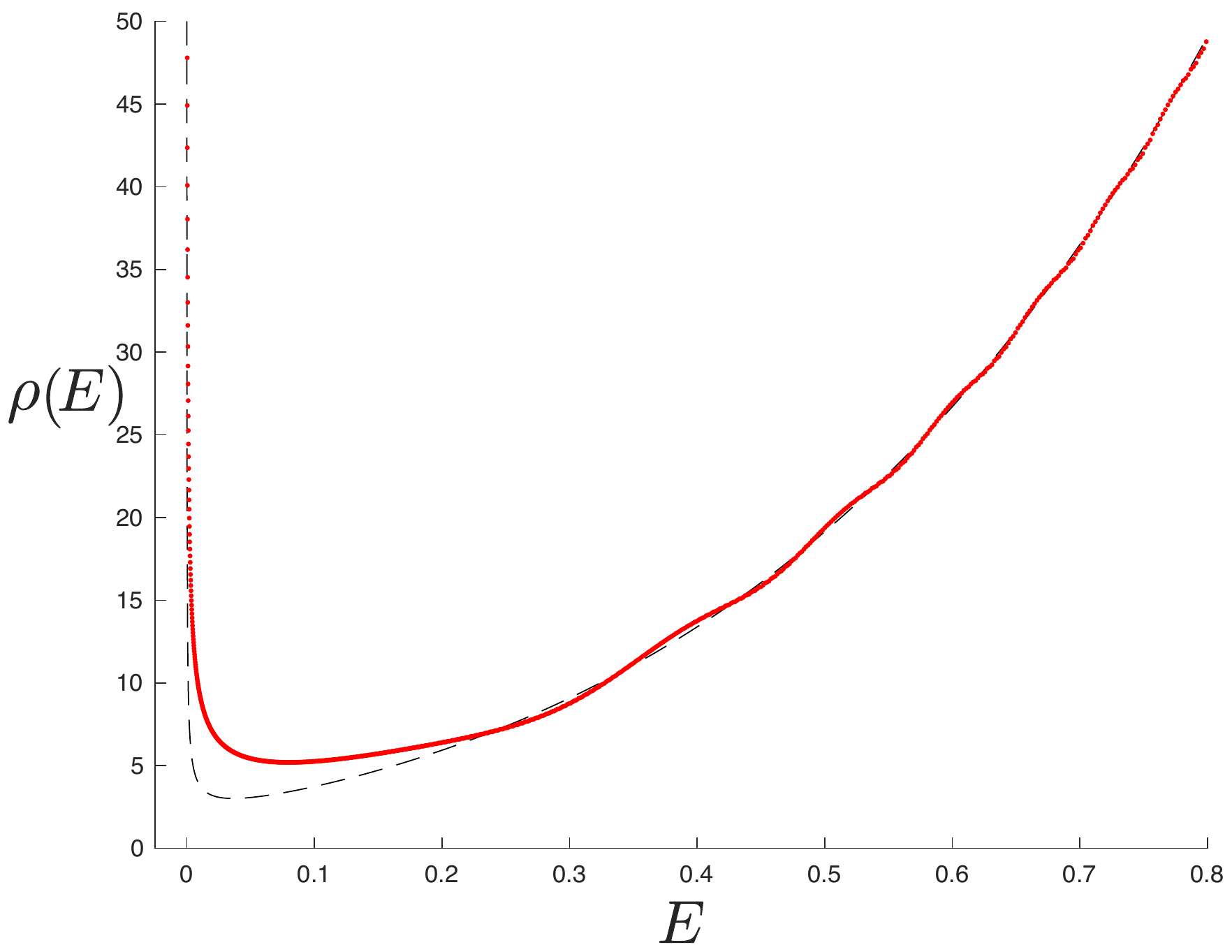}
\caption{\label{fig:non-pert-sd2} The full spectral density (made out of red dots), computed using the methods of this paper, for the (0,2) model of JT supergravity. The dashed blue line is the disc level result of equation~(\ref{eq:spectral_SJT}). Here, $\hbar{=}1$. The companion result for the (2,2) model is in figure~\ref{fig:non-pert-sd1}.}
\end{figure}

The outcome of the numerical spectrum solving was approximately 1000 accurate normalized wavefunctions and their energies, for use in constructing the density in this section, and the spectral form factor in the next. Using them, the spectral density can be  constructed using a simple trapezoidal integration to implement equation~(\ref{eq:spectral-density-integration})  and the result (for $\mu{=}1$) is shown in figure~\ref{fig:non-pert-sd1} for $\Gamma{=}\frac12$ (shown on page~\pageref{fig:non-pert-sd1}) and figure~\ref{fig:non-pert-sd2} for $\Gamma{=}{-}\frac12$. 
Plotted alongside the (dense) line of dots (the computed spectral data) is a dashed line showing the disc level spectral density. Strikingly,  the $1/\sqrt{E}$ divergence present at disc level is erased entirely by non--perturbative effects in the $\Gamma{=}\frac12$ (2,2) case, but not in the $\Gamma{=}{-}\frac12$ (0,2) case. As discussed in refs.~\cite{Stanford:2019vob,Johnson:2020heh}, for $\Gamma{=}{\pm}\frac12$, there are no perturbative corrections to the spectral density beyond the disc, so all the differences here are due to non--perturbative physics, which makes these JT supergravity cases particularly interesting to study when investigating the results of non--perturbative physics. This stark difference will be reflected in comparisons of the spectral form factor, to be studied in the next section. 

Satisfyingly, what has appeared here for the full (2,2) and (0,2) cases are actually  grown--up versions of what was discussed for two baby (Bessel) models in ref.~\cite{Stanford:2019vob}, and so this constitutes a nice  consistency check of the methods of this paper.  The Bessel special cases are models of the very low energy tip of the spectral density---the part that (classically) pushes up against the $E{=}0$ wall, as mentioned earlier.    Generalizing to include arbitrary positive~$\mu$ (see also refs.~\cite{Johnson:2019eik,Johnson:2020heh}), the full non--perturbative Bessel spectral density is (after a change of conventions from those papers):
\be
\label{eq:bessel-spectral-mu}
\rho(E) = \frac{\mu}{\pi\hbar\sqrt{E}} + \frac{\sin\left(-\frac{\pi}{2}\alpha+\frac{2\mu\sqrt{E}}{\hbar}\right)}{2\pi E}\ ,
\ee
where (as a reminder)  $\alpha{=}2\Gamma{+}1$ takes the values $0$ and~$2$. The leading part is the disc contribution and all other perturbative contributions vanish. The oscillating term is the full non--perturbative contribution. At $E{=}0$ for $\alpha{=}2$,  ($\Gamma{=}\frac12$) there is the aforementioned cancellation between the divergent disc contribution and the $E{\to}0$ piece of the oscillating non--perturbative part. This is what is  now seen to be present (as it ought to be) in the full JT supergravity spectral density computed explicitly above. 

\subsection{A Special Formula, and a Tale of Instantons}
Actually, it is possible to go considerably further than the Bessel tail comparison.  The structure of the Bessel models above (along with various other technical features of the matrix model recursion relations)
led Stanford and Witten to suggest (see appendix~E of ref.~\cite{Stanford:2019vob}) that more generally, the JT supergravity spectral density for the (0,2) and (2,2) cases should closely approximate  the form\footnote{The author thanks Douglas Stanford for asking a helpful question about this issue after the 1$^{\rm st}$  version of this manuscript appeared.}:
\be
\label{eq:spectral-density-conjecture}
\rho(E) \approx \rhoo(E)+ \frac{\sin\left(-\frac{\pi}{2}\alpha+\pi\int_0^E \rhoo(E^\prime) dE^\prime\right)}{2\pi E}\ .
\ee
(Note that their conventions were adapted to match  those of the current paper.) Again, the perturbative terms beyond the disc vanish.  The argument of the sinusoidal non--perturbative piece is determined in terms of the disc contribution\footnote{In fact, it is easy to guess generalizations of this formula for other half--integer $\Gamma$, which seem to give sensible physics using the present methods. It would be interesting to test them.}. Crucially, they note that this does not include possible instanton corrections, non--perturbative physics of a form not accounted for by the sinusoidal modulation (hence the use of the $\approx$ sign in the above). 

\begin{figure}[h]
\centering
\includegraphics[width=0.48\textwidth]{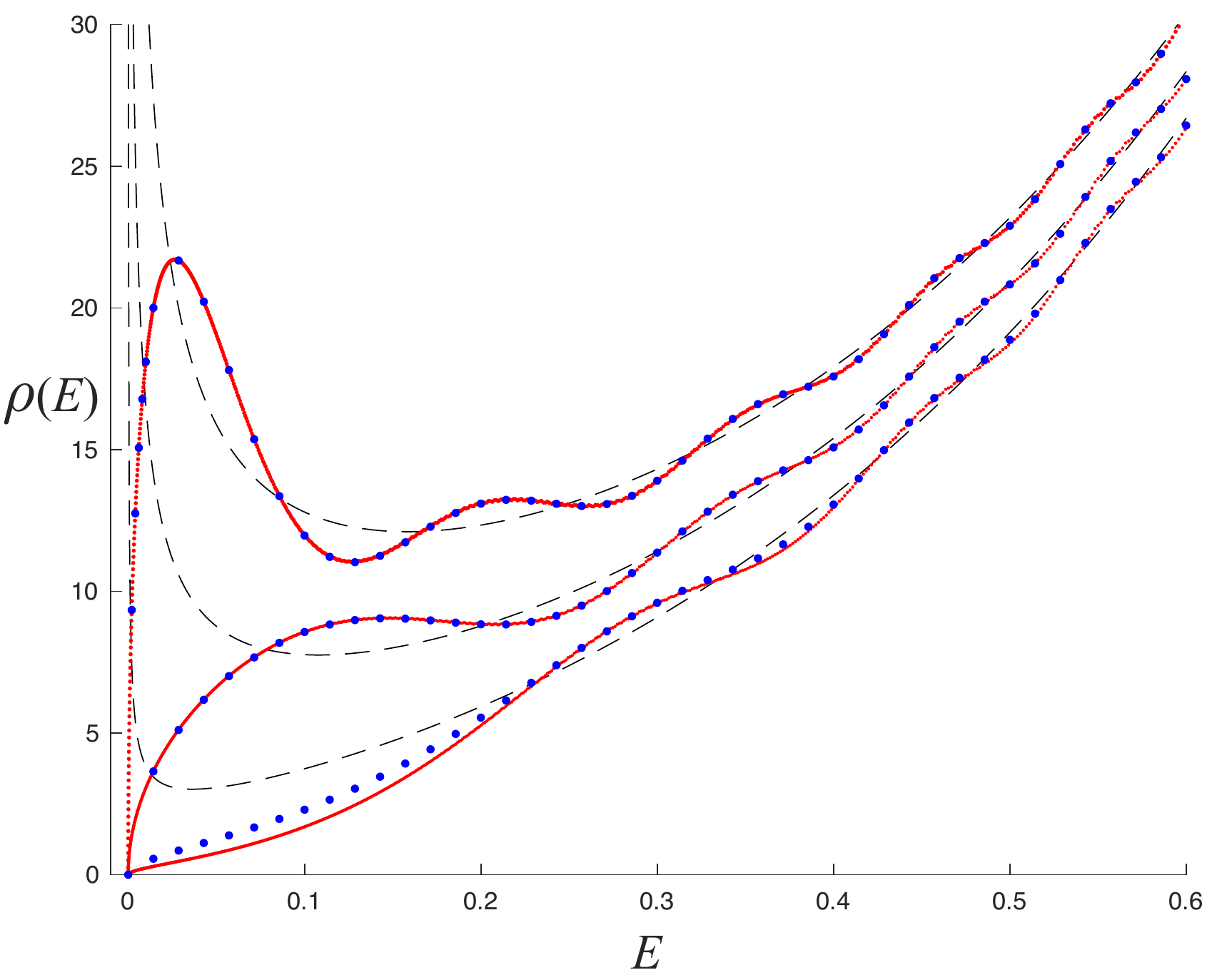}
\caption{\label{fig:non-pert-sd-compare-1} Comparison of the full spectral density for the (2,2) model of JT supergravity, with   equation~(\ref{eq:spectral-density-conjecture}) for the leading form (dark blue dots). The dashed  line is the disc level result~(\ref{eq:spectral_SJT_mu}). The cases~$\mu{=}1,5$ and 10 are shown (lower to upper). Here $\hbar{=}1$.}
\end{figure}

\begin{figure}[h]
\centering
\includegraphics[width=0.48\textwidth]{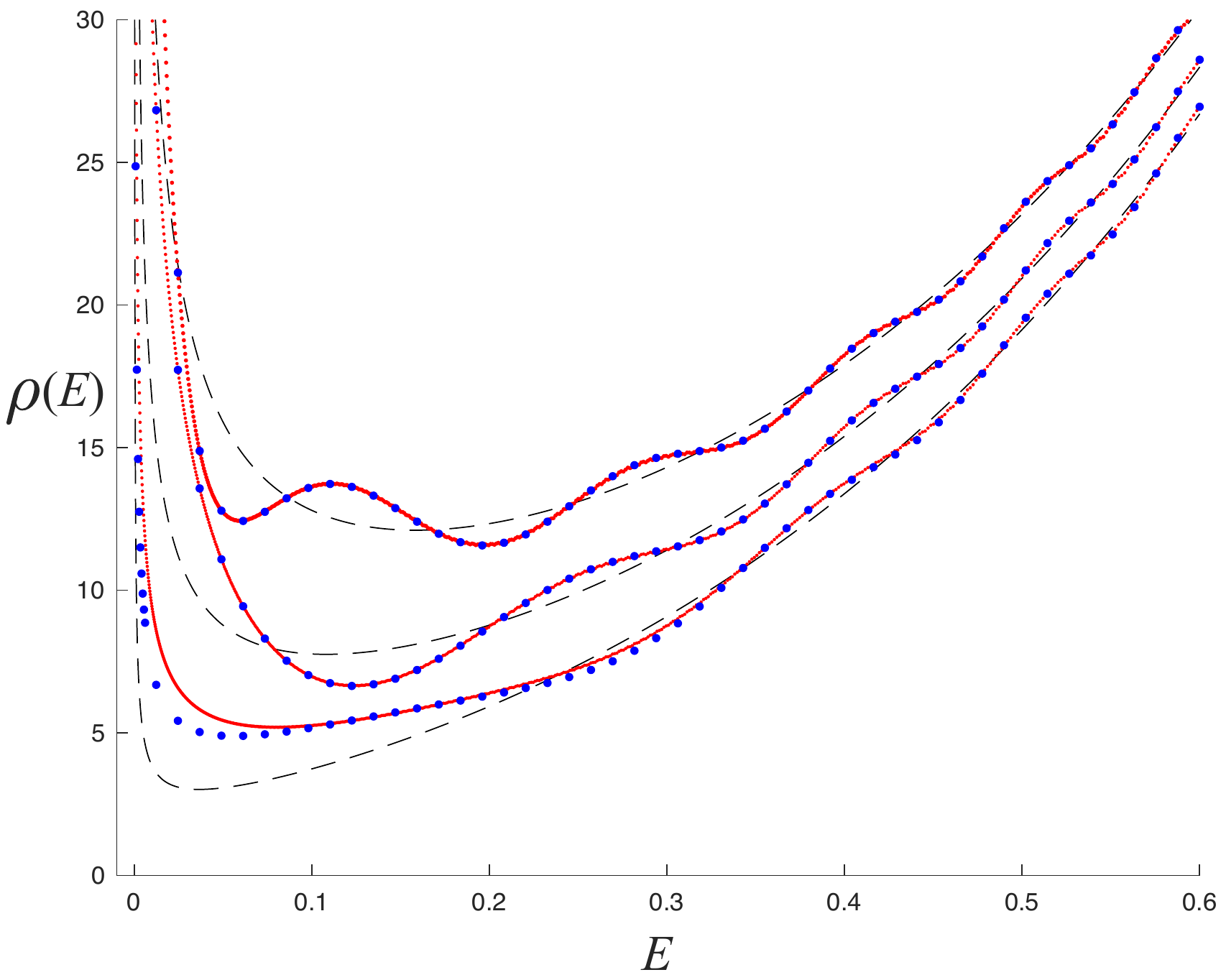}
\caption{\label{fig:non-pert-sd-compare-2} Comparison of the full spectral density for the (0,2) model of JT supergravity, with  equation~(\ref{eq:spectral-density-conjecture}) for the leading form (dark blue dots). The dashed  line is the disc level result~(\ref{eq:spectral_SJT_mu}). The cases~$\mu{=}1,5$ and 10 are shown (lower to upper). Here $\hbar{=}1$.}
\end{figure}

While the expression above was written for the specific value $\mu{=}1$ ($\mu$ is not a variable parameter in ref.~\cite{Stanford:2019vob}), other values of~$\mu$ (see below equation~(\ref{eq:spectral-density-integration}) for its explanation) are quite readily incorporated by this form. This is already obvious for the  simple Bessel prototype~(\ref{eq:bessel-spectral-mu}), but further evidence comes from inserting into it the expression~(\ref{eq:spectral_SJT_mu}) for $\rhoo(E,\mu)$ written earlier for the disc level JT supergravity density for other $\mu$. Plotting this analytical result against  the numerical results for the spectral density  obtained using this paper's methods yields a remarkable agreement, as shown in figures~\ref{fig:non-pert-sd-compare-1} and~\ref{fig:non-pert-sd-compare-2} for $\Gamma{=}\frac12$ and $\Gamma{=}{-}\frac12$, for the values $\mu{=}1,5,$ and~$10$. As a reminder, the red lines are each made of $\sim$1800 points individually computed using the truncation and numerical methods outlined, while the blue dots are samples of the analytical formula~(\ref{eq:spectral-density-conjecture}). It is very striking how well they agree.  

The areas of {\it disagreement} suggest interesting physics, in fact. For smaller $\mu$ ({\it e.g.}, see the case of $\mu{=}1$ in each figure) it is easy to see  disagreement. It is especially quite visible  at smaller energies (although at $E{=}0$ they  agree as they must, since this is covered by the Bessel cases above).  The clear pattern in the discrepancy is suggestive of physics and not numerical inaccuracies. 

For example, the fact that the disagreement rapidly disappears as $\mu$ increases is striking. Nothing about the numerical methods should show this
sort of systematic dependence. As a further example, the form of the disagreement suggests that there is a change in the amplitude of the sinusoidal modulation,
 again a pattern that is hard to reproduce with mere numerical systematics. A natural guess is that there are further multiplicative factors in the non--perturbative parts that are not present in the analytic formula. They must be small, and of an instanton form (as already anticipated in remarks in ref.~\cite{Stanford:2019vob}). 
 
In the $k$th minimal model the height of the effective potential barrier for one eigenvalue in the background of all the others yields the action of the instanton effects~\cite{Shenker:1990uf} (they are ``ZZ--branes'' in the modern parlance~\cite{Zamolodchikov:2001ah}). Alternatively, it can be computed using a WKB analysis of the string equation itself. This was all worked out long ago for the complex matrix models in ref.~\cite{Dalley:1992vr} (see ref.~\cite{Johnson:2004ut} for a review and more contemporary presentation), and the dependence of the action on $\mu$ and $\hbar$ is $(\frac{4k}{2k+1})\mu^{1+\frac{1}{2k}}/\hbar$. Therefore  a dependence $\exp(-2\mu/\hbar)$ should be expected for instanton corrections to JT supergravity constructed out of minimal models in the manner of this paper. 

To test this, a rough estimate of the dependence was done by measuring the size of the deviations of the type seen in the figures (\ref{fig:non-pert-sd-compare-1} and~\ref{fig:non-pert-sd-compare-2}) for some sample cases,  as~$\mu$ varied from $0$ to $5$. The deviations fall rapidly in that range (as is evident from the figures), and a logarithmic plot of them against $\mu$ yielded a straight line of negative unit slope to good (better than~1\%) accuracy. While this can be made more precise, it is already a strong indication that the instanton effects of the expected form are present and accounted for. It will be interesting to explore this further, but this will be left for later work.

\subsection{Going to Weak Coupling}

Before moving on it is worth showing the effect of going to weak coupling by reducing $\hbar$. The spectra shown so far are for $\hbar{=}1$, and working in this regime has the advantage of making much of the non--perturbative physics readily visible. Nevertheless, it is useful to be able to gain access to weaker coupling, not the least because  certain dual systems of interest (such as SYK--type models, or large black holes (where $S_0$ is large)) may be in that regime. 

\begin{figure}[h]
\centering
\includegraphics[width=0.5\textwidth]{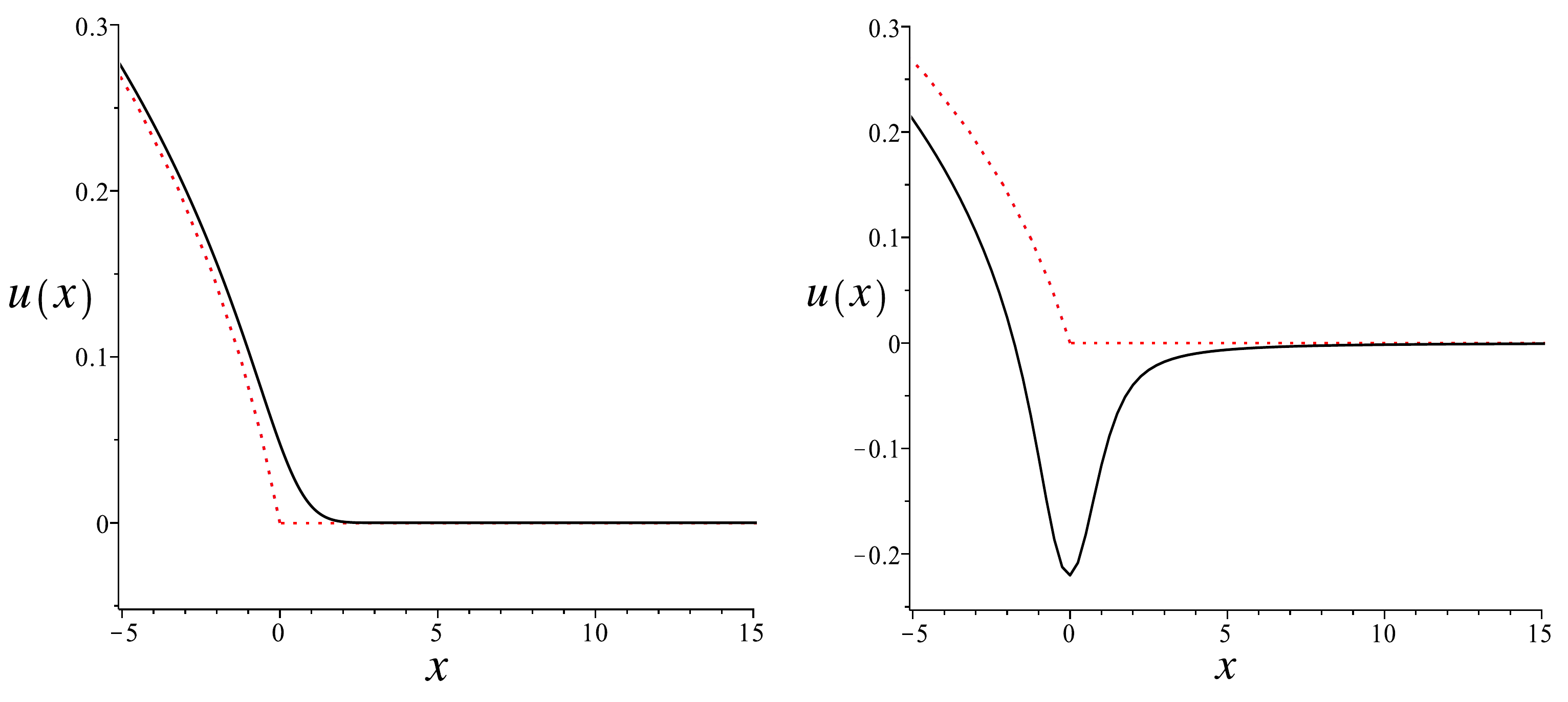}
\caption{\label{fig:truncation-examples-X}  The solid lines  show close--ups of the $\Gamma{=}\frac12$ or (2,2) solution (left) and the $\Gamma{=}{-}\frac12$ or (0,2) solution (right) of the string equation for truncation up to $t_6$, for $\hbar{=}\frac15$. For comparison, the full classical solution  is shown too (dotted).}
\end{figure}

It is possible to reduce the value of $\hbar$, but it is at the expense of not being able to access as high energies in the spectrum for a given truncation, unless there is a compensation in terms of increased computational effort (working on a finer grid and producing more wavefunctions). Additionally, solving the string equation takes more care, as already mentioned, because the derivatives are more in play. Nevertheless, progress is possible. For example, dialling  down to {\it e.g.,} $\hbar{=}\frac15$, the string equations can be coaxed to produce the solutions shown (just as close--ups, for brevity) in figure~\ref{fig:truncation-examples-X}. These should be compared to what was shown in the insets of figures~\ref{fig:truncation-examples-A} and~\ref{fig:truncation-examples-B}. For the (2,2) case, the curve turns the corner more sharply (showing the increased role of the derivatives), while for the (0,2) case, the well is far deeper and narrower (these two features come together to continue to ensure there are no bound states). 

It is easy to imagine how these developments continue in order to get to the classical limit $\hbar{\to}0$ (the red dotted line in the figures). The (2,2) curve turns the corner ever more sharply, and for the (0,2) curve the well deepens and narrows and eventually is entirely squeezed away. 

\begin{figure}[h]
\centering
\includegraphics[width=0.48\textwidth]{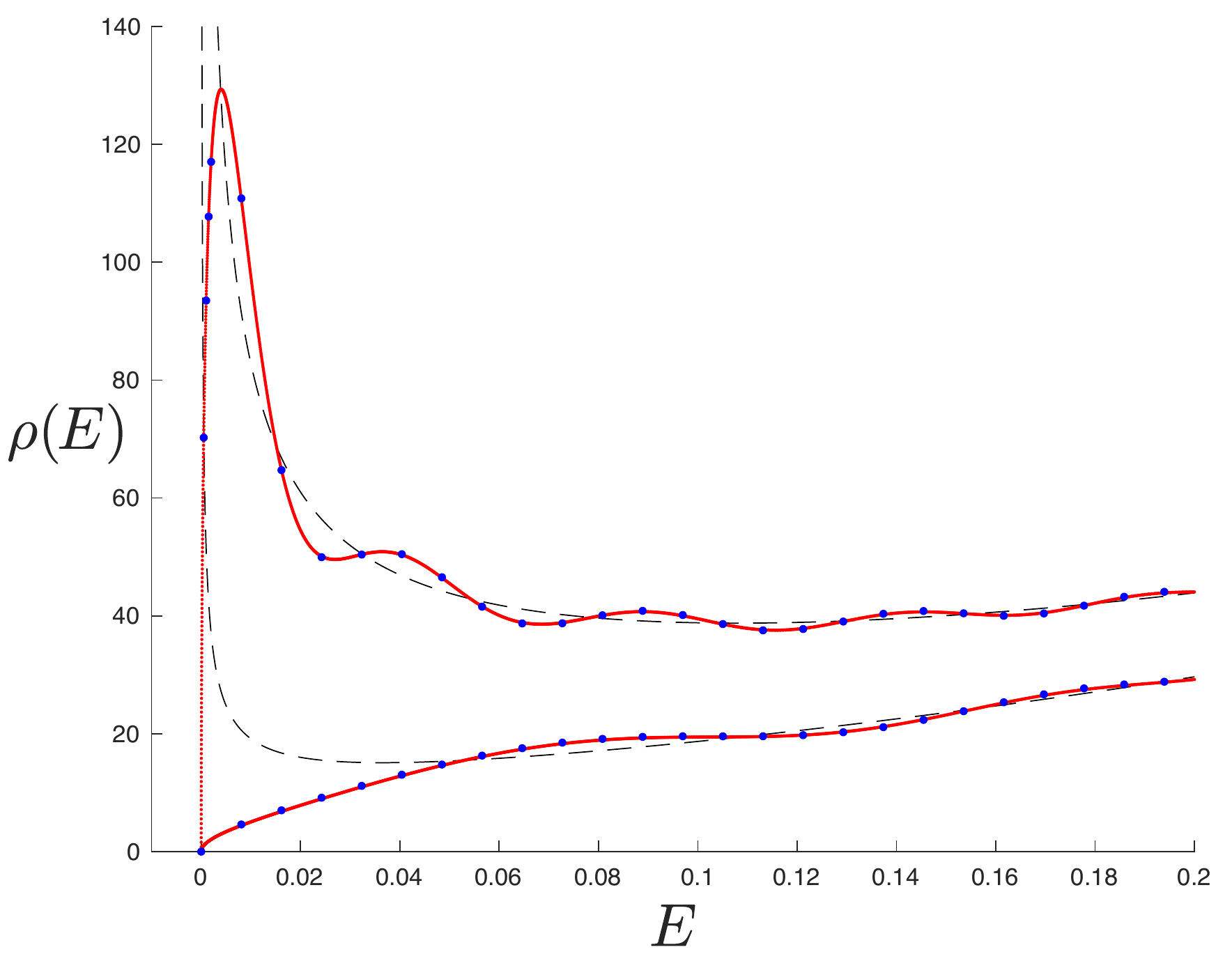}
\caption{\label{fig:non-pert-sd1-alt} Two examples ($\mu{=}1$, lower; $\mu{=}5$, upper) of the full spectral density for the (2,2) model of JT supergravity, for $\hbar$ reduced to $\frac15$. The (blue) dots are points from the equation~(\ref{eq:spectral-density-conjecture}). In each case the dashed  line is the disc  result of equation~(\ref{eq:spectral_SJT_mu}). {\it c.f.} the $\hbar{=}1$ cases of figure~\ref{fig:non-pert-sd-compare-1}.}
\end{figure}

The  techniques already described in the previous two sections can be carried out to study the spectral density, and the result in figure~\ref{fig:non-pert-sd1-alt} displays the results (red dots) for the $\Gamma{=}\frac12$ or (2,2) case, for the cases $\mu{=}1$ and~5. (The $\mu{=}10$ case, which rises to 508 vertically, was omitted for clarity.) Again, the blue dots are samples of the analytic formula~(\ref{eq:spectral-density-conjecture}) with~(\ref{eq:spectral_SJT_mu}) input for $\rhoo(E,\mu)$ (now $\hbar{=}\frac15$ is inserted), showing again how well this procedure works. Notice that, in accord with the fact that the coupling is weaker, the $\exp(-\mu/\hbar)$ instanton effects described in the previous section, which would appear as deviations of the red curves from the blue dots, are  too small to seen here. 

Comparing  to the $\hbar{=}1$ curves in  figure~\ref{fig:non-pert-sd-compare-1}, the quantum undulations are smaller, for generic $E$, and the curves stray less from the classical curve, beginning marked deviations only at  lower energies compared to the $\hbar{=}1$ situation, until the curve falls quickly to zero. This is as it should be.

\section{The Spectral Form Factor}
\label{sec:spectral-form-factor}

\subsection{General Remarks}
As mentioned  in the brief review of section~\ref{sec:JT-gravity-summary}, the spectral form factor  is derived from the  two point function of $Z(\beta)$. This has two parts, a disconnected piece $\langle Z(\beta)\rangle\langle Z(\beta^\prime)\rangle$ and a connected piece $\langle Z(\beta)  Z(\beta^\prime)\rangle$. In the old matrix model  language, the connected piece is the connected correlator of two ``macroscopic loops'', and  this is readily written down as~\cite{Banks:1990df,Ginsparg:1993is}:
\begin{eqnarray}
\label{eq:correlator-connected}
\langle Z(\beta) Z(\beta^\prime)\rangle &=& {\rm Tr}(e^{-\beta{\cal H}}(1-{\cal P}) e^{-\beta^\prime{\cal H}}{\cal P}) \\
&=& {\rm Tr}(e^{-(\beta+\beta^\prime){\cal H}}) - {\rm Tr}(e^{-\beta{\cal H}}{\cal P} e^{-\beta^\prime{\cal H}}{\cal P})\nonumber \\ 
&=&  Z(\beta{+}\beta^\prime) -\!\! \int \!\!dE\!\int \!\!dE^\prime \rho(E,E^\prime) \rho^*\!(E^\prime,E)
\ ,\nonumber
\end{eqnarray}
where ${\cal P}$ is discussed below equation~(\ref{eq:partition-function-from-H}) and 
\be
\rho(E,E^\prime) = \int_{-\infty}^\mu\!\!  dx \, \psi^*\!(x,E)\psi(x,E^\prime) \ .
\ee
The quantity $\mu$, discussed in the previous section, will be set to unity henceforth, corresponding to having the disc level spectral density given in equation~(\ref{eq:spectral_SJT}) (without the overall $\sqrt{2}$).

\subsection{A Phase transtition}
Set $\beta{=}\beta^\prime$ for a while. Generically, the disconnected part (corresponding to two black holes) and the connected part (a wormhole) are worth studying in their own right as distinct sectors of the quantum gravity that compete for dominance~\cite{Maldacena:2018lmt} as a function of $\beta$. The disconnected part, being the square of the partition function, rapidly decreases with increasing $\beta$ while the connected part increases. At some point $\beta_{\rm cr}$ there is a transition, and the connected diagram becomes more dominant. This is also true in the complete (not just perturbative) theory discussed here.  
\begin{figure}[h]
\centering
\includegraphics[width=0.48\textwidth]{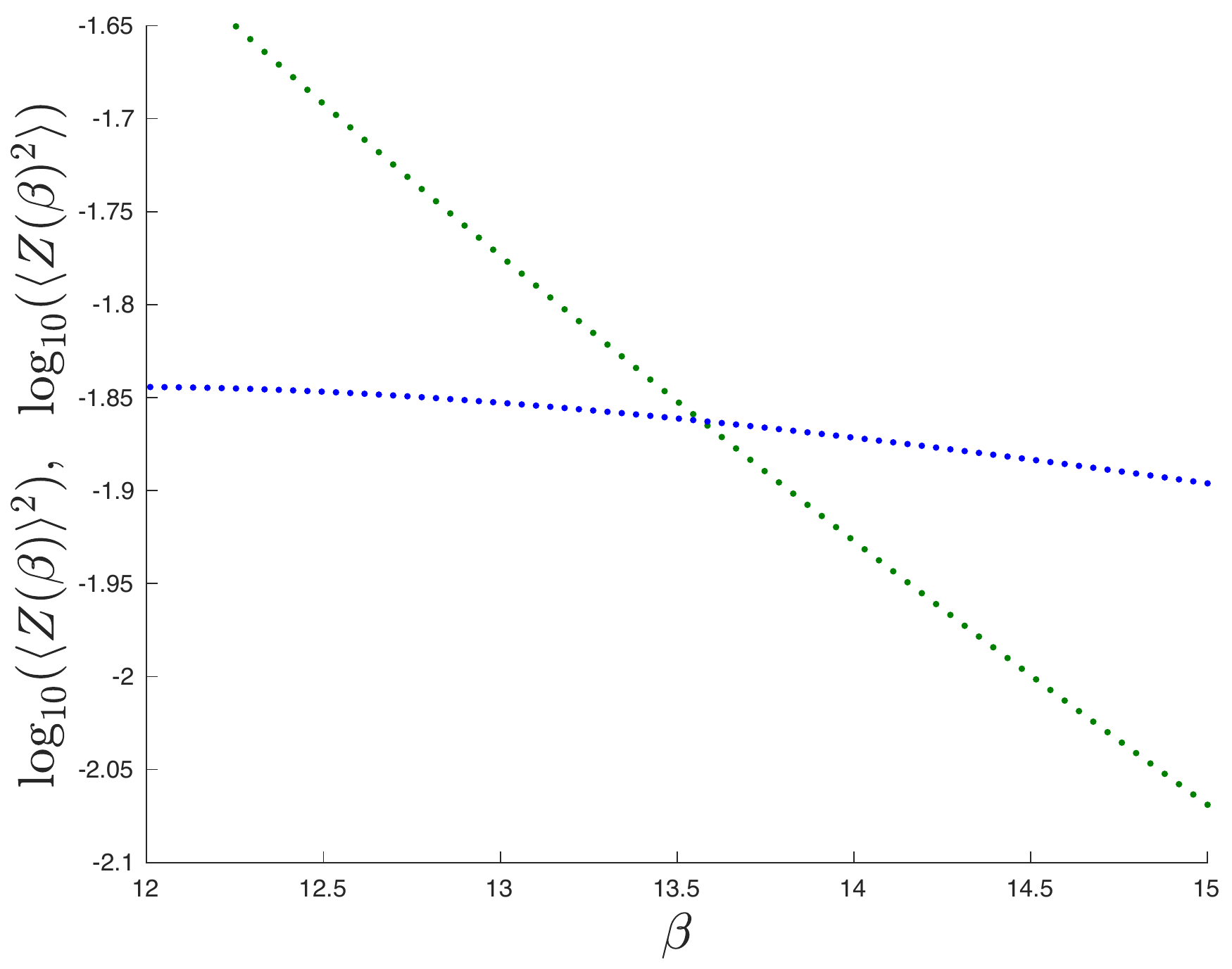}
\caption{\label{fig:phase-transition} The disconnected (starting higher at the left) {\it vs.} the connected (lower) two--point function of the partition function as a function of $\beta$ at $\hbar{=}1$, showing a phase transition at $\beta_{\rm cr}{\simeq}13.56$.}
\end{figure}
This is all nicely under control in the current definitions of JT supergravity. The spectrum has been computed in the previous section and so all the elements in equation~(\ref{eq:correlator-connected}) are readily computable to the desired accuracy. Figure~\ref{fig:phase-transition} shows a plot of the (log$_{10}$ of the) connected and disconnected pieces {\it ---with all perturbative and non--perturbative contributions included---} as a function of $\beta$, for the (2,2) JT supergravity case, showing the transition at $\beta_{\rm cr}{=}13.56$. A similar computation (yielding a similar graph, omitted) shows that $\beta_{\rm cr}{=}28.30$ for the (0,2) JT supergravity case.  

Notably, for the $(0,2)$ case the amplitudes are over an order of magnitude larger. This is a striking effect attributable entirely to  non--perturbative effects. The (2,2) JT supergravity has, as mentioned in the previous section~\ref{sec:spectral-density}, non--perturbative effects that cancel the $1/\sqrt{E}$ behaviour at low energy, coming from the leading disc amplitude. The (0,2) version does not cancel this away, and so while to larger $E$ the two models' spectra are roughly similar (see figures~\ref{fig:non-pert-sd1} and~\ref{fig:non-pert-sd2}),  there is an enhancement at low~$E$ that means  much larger contributions to the partition function for any fixed $\beta$. This marked difference between the non--perturbative physics will make a major appearance in  the temporal behaviour of the spectral form factor too, studied next.

 \subsection{Time Dependence}

\begin{figure}[h]
\centering
\includegraphics[width=0.48\textwidth]{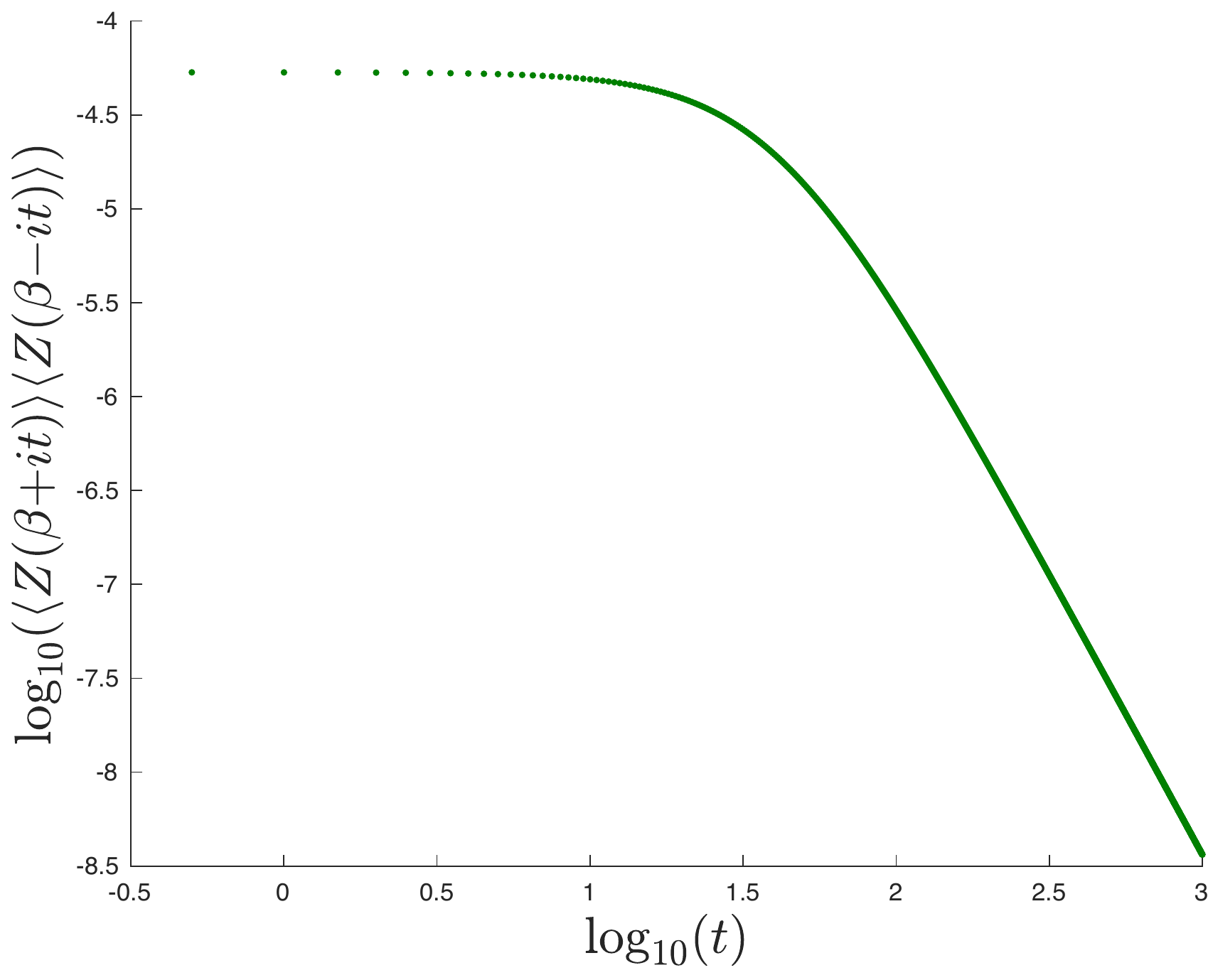}
\caption{\label{fig:disconnected-sff-SJT1} The disconnected part of the  (2,2) JT supergravity spectral density  function {\it vs.} $t$, at $\beta{=}50$ and $\hbar{=}1$, showing the classic slope feature.}
\end{figure}

The above  computation served as a useful guide for what to expect for the spectral form factor, which tracks the correlation function over time. This is done by putting $\beta{\to}\beta{+}it$ and $\beta^\prime{\to}\beta{-}it$, with $\beta$ fixed, studying the dependence on $t$. The fixed $\beta$ can be above or below~$\beta_{\rm cr}$. The disconnected part will start out as the squared partition function and then decrease with $t$. This is the ``slope'' behaviour of the spectral form factor. On the other hand, looking at equation~(\ref{eq:correlator-connected}), it can be seen that the connected part has a $t$--independent part, $Z(2\beta)$, from which is subtracted a positive piece which gets small at large $t$, with only significant contributions from energies that are close to each other. The value of $Z(2\beta)$ therefore sets the height of the universal ``plateau'' feature and the  approach to it, the ``ramp'' has its size set by how rapidly the energy correlations die away at large~$t$. The ``dip'' region is formed by the process of handing over from the decreasing disconnected part to the increasing connected part. The formalism here allows, using the complete package of almost 1000 good wavefunctions and energies, for this all to be computed to good accuracy for JT supergravity (and for a non--perturbatively well--behaved definition of JT gravity in section~\ref{sec:JT-gravity-regular}) for the first time. Using again the JT supergravity example, figure~\ref{fig:disconnected-sff-SJT1} shows the disconnected contribution to the spectral density function, for $\beta{=}50{>}\beta_{\rm cr}$, with $\log_{10}$ axes. 

\begin{figure}[h]
\centering
\includegraphics[width=0.48\textwidth]{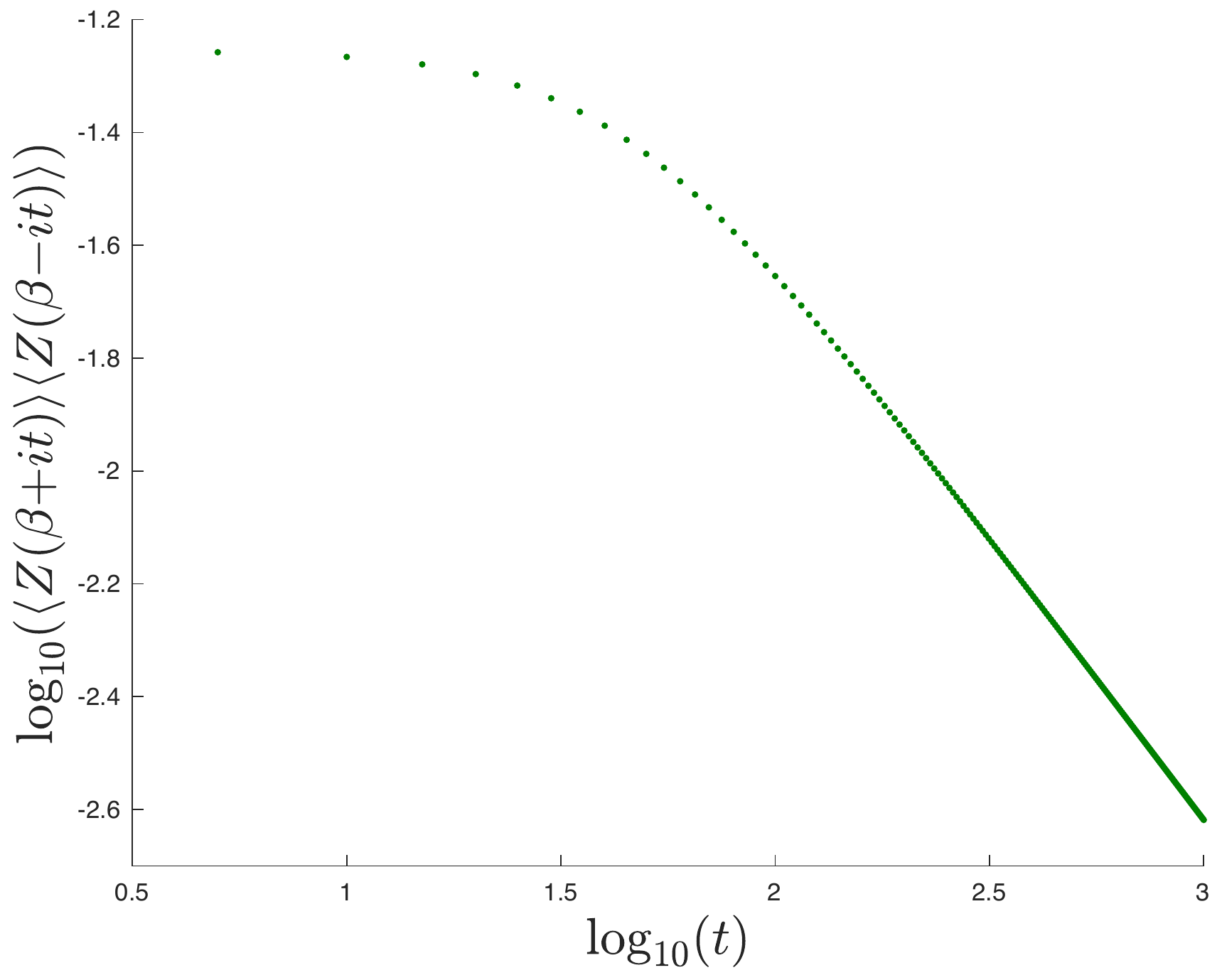}
\caption{\label{fig:disconnected-sff-SJT2} The disconnected part of the  (0,2) JT supergravity spectral density  function {\it vs.} $t$, at $\beta{=}50$ and $\hbar{=}1$.}
\end{figure}

The slope behaviour is quite evident. Since the axes are logarithmic, it is easy to see from  the figure that the slope of it is roughly ${-3}$, suggesting a fall--off of ${\sim}t^{-3}$.  This might seem surprising since it is similar to the perturbative fall off rate for ordinary JT gravity (see ref.~\cite{Cotler:2016fpe}). The reason for this faster rate is clear, and  again attributable to  non--perturbative physics. The slope's fall--off is controlled by the behaviour of the  endpoint of the spectral density. While at the disc level for JT supergravity  $\rho(E){\sim}1/\sqrt{E}$ there,  producing a $t^{-1}$ fall--off~\cite{Hunter-Jones:2017crg}, the fact that in the (2,2) case non--perturbative corrections  remove this $1/\sqrt{E}$ behaviour results in the faster fall off more usually associated with the ordinary JT case (and Hermitian matrix models). There are far fewer states in the vicinity of the endpoint. This reasoning predicts that for the (0,2) supergravity, the slope should be closer to a $t^{-1}$ fall-off.  Indeed,  this is clear from the behaviour of the disconnected piece for (0,2) supergravity, shown in  figure~\ref{fig:disconnected-sff-SJT2}, for $\beta{=}50$.  
There, the slope of the  linear part shows $t^{-1}$ fall off. Continuing in line with these expectations is a computation of the same quantity for the $\hbar{=}1/5$ case mentioned earlier. The spectrum was shown in figure~\ref{fig:non-pert-sd1-alt}, and from there it is natural to guess that the fall-off would be even faster since the non--perturbative effects have scooped away even more states near the endpoint. A check showed that this is indeed correct, although another figure will not be presented to display the result, to save repetition.

\begin{figure}[h]
\centering
\includegraphics[width=0.48\textwidth]{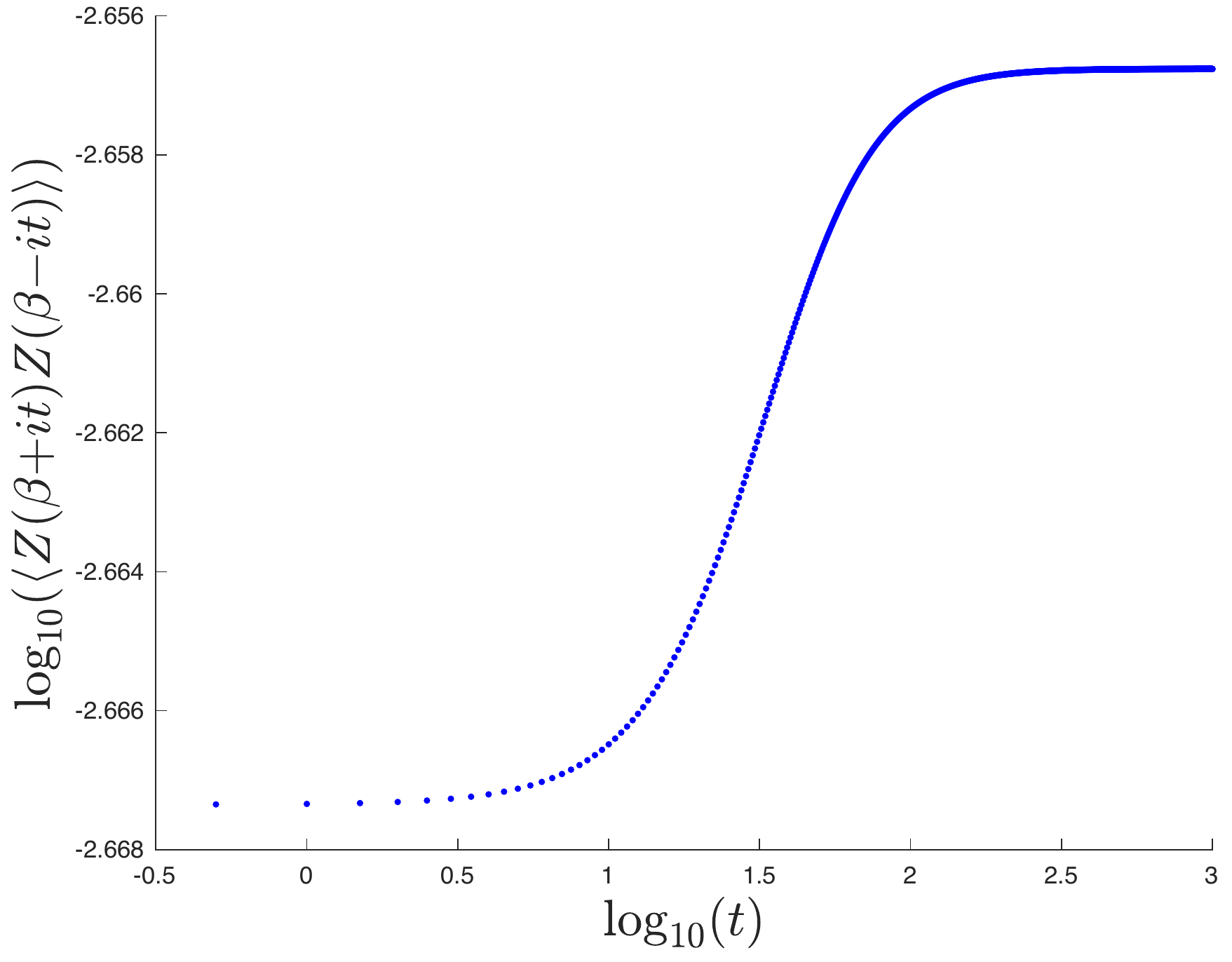}
\caption{\label{fig:connected-sff-SJT1} The connected part of the (2,2) JT supergravity  spectral density  function {\it vs.} $t$, at $\beta{=}50$ and $\hbar{=}1$, showing the classic ramp and plateau features.}
\end{figure}

Moving to the connected contribution's time dependence, the (2,2) case is shown in  figure~\ref{fig:connected-sff-SJT1} for the same value of~$\beta{=}50$.   
The ramp and plateau structures, and the transition between them, are visible. Strikingly, the  rise to the plateau is very short--lived, the ramp regime rising only a small (on the logarithmic scale) amount before transitioning to the plateau. 

The beginning shape of the ramp is already anticipated in the perturbative answer, ${\sim}\beta^{-1}\sqrt{\beta^2+t^2}$,  long known for two--macroscopic--loop correlators~\cite{Ginsparg:1993is} (appropriately continued to yield the $t$--dependence~\cite{Saad:2019lba}), but there are strong non--perturbative corrections such that before the long--time linear part can manifest, the other effects turn the ramp over into the plateau.    

\begin{figure}[h]
\centering
\includegraphics[width=0.48\textwidth]{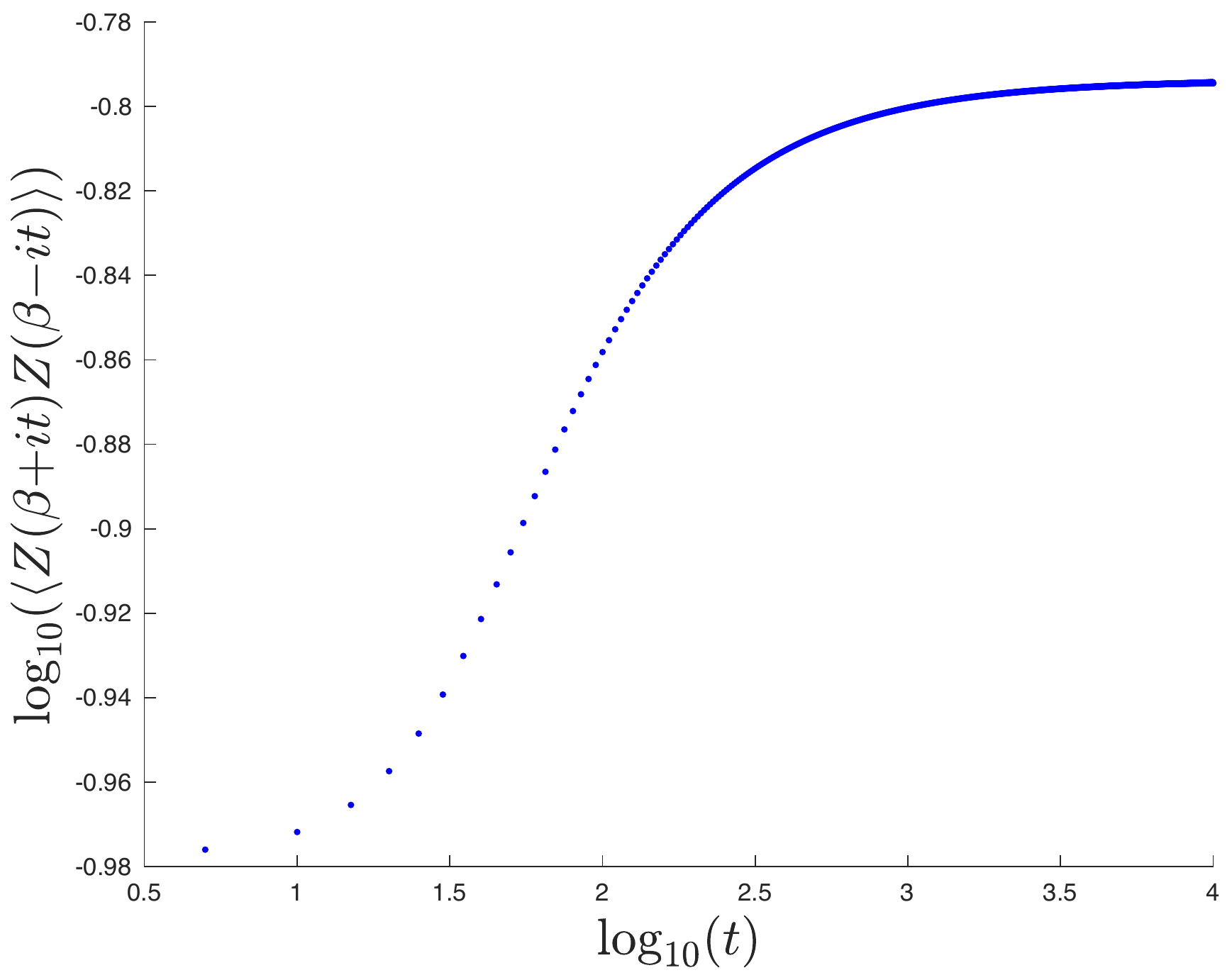}
\caption{\label{fig:connected-sff-SJT2} The connected part of the (0,2) JT supergravity  spectral density  function {\it vs.} $t$, at $\beta{=}50$ and $\hbar{=}1$.}
\end{figure}

A similar story is told, initially, by the  ramp shape seen for the (0,2) case, shown in figure~\ref{fig:connected-sff-SJT2} (again for $\beta{=}50$). However, there's a new feature. The rise is indeed slow, but it is remarkably slow. After almost two orders of magnitude more time has elapsed, as compared to the (2,2) case, the saturation to the plateau has still not quite completed. 

This is a rather novel feature of this case, and worth further investigation. 
The origin of this physics is most likely again to be attributed to the peculiar pileup of states in the vicinity of $E{=}0$ that this model has. There's an endless supply of closely spaced low--lying states contributing to the part of the form factor that subtracts from the saturation value $Z(2\beta)$ (see equation~(\ref{eq:correlator-connected})). At longer and longer times there are even more low lying states to contribute, and  still closely spaced, maintaining their effect of slowing the saturation.

Turning back to the ramp itself, note that this was (for both (2,2) and (0,2))  for the case $\hbar{=}1$. At smaller values of $\hbar$ there is  more time for the ramp to develop, with an increased rate of rise before the turnover.  Note however that   it is only for  extremely small $\hbar$ that the linear behaviour of the ramp has a chance to appear.  

There are two important lessons here. The first is that associating the ramp with linear behaviour (as is sometimes done in the literature) is maybe not the most  accurate descriptor. The second is that non--perturbative effects can enhance the appearance of the ramp in the JT supergravity case (in the full spectral form factor made by taking the sum of disconnected and connected parts, even though perturbative expectations might have suggested a reduction~\cite{Hunter-Jones:2017crg}. The potential reduction of the ramp feature is based on the idea that a slow $t^{-1}$ rate of fall might not give the ramp time to develop before the plateau sets in. In fact, non--perturbative effects are seen  here to produce a rapid fall (sometimes faster even than the perturbative bosonic $t^{-3}$) giving plenty of time to develop a sharp ``dip'', a clear ramp, and a smart turnover into the plateau  for the (2,2) case.

\begin{figure}[h]
\centering
\includegraphics[width=0.45\textwidth]{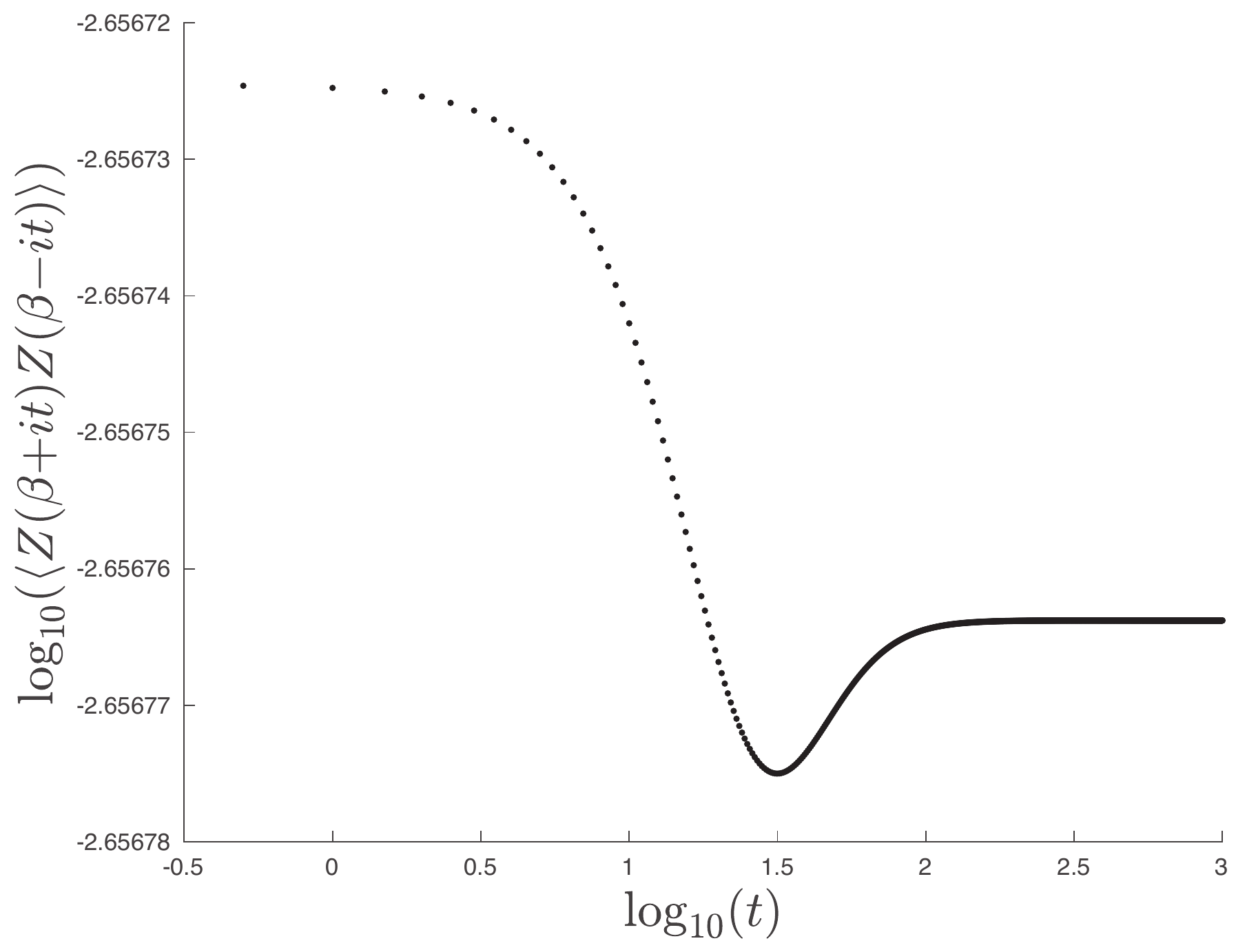}
\caption{\label{fig:combined-sff-SJT1} The full spectral form factor for the (2,2) model of JT supergravity at $\beta{=}50$, $\hbar{=}1$. (The case of $\hbar{=}1/5$ is in figure~\ref{fig:combined-sff-SJT1-alt}).}
\end{figure}

The sum of the connected and disconnected pieces gives, for (2,2) supergravity, the classic saxophone shape\footnote{The shape deserves a name, and saxophone seems a good choice, to balance out the many uses of the name ``trumpet'' in other aspects of JT gravity.} known from studying spectral density functions in a wide range of contexts. It is displayed in figure~\ref{fig:combined-sff-SJT1}.
This is for $\hbar{=}1$, and the case of   $\hbar{=}1/5$ was already presented in figure~\ref{fig:combined-sff-SJT1-alt}, at  $\beta{=}35$. The latter, being at smaller~$\beta$
and $\hbar$, is larger overall and develops a wider variation. The linear part of the ramp would be even more visible for lower values of these parameters.
 
\begin{figure}[h]
\centering
\includegraphics[width=0.48\textwidth]{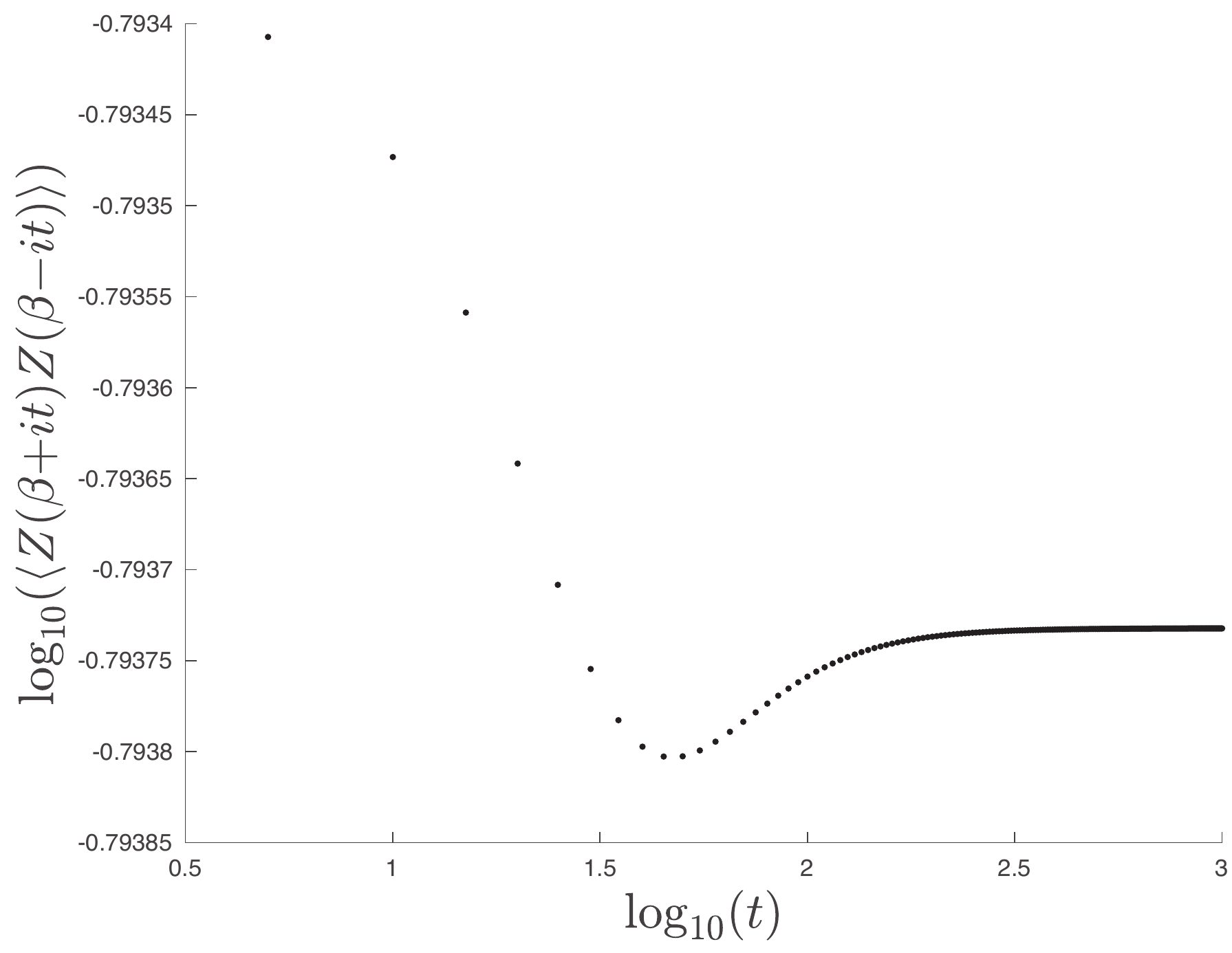}
\caption{\label{fig:non-pert-sff-SJT2} The full spectral form factor for (0,2) JT supergravity. Here, $\beta{=}50$ and $\hbar{=}1$. {\it c.f.} the (2,2) case  in figure~\ref{fig:combined-sff-SJT1}.  }
\end{figure}

For   $(0,2)$ JT supergravity, figure~\ref{fig:non-pert-sff-SJT2}  shows the resulting $\hbar{=}1$ full spectral form factor, again at $\beta{=}50$. It is, as to be anticipated, almost two orders of magnitude larger than for the (2,2) case, because of non--perturbative effects (already discussed). 
%
There is (on the logarithmic scale) a ramp--to--plateau transition (although it is, from the discussion above, much slower than for (2,2)). Also visible is the  slower slope--to--dip time  (due to its slower decay rate), as already discussed.

\subsection{Temperature Dependence}

\begin{figure}[h]
\centering
\includegraphics[width=0.48\textwidth]{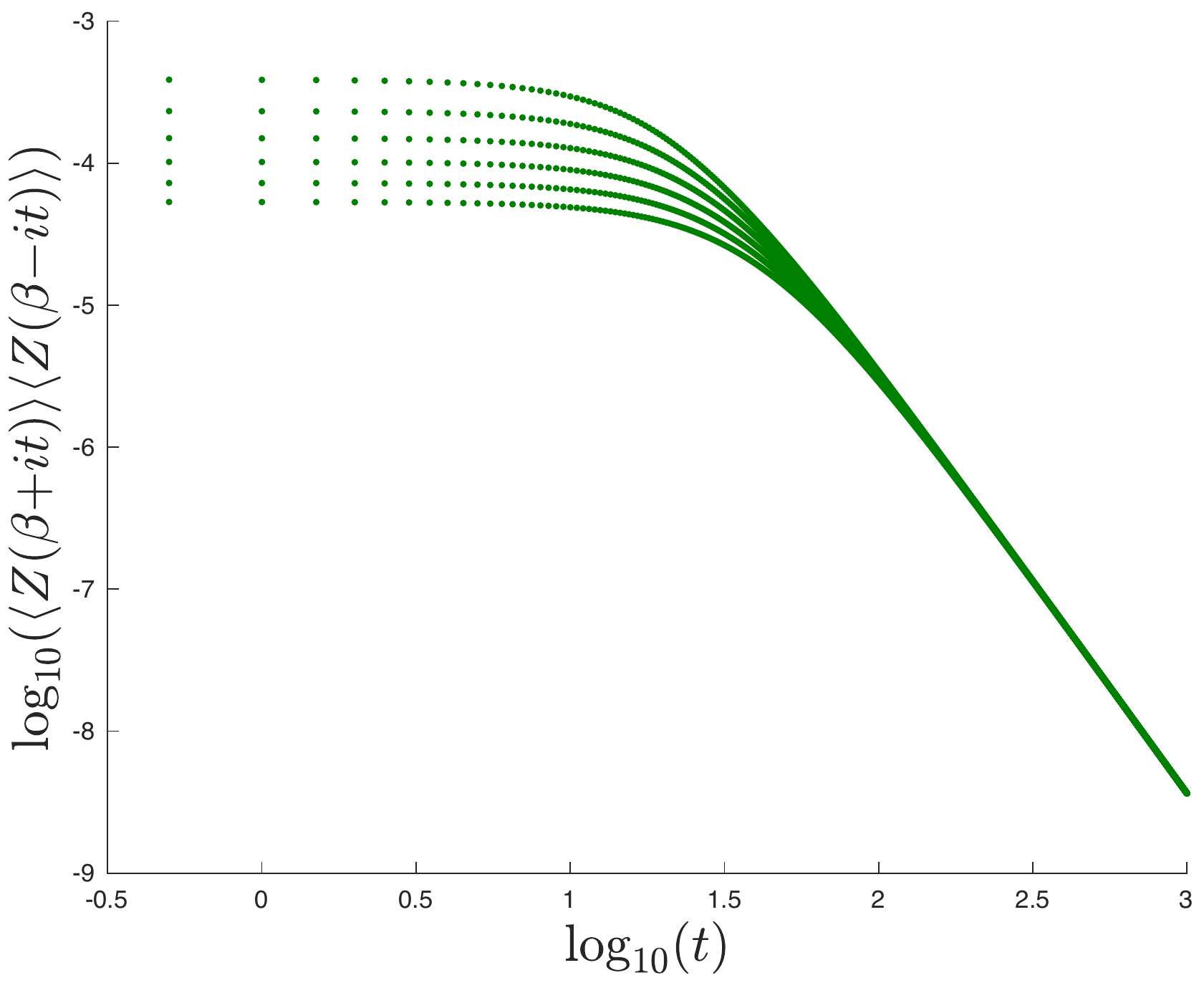}
\caption{\label{fig:non-pert-sff-JT1-series-disconnected} A series showing the evolution, with $\beta$, of the disconnected part of the spectral form factor, for (2,2) JT supergravity. Here $\hbar{=}1$. }
\end{figure}

It is of interest to see how the spectral form factor  evolves as a function of temperature. The results for a  series of increasing temperatures, $\beta{=}50,46,42,38,34$ and~$30$ are presented for the disconnected part of the spectral density in figure~\ref{fig:non-pert-sff-JT1-series-disconnected}. %

\begin{figure}[h]
\centering
\includegraphics[width=0.48\textwidth]{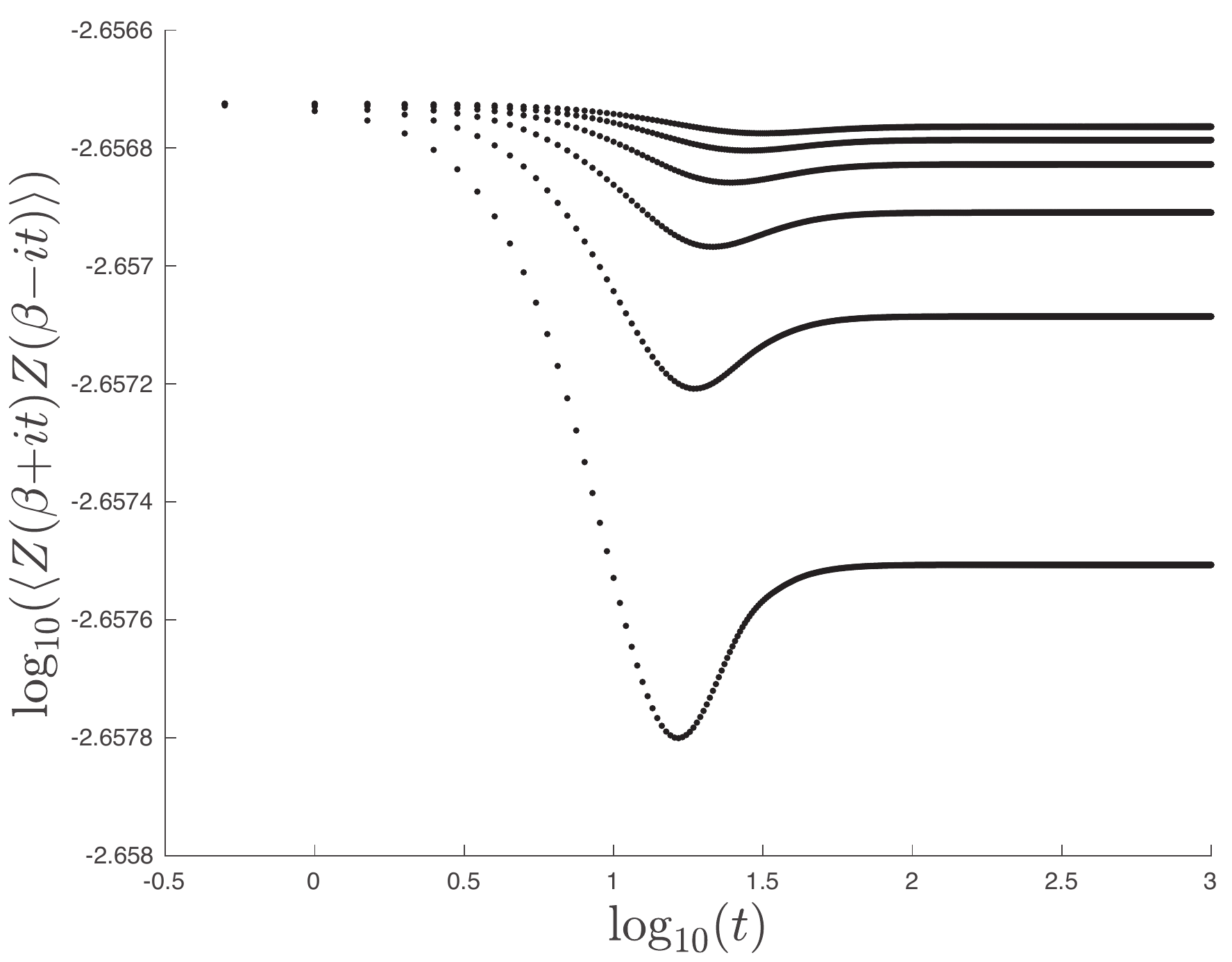}
\caption{\label{fig:non-pert-sff-JT1-series-combined} A series showing the evolution, with $\beta$, of the full spectral form factor, for (2,2) JT supergravity. Here $\hbar{=}1$.  For comparison, the curves have been rescaled to start out at the same height as the highest temperature curve.}
\end{figure}

\bigskip
The highest temperature curve is at the top. Strikingly, the curves soon merge into each other and follow the  fall--off already discussed, regardless of the starting temperature.  In figure~\ref{fig:non-pert-sff-JT1-series-combined},  there is a series of the full spectral density, for the same set of temperatures.  Crucially, for comparison purposes, the curves are all uniformly scaled (on the vertical axis) to have the same initial height as the highest temperature ($\beta{=}30$) case, which is the lowermost curve. Therefore, in this figure, relative slopes should not be taken literally. 
This scaling allows for ready access to some of the more meaningful comparisons to be made, such as the relative size of the curves: Higher temperature (smaller $\beta$) gives a vertically larger curve: Higher temperature ``shakes" the system up more, resulting in a wider amplitude of deviation from the initial value before it settles down. Interestingly though,  the dip time increases slightly with higher temperature, although not dramatically. The rapid ramp time also changes very slowly  with $\beta$.

\begin{figure}[h]
\centering
\includegraphics[width=0.48\textwidth]{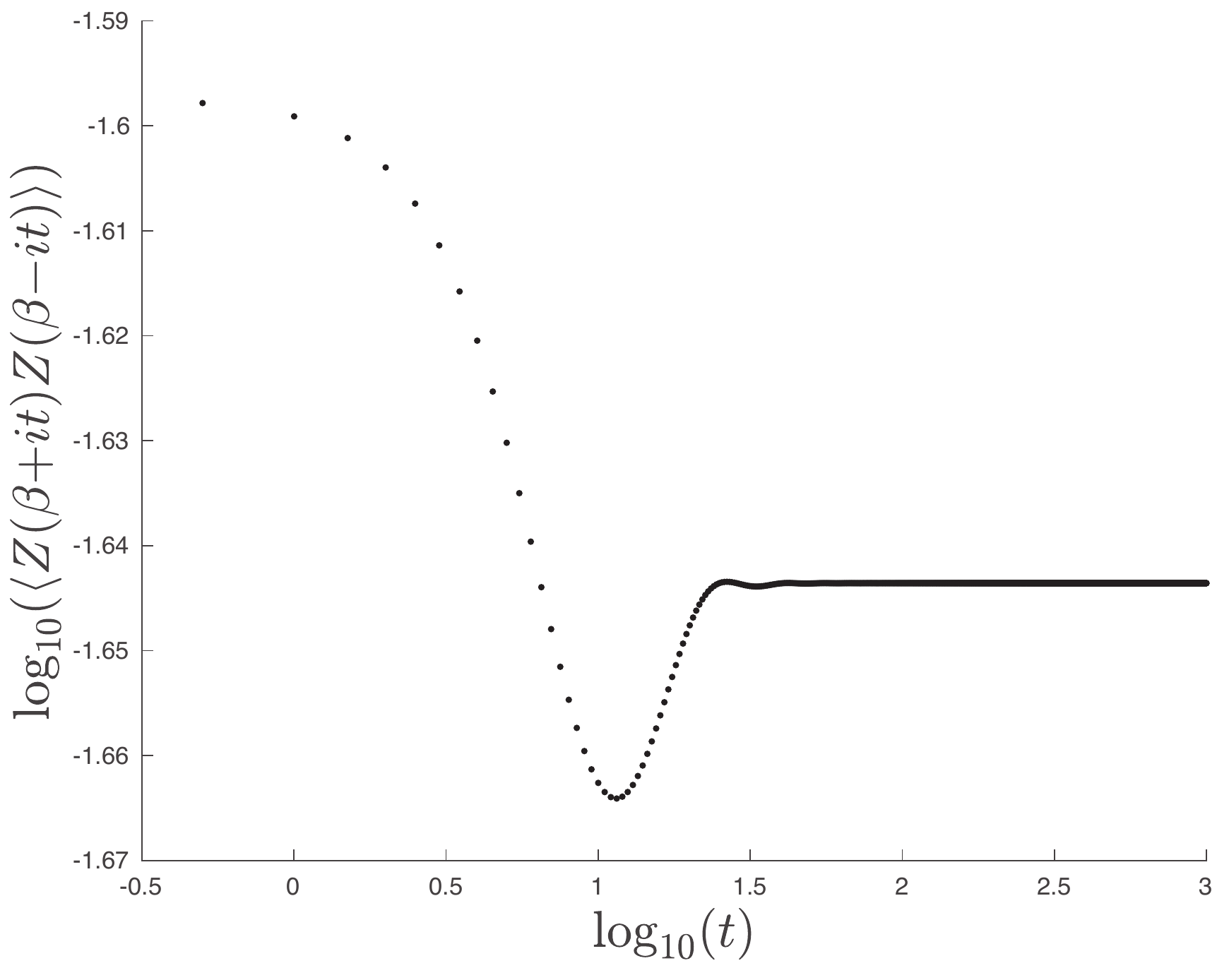}
\caption{\label{fig:damped-wobble-sff} The full spectral form factor at $\beta{=}14$ $\hbar{=}1$,  for (2,2) JT supergravity. A (relatively) high temperature feature appears near the cross--over from  ramp to plateau. See text for discussion.}
\end{figure}

Heading toward smaller (but still moderate) values of~$\beta$ (in the region of $\beta{\simeq}\beta_{\rm crit}$) there are small modulations in both the disconnected and  connected parts of the form factors, accumulating (in the latter) near the cross--over from ramp to plateau. The combined result of these higher temperature structures is that there is a damped wobble as the ramp merges into the plateau. An example of the full spectral form factor showing this feature  is given in figure~\ref{fig:damped-wobble-sff} for the case $\beta{=}14$. 
Whether this is interesting physics or not is not clear~\footnote{It is reminiscent of features of an interesting exact expression derived in ref.~\cite{Lau:2020qnl} in the context of an SYK model with source terms.}. The value of the temperature at which this can be seen seems comfortably below the highest energy allowed by the truncation.

\begin{figure}[h]
\centering
\includegraphics[width=0.48\textwidth]{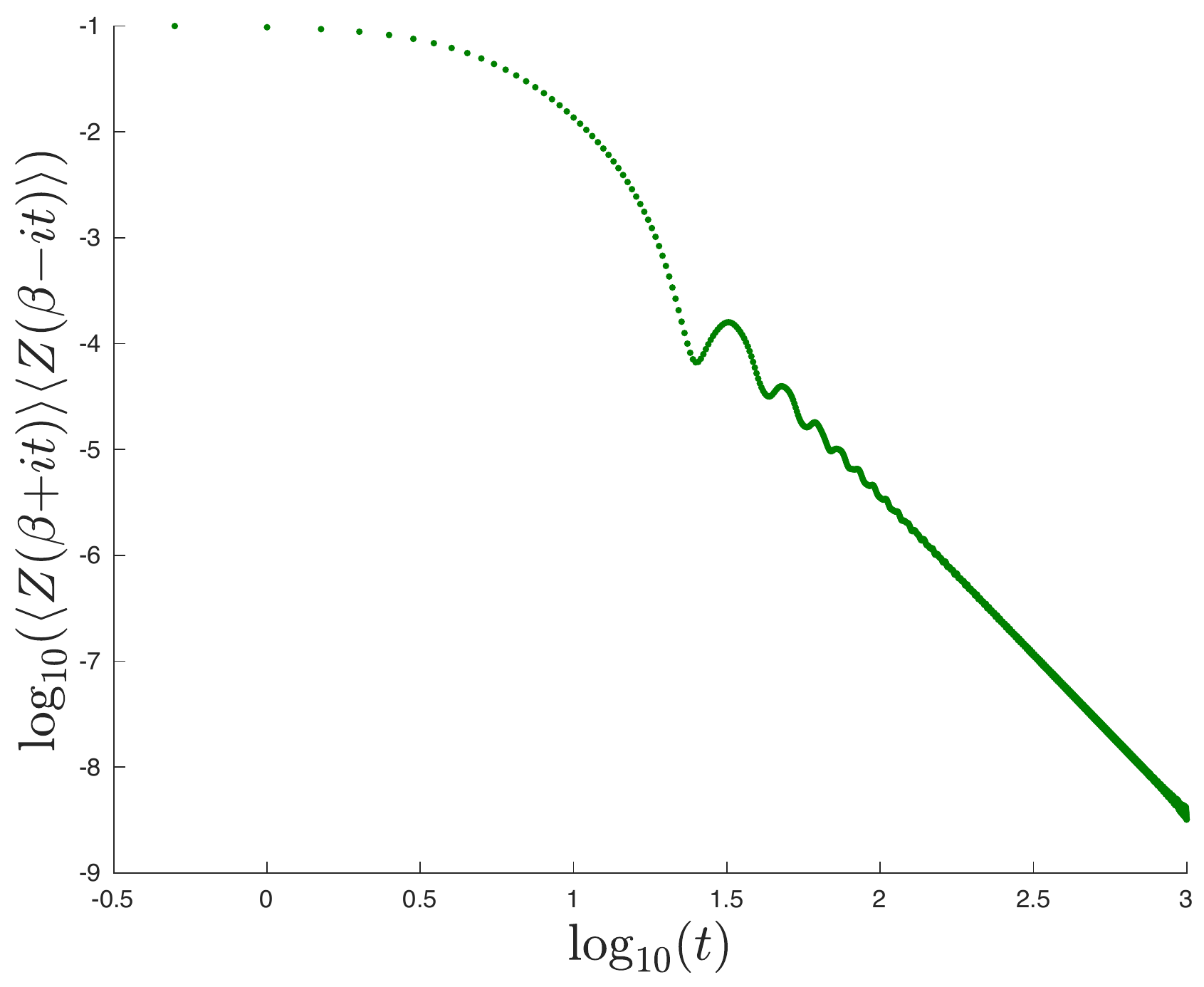}
\caption{\label{fig:high-T-sff} Disconnected part of the spectral form factor at $\beta{=}9$ $\hbar{=}1$,  for (2,2) JT supergravity. The growing dips are consistent with a pattern of zeros developing at infinite temperature.}
\end{figure}

Some other other  fascinating structures become apparent in the very high temperature regime. How useful they are for the physics in question is debatable since this whole context (the Schwarzian, the connection to black holes, SYK, {\it etc.,}) is in a low energy limit. Moreover, high temperature also begins to go beyond the energies for which the truncation of the string equation remains reliable. However it is interesting to observe the features anyway, and could well be instructive for understanding JT models with a cutoff~\cite{Gross:2019uxi,Gross:2019ach,Iliesiu:2020zld,Stanford:2020qhm}. Looking at the disconnected part of the spectral form factor for the (2,2) supergravity case, a series of dips evolve, becoming more pronounced toward higher temperature (smaller $\beta$).  See figure~\ref{fig:high-T-sff}.  

They have a clear pattern and structure and are consistent with observations made for large $N$ random matrix systems (even without double scaling). (They are also analytically obtainable in the exact Airy example reviewed in Appendix~\ref{app:airy-model}.) 

In fact, the disconnected function is beginning to resemble the form $J_1(t)^2/t^2$ that has been derived analytically for the infinite temperature case. After taking the logarithm, the zeroes of the Bessel function~$J_1$ become the dips in the logarithmic plot.  It would be interesting to show that this (or a variation thereof) analytical form emerges in this JT supergravity context as well. Following the numerics to smaller~$\beta$ seems to confirm this (a zero in the connected function also appears, in some examples), although eventually numerical inaccuracies begin to overwhelm the results, presumably because the correct physics needs to include contributions from energies that lie beyond the  cutoff on the spectrum up to which  the truncated equations are valid.

\section{Non--Perturbative JT Gravity}
\label{sec:JT-gravity-regular}
This section presents results analogous to those shown in earlier sections for the non--perturbative completion of ordinary JT gravity presented in ref.~\cite{Johnson:2019eik}. It might seem odd to have studied the JT supergravity examples first, leaving this case for last, but there is good reason. The non--perturbative physics of this case is more subtle. JT gravity was shown, in ref.~\cite{Saad:2019lba}, to be {\it perturbatively} (in the topological expansion) equivalent to  a double--scaled Hermitian matrix model, {\it i.e.} classified in the Gaussian Unitary ensemble ($\beta{=}2$ in the Dyson--Wigner series). On general grounds, such double--scaled Hermitian matrix models are known to sometimes have non--perturbative  (in topology) instabilities, and so it is possible that the JT gravity definition inherits them. More specifically,  thinking about the model in terms of constituent minimal models, as in section~\ref{sec:minimal-model-deconstruction}, it is made up of an interpolating family of minimal models that have the  $x{\to}{-}\infty$ boundary condition  in equation~(\ref{eq:boundary-conditions}) in both directions, and  instead solve the string equation ${\cal R}{=}0$. (Recall that~${\cal R}$ is given in equation~(\ref{eq:flow-object})). These are the $(2k{-}1,2)$ bosonic minimal string models. For $k$ even, these models are non--perturbatively unstable, as has been known for some time~\cite{Banks:1990df,David:1990ge,Douglas:1990xv,Dalley:1991zr}. From the point of view of the spectrum, all the models, when non--perturbative effects are taken into account, have contributions from arbitrarily negative energy sectors. Even though exponentially suppressed, for even $k$ the effective potential turns downward for states at these energies, signalling  the system's wish to tunnel to an entirely new solution that is quite different from the one around which perturbation theory was developed. From the perspective of this paper (solving string equations non--perturbatively), this means that for each of those ($k$ even) models, there simply are no real smooth solutions of the $k$th equation with those conditions~\cite{Banks:1990df,Douglas:1990xv,Moore:1990mg,Moore:1990cn}. Since JT gravity is made up of, in equal measure, even and odd $k$ models, this strongly suggests that it inherits these problems, as already noted in ref.~\cite{Saad:2019lba}.\footnote{An earlier version of this manuscript contained an idea for a possible evasion of this reasoning. Since individual odd $k$ members of the $(2k{-}1,2)$ minimal series are better non-perturbatively defined, the idea was that perhaps one could define the model as an odd $k$ model (for $k$ large) within which all the lower order models are turned on. This was thought to possibly yield good non--perturbative solutions to the ${\cal R}{=}0$ differential equation, as deformations of the known good solutions for $k$ odd. This does not seem to work.} 

The route that ref.~\cite{Johnson:2019eik} took to supply a non--perturbative definition of JT gravity was to embed it into a larger problem. The minimal models used in previous sections for JT supergravity, also indexed by $k$, have the same $x{\to}{-}\infty$ boundary condition as the bosonic minimal models, and in fact when $\Gamma{=}0$ they have {\it identical} perturbation theory as solutions to the differential equation~(\ref{eq:string-equation-2}). Put differently they solve ${\cal R}{=}0$ perturbatively at large ${-}x$. This means that if used to construct a JT gravity model, they will yield the same physics at high energies $E$, but yield different physics at lower energies that is untroubled by the stability issues. The combination of models needed (that will give the Schwarzian spectral density~(\ref{eq:spectral_JT})) at high $E$ is as follows:
\be
\label{eq:teekay_JT}
t_k=\frac12\frac{\pi^{2k-2}}{k!(k-1)!}\ .
\ee
(This relation was first derived in ref.~\cite{Okuyama:2019xbv}, but with different normalization.) 
In fact it is possible to  integrate the $f(u_0)$ that results from this combination  to find the explicit classical potential $u_0(x)$ that yields the Schwarzian density, through:
\be
 \label{eq:class-pot-JT}
 x= -\frac{\sqrt{u_0}}{\pi}I_1(2\pi\sqrt{u_0})\ ,
 \ee the analogue of the case~(\ref{eq:class-pot-SJT}) for the JT supergravity found earlier.
 Then the $u(x)$ for JT gravity is constructed fully non--perturbatively using equation~(\ref{eq:string-equation-2}) with $\Gamma{=}0$.
This non--perturbative taming of JT gravity can be explored extensively along the same lines as done for the JT supergravity models. The truncation scheme works along the similarly, and so there is no need to retread the ideas again. 

\begin{figure}[h]
\centering
\includegraphics[width=0.45\textwidth]{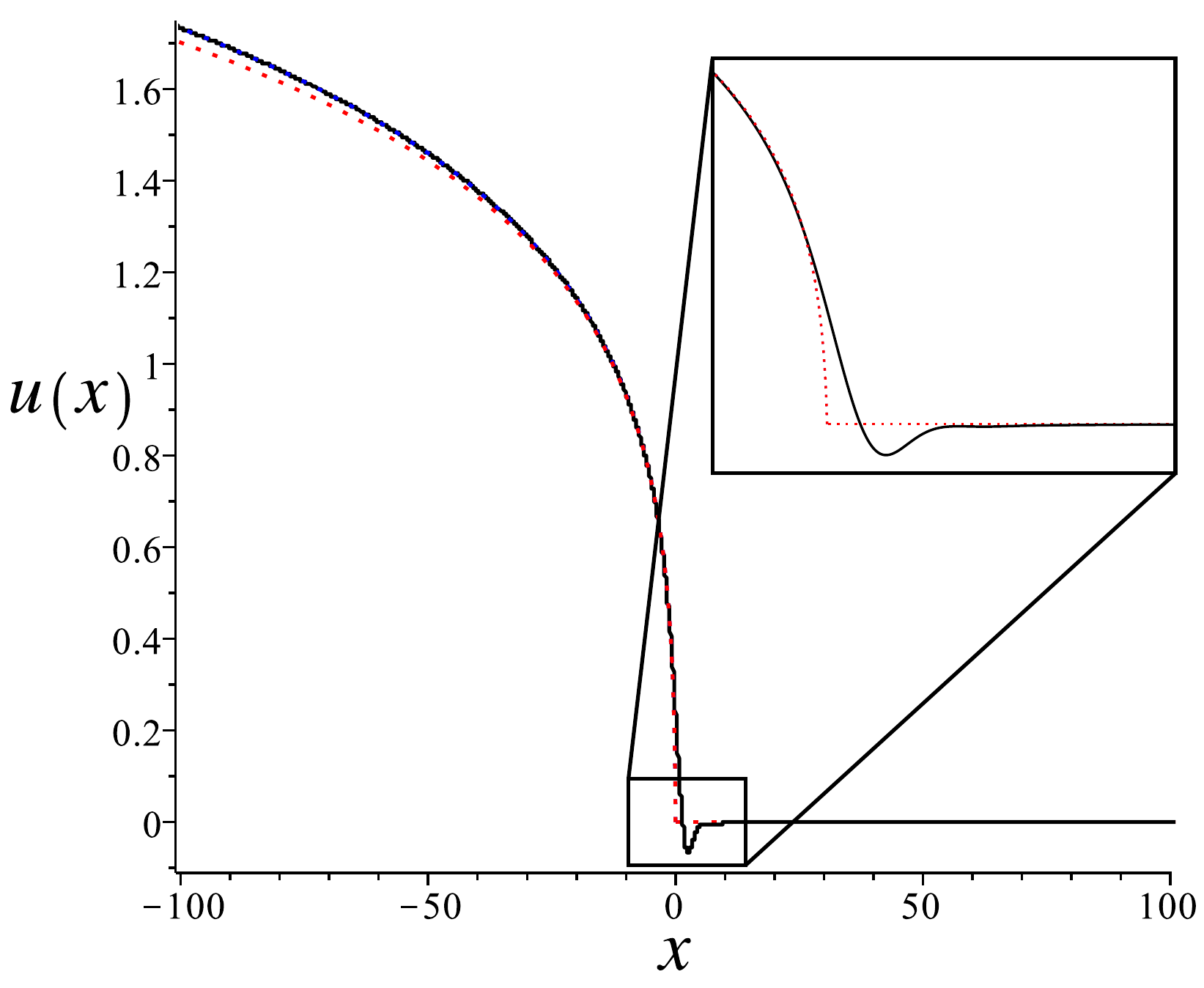}
\caption{\label{fig:truncation-examples-C} The  solution (solid line) of the string equation for truncation up to $t_7$. The inset shows the well that developed in the interior.  The full classical solution  is shown too (dotted).}
\end{figure}

A solution to the string equation~(\ref{eq:string-equation-2})  with the bosonic JT combination of minimal models~(\ref{eq:teekay_JT}) was found with the first $7$ models turned on, constituting a very good truncation where energies up to $E{\sim}1.5$ can be trusted. A combination of numerical and analytic methods was used to find the solution to the 15th order differential equation. See appendix~\ref{app:notes-and-tips-1} for some tips on how this was done. Figure~\ref{fig:truncation-examples-C} shows the solution, with the classical (disc level) potential that gives the  JT spectral density~(\ref{eq:spectral_JT}) displayed as a dashed line. 

Again there is a small well (not deep enough to support bound states) in the interior, and then $u(x)$ settles to zero to the right. These are features shared by the JT supergravity models, as should be expected since their components are being used here as a non--perturbative low energy ``regulator'' in a sense, removing the leakage to negative energy. 

\begin{figure}[h]
\centering
\includegraphics[width=0.45\textwidth]{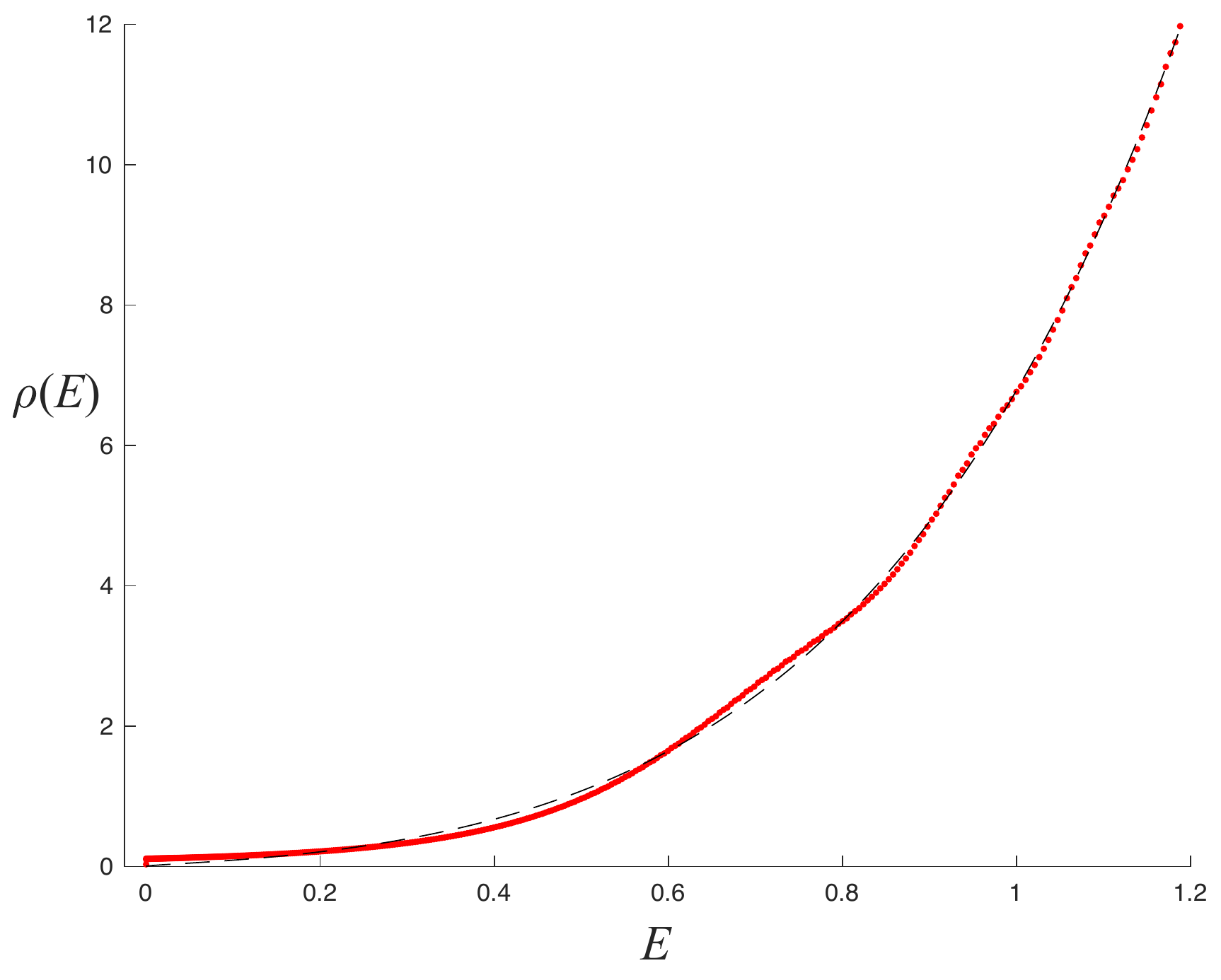}
\caption{\label{fig:non-pert-sd3} The full spectral density, for   refs.\cite{Johnson:2019eik}'s non--perturbative completion of JT gravity. The dashed blue line is the disc level result of equation~(\ref{eq:spectral_JT}).}
\end{figure}

The methods of section~\ref{sec:spectral-density} then allowed for the spectrum to be computed. For this definition, $\mu{=}0^{-}$ is used for the upper limit in equation~(\ref{eq:spectral-density-integration}), matching to the $x{<}0$ perturbation theory that builds JT gravity (see the discussion in ref.~\cite{Johnson:2020heh} for more discussion about how to efficiently extract perturbation theory using the Gel'fand--Dikii resolvent equation), so in contrast to the JT supergravity cases, there is no $1/\sqrt{E}$ classical contribution to the spectrum. Figure~\ref{fig:non-pert-sd3} displays, for the first time, the spectral density of a non--perturbative completion of JT gravity with all perturbative and non--perturbative corrections included (up to this energy). 
It is clear from the figure that the non--perturbative ripples have already begun to die away and merge into the classical smooth region, showing that this truncation has captured the key physics that is affected by non--perturbative contributions. Notice that the spectrum is naturally  bounded below by $E{=}0$, as already shown explicitly in preliminary studies in ref.~\cite{Johnson:2019eik}. Note that ref.~\cite{Johnson:2019eik}  generalized the construction by  turning on the parameter $\mu$ that was discussed in the supergravity context in section~\ref{sec:spectral-density}. This is straightforward to do, and there were no additional insights to be gained, and so results are not presented here. The tail of the resulting distribution, not surprisingly, resembles the tails already displayed in ref.~\cite{Johnson:2019eik}, including the interesting feature that non--perturbative effects generate a non--zero $\rho(E{=}0)$.

\begin{figure}[h]
\centering
\includegraphics[width=0.48\textwidth]{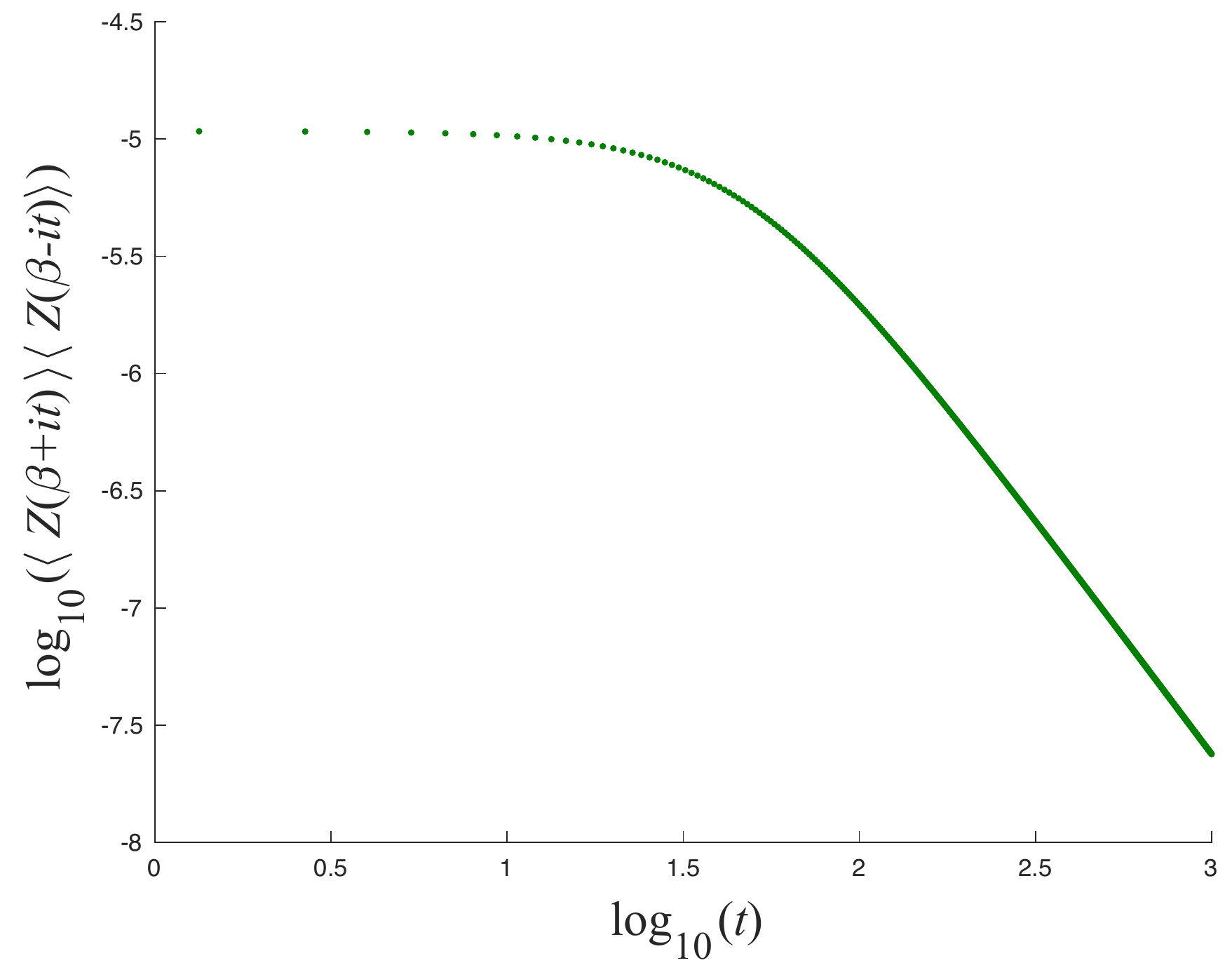}
\caption{\label{fig:disconnected-sff-JT} The disconnected part of the  JT gravity  spectral form factor  {\it vs.} $t$, at $\beta{=}50$ and $\hbar{=}1$, for the non--perturbative scheme of ref.~\cite{Johnson:2019eik}.}
\end{figure}

\begin{figure}[h]
\centering
\includegraphics[width=0.48\textwidth]{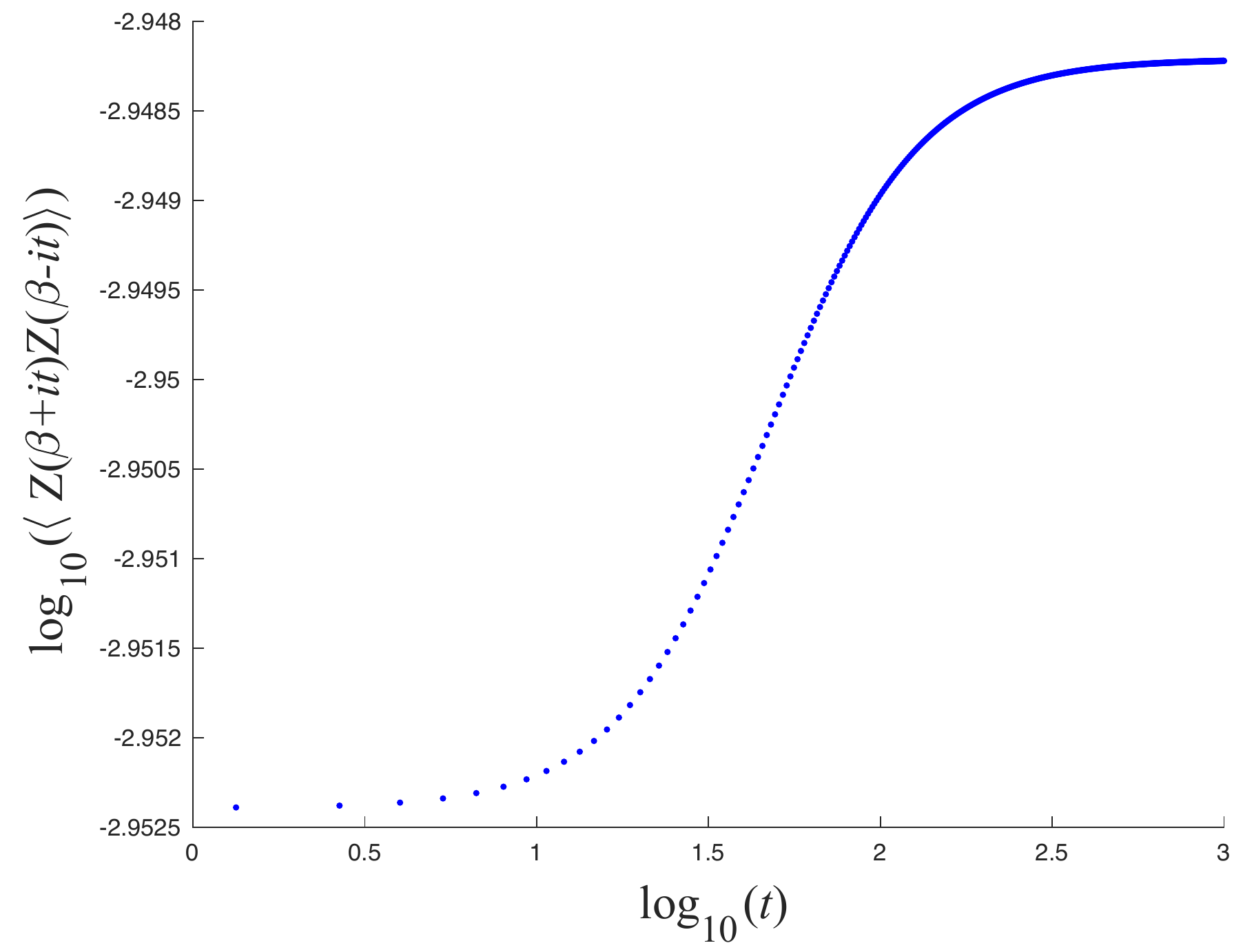}
\caption{\label{fig:connected-sff-JT} The connected part of the JT gravity  spectral form factor  {\it vs.} $t$, at $\beta{=}50$ and $\hbar{=}1$, for the non--perturbative scheme of ref.~\cite{Johnson:2019eik}.}
\end{figure}

\begin{figure}[h]
\centering
\includegraphics[width=0.48\textwidth]{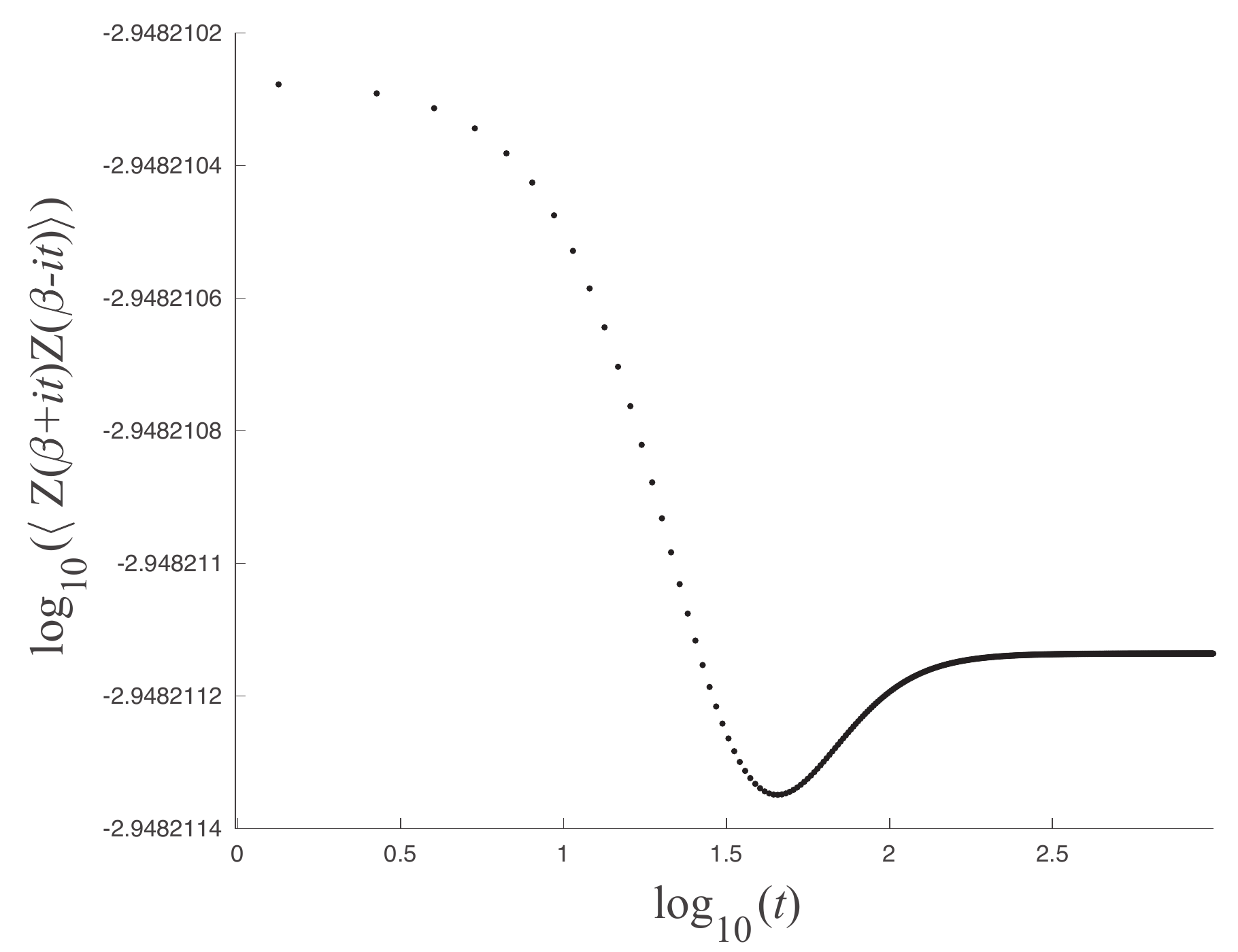}
\caption{\label{fig:combined-sff-JT} The full  JT gravity  spectral form factor  {\it vs.} $t$, at $\beta{=}50$ and $\hbar{=}1$, for the non--perturbative scheme of ref.~\cite{Johnson:2019eik}.}
\end{figure}

Of course, with the spectrum in hand (approximately 1800  normalized wavefunctions and their energies) the next natural step is to compute the spectral form factor, using the methods of section~\ref{sec:spectral-form-factor}. The correlator of two boundaries is readily computed, and the phase transition where the disconnected part (two black holes) hands over to the connected part (wormhole) happens at $\beta_{\rm cr}$. (A  figure similar to figure~\ref{fig:phase-transition} is omitted here to avoid repetition.)  For $\beta{=}50$, the disconnected, connected, and combined spectral form factor are shown in figures~\ref{fig:disconnected-sff-JT},~\ref{fig:connected-sff-JT}, and~\ref{fig:combined-sff-JT}, respectively.

The most striking feature overall is in the disconnected portion of the form factor, controlling the initial slope. As might be expected from the absence of the $1/\sqrt{E}$ low energy behaviour of the regulating models, the  time dependence of the slope is not $t^{-1}$ (as it is for the (0,2) JT supergravity, but nor is it the $t^{-3}$ expected from the classical low energy physics to be read off from the $\beta^{-3/2}$ dependence of the partition function~(\ref{eq:disc-partition-function-JT}). Rather, it interpolates between them, and is ${\sim}t^{-2}$, to the nearest integer. From what was learned from the supergravity cases of last section, the origin of this is clear. The endpoint of the distribution has a different structure (see figure~\ref{fig:non-pert-sd3}), with some non--zero $\rho(0)$ at the end. There are far fewer states than for the (0,2) supergravity case but more than the (2,2) case, and hence the fall--off rate (at least for this value of~$\hbar$) is between that of those two cases.

Again, as seen in all the models studied in this paper (and also the special Airy model recalled in Appendix~\ref{app:airy-model}) non--perturbative effects hasten the transition from dip to ramp to plateau such that the linear  part of the classical contribution to the ramp that emerges at long times simply does not have time to develop, for moderate values of $\hbar$.

\section{Closing Remarks}
\label{sec:closing-remarks}

The purpose of this paper was to explicitly uncover and examine the non--perturbative physics for JT gravity and supergravity that is accessible if they are formulated using  minimal model building blocks. This construction is not a simple large $k$ limit of a minimal model, but a more refined affair involving coupling them together in a particular combination, as suggested perturbatively in ref.~\cite{Okuyama:2019xbv}, and extended to non--perturbative physics in refs.~\cite{Johnson:2019eik,Johnson:2020heh}. The principal applications that demonstrated the facility of the technique was the explicit computation, for the first time,  of the full non--perturbative spectral densities of various JT gravity and supergravity models, and the use of these spectral densities to compute the spectral form factor in each case, showing how the non--perturbative effects affect the shape (sometimes dramatically) of this important diagnostic quantity. Having explicit access to the non--perturbative features in this manner turned out to be rather instructive, as extensively discussed in the body of the paper. 

 Techniques to allow such non--perturbative properties to be extracted, in a consistent and well--defined scheme (for generic values of $\beta$ and $\hbar$), have not been presented in the literature before, and it is hoped that these methods and results will go some way to helping uncover more of the fascinating web of interconnections between geometry, quantum mechanics, gravity, and chaos that seems to be emerging from these studies.

It is likely that other models of JT gravity can be ``deconstructed" in terms of minimal models in a way analogous to what was done here, and thereby be given a non--perturbative definition, not just in principle, but (as shown here) in useful accessible terms.    It could possibly also encompass some of the new kinds of matrix model descriptions of JT gravity black holes mentioned recently in refs.~\cite{Witten:2020ert,Maxfield:2020ale}. Perhaps  the results of explorations along these lines will be reported soon.

\appendix

\section{Numerical Testbed:  The Airy Model}
\label{app:airy-model}

As a means of sharpening understanding of some of the key features of the spectral form factor, and for modelling what kinds of physics can be reliably captured by the numerical approaches used in the main body of the paper, this Appendix presents a numerical exploration of the exactly solvable Airy model, which is the double--scaled limit of the simple  Hermitian matrix model with Gaussian potential, obtained by magnifying the infinitesimal region at the edge of Wigner's semi--circle~\cite{Ginsparg:1993is}.  It is the $k{=}1$ model of the $(2k{-}1,2)$ minimal string series, and as such is also a model of the extreme low energy tail of ref.\cite{Saad:2019lba}'s matrix model of JT gravity. 

\begin{figure}[h]
\centering
\includegraphics[width=0.45\textwidth]{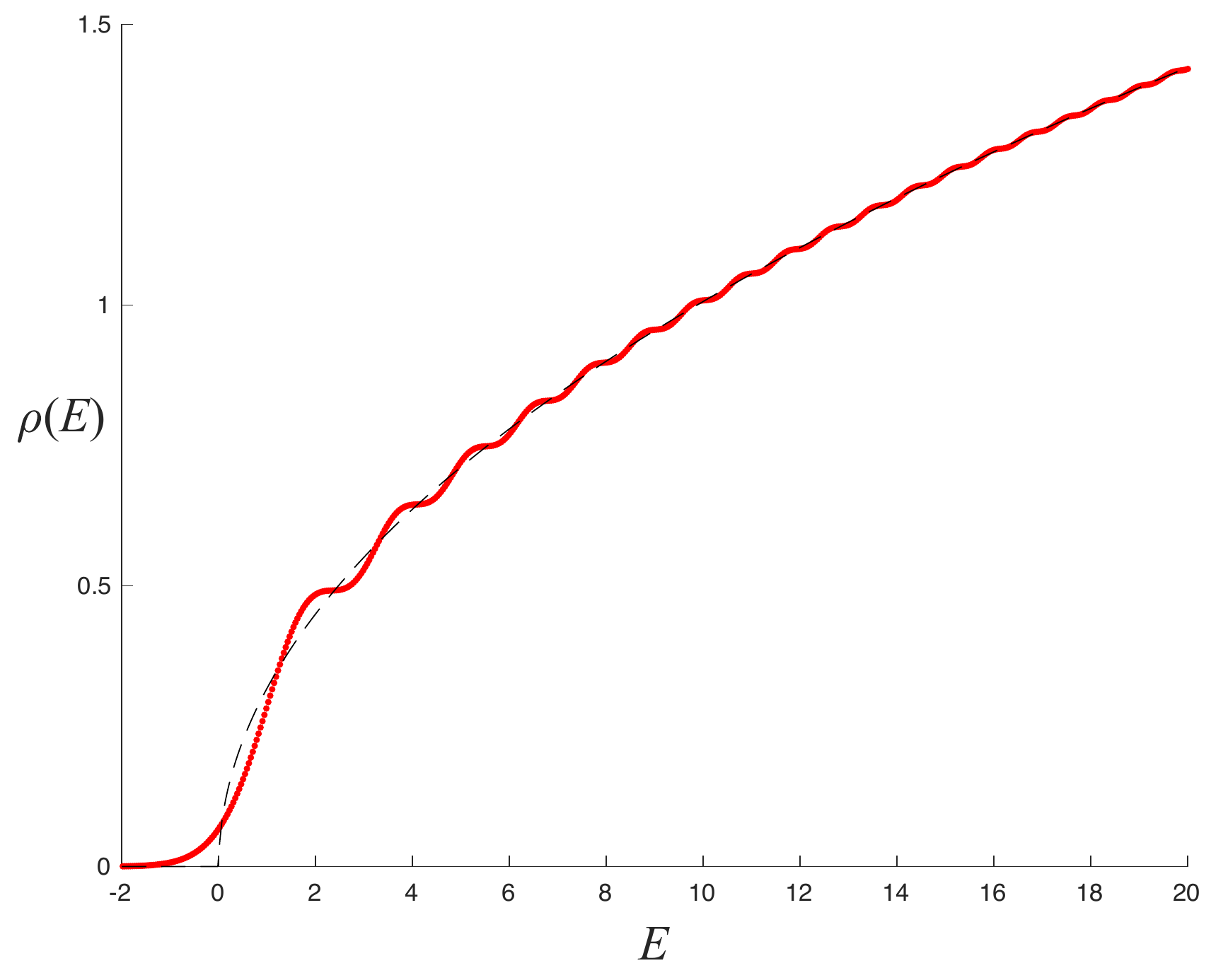}
\caption{\label{fig:non-pert-sd4} The full spectral density for the exactly solvable Airy model, at $\hbar{=}1$. The dashed  line is the disc level result.}
\end{figure}

In the language of this paper, it comes from using the simple linear potential $u(x){=}{-}x$ in the Hamiltonian~(\ref{eq:schrodinger}), and the resulting  equation to solve for the spectrum is simply (after a change of variables) Airy's differential equation. The wavefunctions for energy $E$ are:
\be
\label{eq:airy-wavefunctions}
\psi(E,x)=\hbar^{-\frac23}{\rm Ai}(-\hbar^{-\frac23}(E+x))\ ,
\ee
and the spectral density that results is:
\bea
\rho(E)=\!\!\int_{-\infty}^0\!\!|\psi(E,x)|^2dx=\hbar^{-\frac23}[{\rm Ai}^\prime(\zeta)^2-\zeta{\rm Ai}(\zeta)^2]\ ,\nonumber\\
\eea
where $\zeta{=}{-}\hbar^{-\frac23}E$. See figure~\ref{fig:non-pert-sd4} for a plot of the spectrum, showing the exponential tail running to negative~$E$. At large $E$, the non--perturbative oscillations of the Airy function die out, leaving the classical (disc) contribution
$\rhoo(E){=}(\pi\hbar)^{-1}\sqrt{E}$. This is shown as a dashed line.

The correlator of two boundaries can be computed exactly using properties of the Airy functions out of which the wavefunction is built\footnote{In fact, ref.\cite{okounkov2001generating} writes down expressions for correlators of multiple loops in this model.}. The disconnected part is simply the square of the partition function, which can be evaluated by Laplace transform, remembering to include negative energies to incorporate the full non--perturbative spectrum:
\be
Z_{\rm Ai}(\beta)=\int_{-\infty}^{+\infty} \rho_{\rm Ai}(E)e^{-\beta E}dE = \frac{e^{\frac{\hbar^2}{12}\beta^3}}{2\pi^{1/2}\hbar\beta^{3/2}}\ .
\ee 
 This gives 
 \be
 \label{eq:disconnected-airy}
 \langle Z(\beta)\rangle\langle Z(\beta^\prime)\rangle = \frac{e^{\frac{\hbar^2}{12}(\beta^3+{\beta^\prime}^3)}}{4\pi\hbar^2(\beta\beta^\prime)^{3/2}}\ ,
 \ee 
 while implementing equation~(\ref{eq:correlator-connected}) yields the connected piece to be:
 \bea
 \label{eq:connected-airy}
 \langle Z(\beta)Z(\beta^\prime)\rangle &=&\\
 &&\hskip-1.5cm \frac{e^{\frac{\hbar^2}{12}(\beta+\beta^\prime)^3}}{2\pi^{1/2}\hbar(\beta+\beta^\prime)^{3/2}}{\rm Erf}\left(\frac{1}{2}\hbar\sqrt{\beta\beta^\prime(\beta+\beta^\prime)}\right)\ ,\nonumber
 \eea
 These are exact expressions, {\it i.e.}, both perturbative and  non--perturbative parts are incorporated (see refs.~\cite{Ginsparg:1993is,okounkov2001generating} and the review in the Appendix of ref.~\cite{Okuyama:2019xbv}). Nevertheless, it is instructive to pretend that only a finite number of wavefunctions are known, (only numerically),  for a discrete set of energies up to some maximum energy. The question is then how well the exact expressions can be reproduced. This is the situation of the body of the paper, resulting from the controlled truncation of the infinite order string equation.
 
\begin{figure}[h]
\centering
\includegraphics[width=0.48\textwidth]{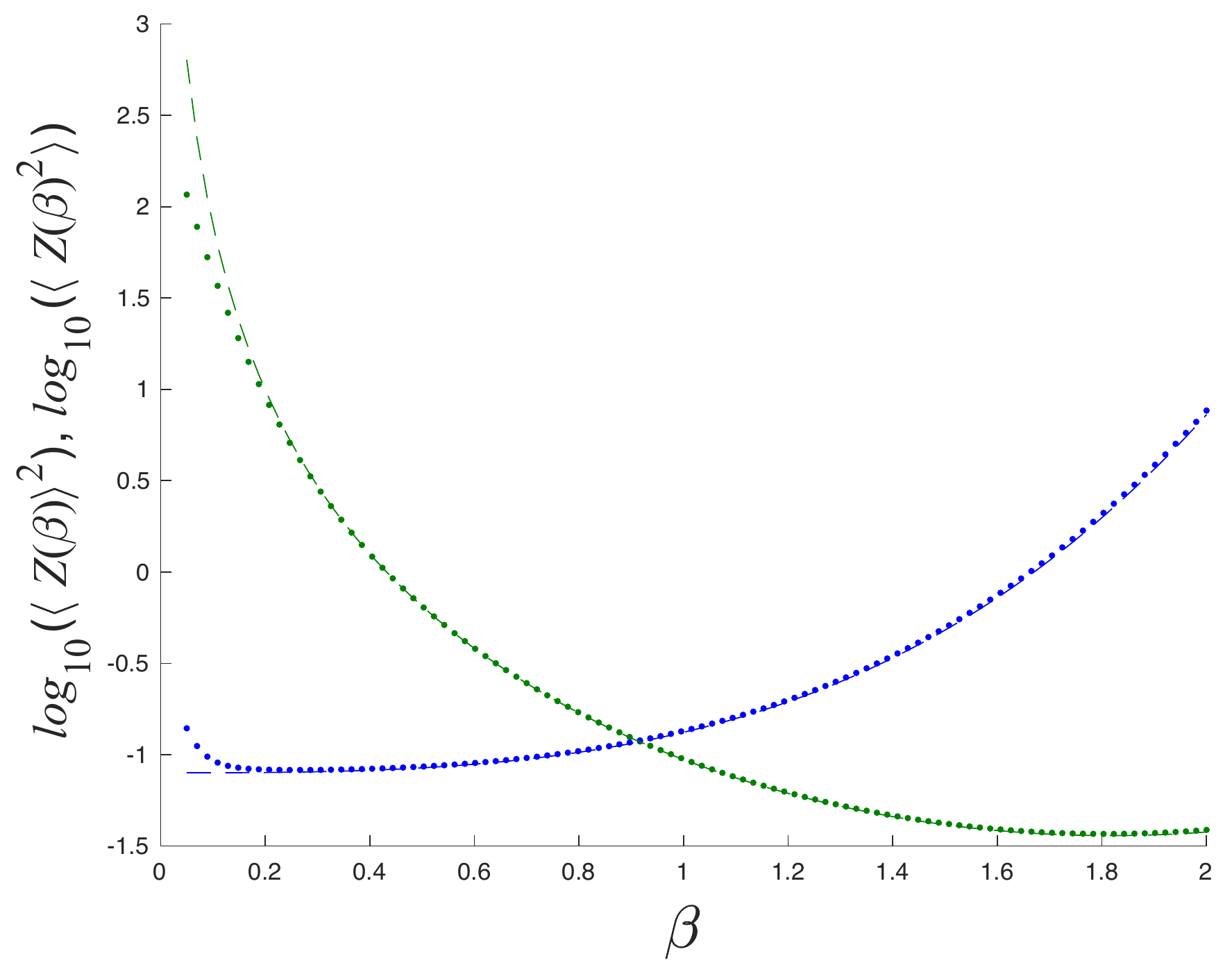}
\caption{\label{fig:phase-transition-airy} The disconnected (starting higher at the left) {\it vs.} the connected (lower) two--point function of the Airy model's  partition function as a function of $\beta$, showing a phase transition at $\beta{=}\beta_{\rm cr}{\simeq}0.92$. A dashed line shows the exact result, dots are for a numerical truncation.}
\end{figure}

\begin{figure}[h]
\centering
\includegraphics[width=0.48\textwidth]{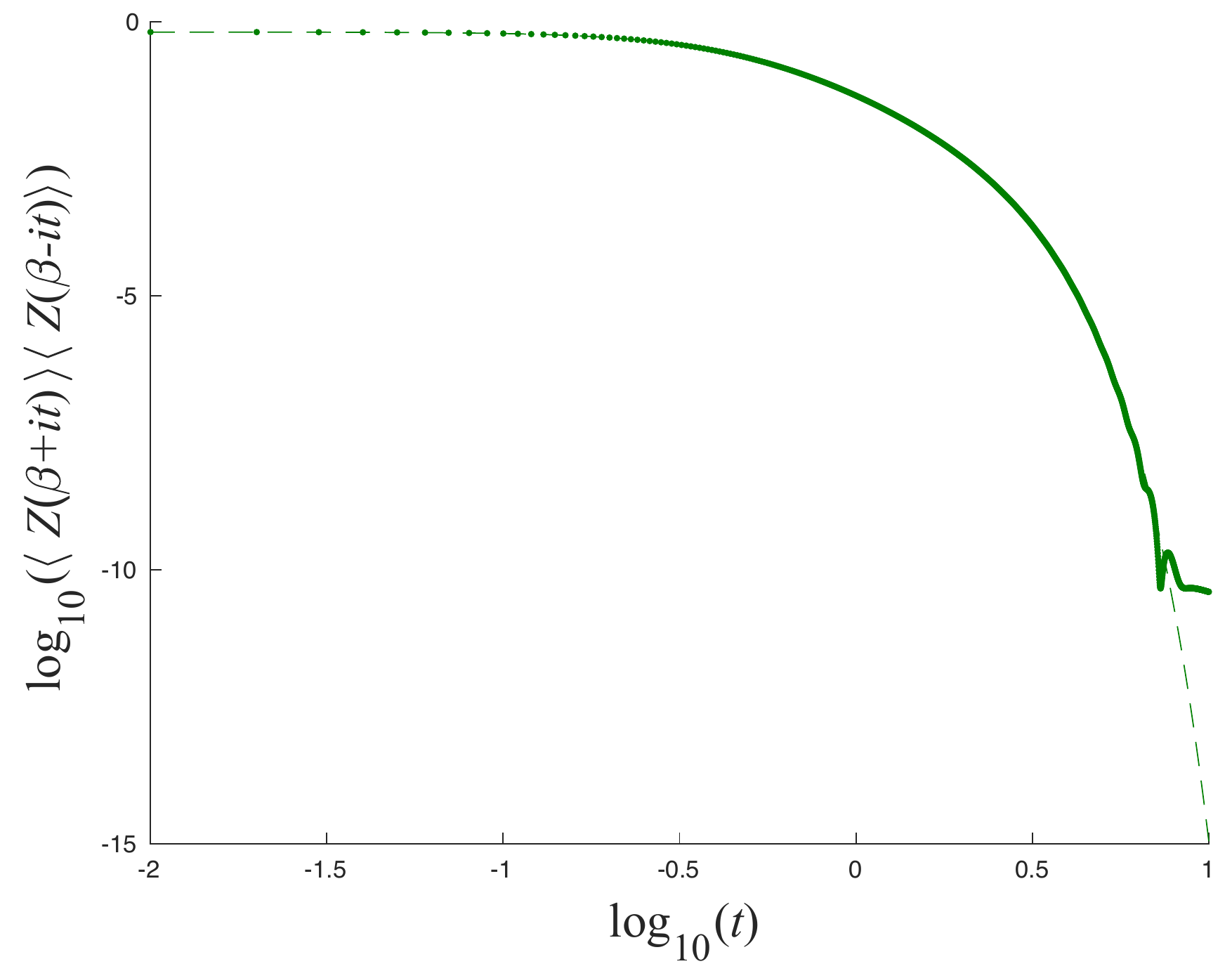}
\caption{\label{fig:disconnected-sff-Airy} The disconnected part of the  Airy model's  spectral form factor  {\it vs.} $t$, at $\beta{=}1/2$, showing the classic slope feature. A dashed line shows the exact result, dots are for a numerical truncation. (The undulations at the end are numerical errors at ultra--small values and so should be ignored.)}
\end{figure}

\begin{figure}[h]
\centering
\includegraphics[width=0.48\textwidth]{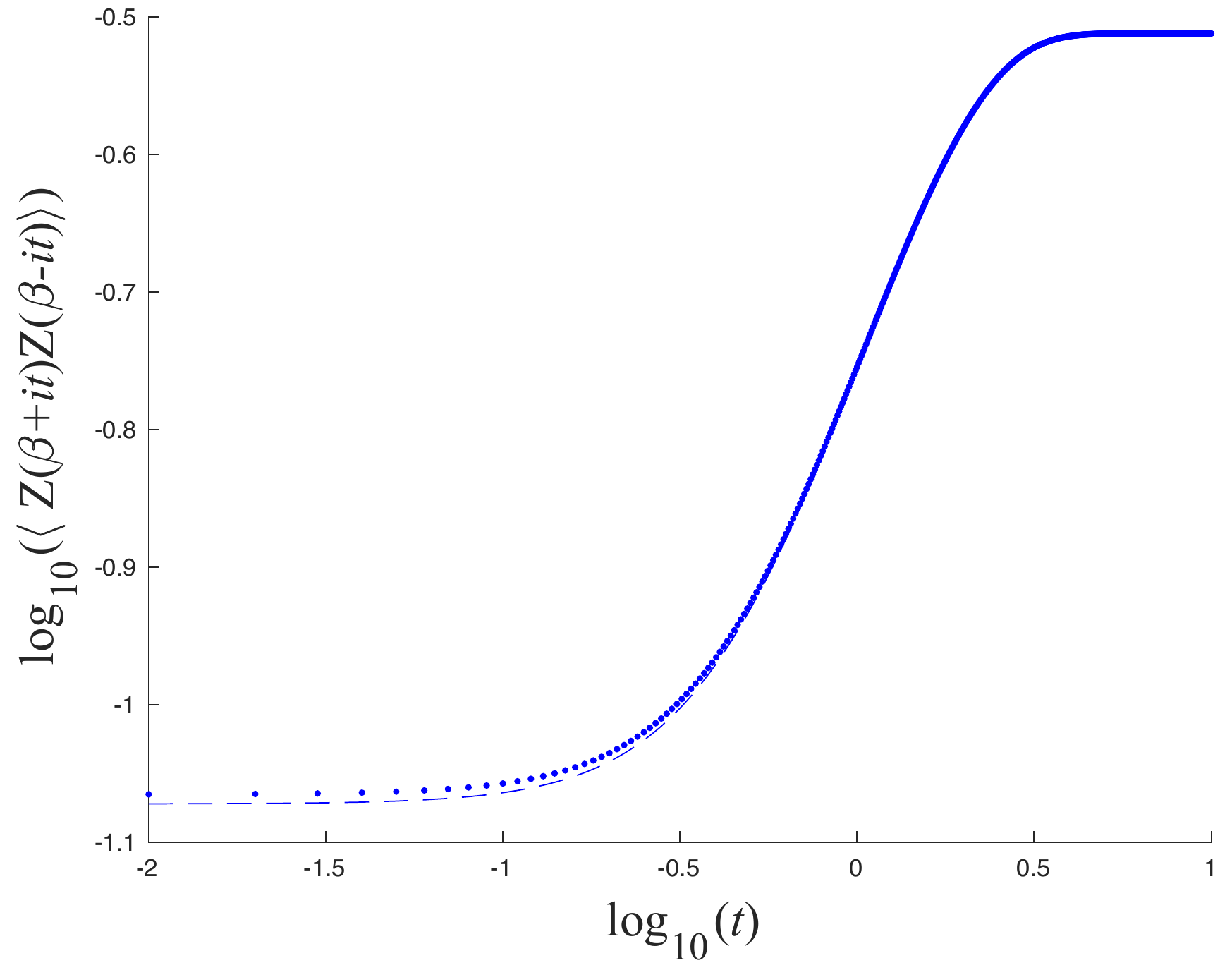}
\caption{\label{fig:connected-sff-Airy} The connected part of the   Airy model's  spectral form factor  {\it vs.} $t$, at $\beta{=}1/2$. A dashed line shows the exact result, dots are for a numerical truncation.}
\end{figure}

\begin{figure}[h]
\centering
\includegraphics[width=0.48\textwidth]{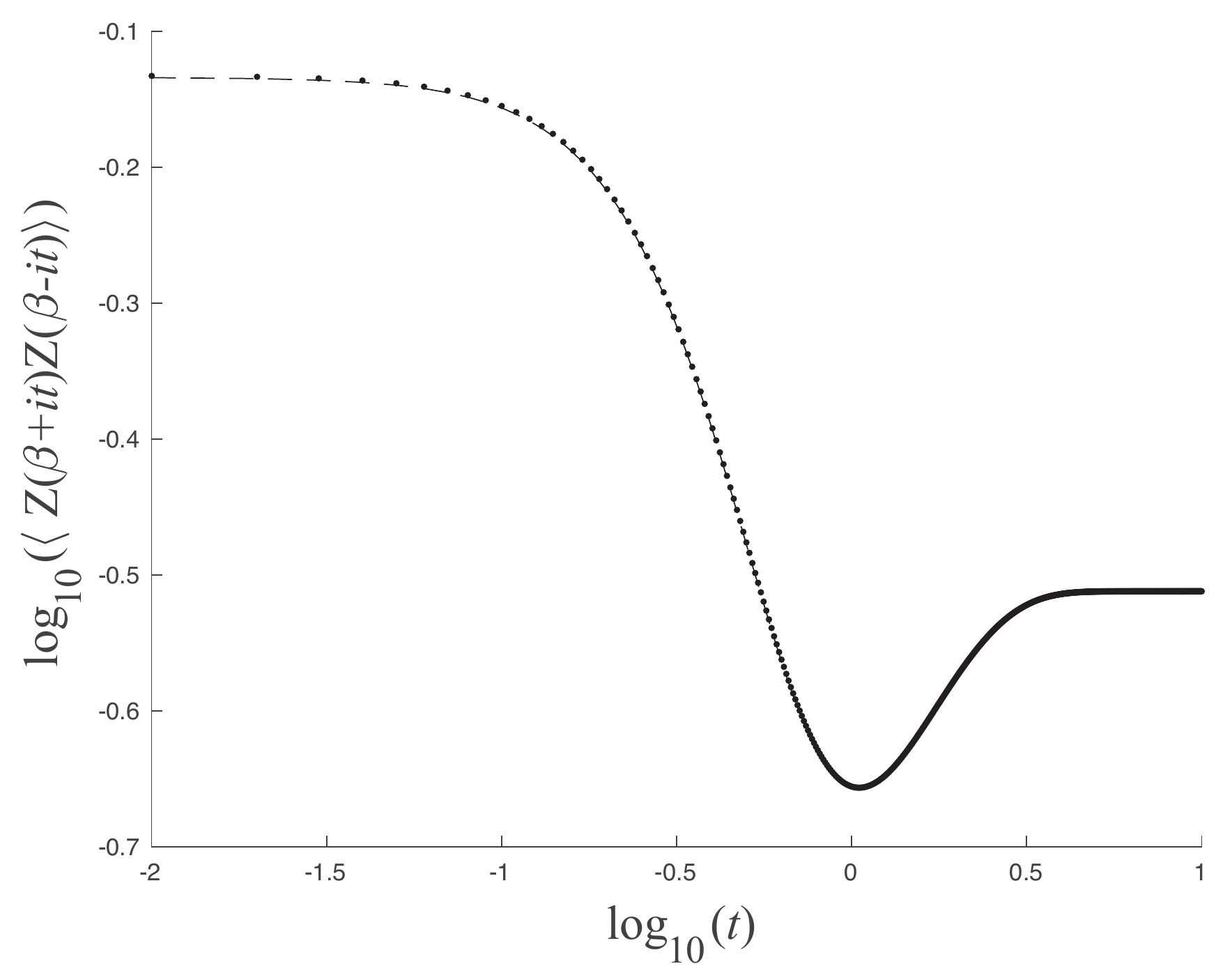}
\caption{\label{fig:combined-sff-Airy} The full spectral form factor of the Airy model   {\it vs.} $t$, at $\beta{=}1/2$.}
\end{figure}

 The answer to the question, reassuringly, is that a great deal of the important physics is accessible. To show this, a set of $1000$ of the wavefunctions~(\ref{eq:airy-wavefunctions}) were discretized on the same size grid used in the body of the paper ($x$ is broken up into 20,000 points), for a range of energies ${-}20{\ge}E{\ge}20$, and the same code that performed the numerical implementation of the expressions given in equations~(\ref{eq:partition-function-integration}) and~(\ref{eq:correlator-connected}) were carried out for this exact model. The result presented in figure~\ref{fig:phase-transition-airy} shows the crossover between the two portions of the correlator as a function of~$\beta$, 
and figures~\ref{fig:disconnected-sff-Airy},~\ref{fig:connected-sff-Airy} and~\ref{fig:combined-sff-Airy} show the disconnected, connected, and combined parts of the spectral form factor, for temperature $\beta{=}1/2$. The dashed lines are the plots of the exact functions~(\ref{eq:disconnected-airy}) and~(\ref{eq:connected-airy}) while the dots show the results of the numerical computations.

 As might be expected, significant deviations from the dashed lines occur when $\beta$ becomes too small, including patterns of zeros representing finite size effects. This signals temperatures that excite higher energies that are not included in the numerical scheme (but are in the exact expressions) and hence the results deviate. As long as such extremes are avoided (depending upon the truncation energy chosen), the numerical results are very reliable. This is a good controlled model of the truncation scheme used in the body of the paper, showing that the results obtained are robust.

\section{Numerical Recipes}
\subsection{Suggestions for solving the string equation}
\label{app:notes-and-tips-1}
The string equation that supplies the potential $u(x)$ for a  particular problem is highly non--linear, and of high order ($2k$) in derivatives if truncating to the $k$th model. Even the simplest solution (with the boundary conditions of interest given in equation~(\ref{eq:boundary-conditions})), where all $t_k$ are set to zero except $k{=}1$, requires numerical techniques to extract its explicit form (it is related to a Painl\'eve transcendent, and hence cannot be written in terms of other elementary functions).  Here are some  suggestions for finding numerical solutions, to help the interested reader learn how to extract useful information for themselves. First, {\tt Maple} was used in this case (although {\tt MATLAB} works well too, as probably would other programs). The {\tt dsolve} routine was used, with an error tolerance of $10^{-5}{-}10^{-10}$, depending upon the equation being solved. (In fact  when these equations were first solved~\cite{Dalley:1992qg,Johnson:thesis} it was for the cases $k{=}1,2,$ and~$3$, including cases where the models were non--trivially coupled~\cite{Johnson:1992pu}. Back then, their solution was found by writing a program in $\tt FORTRAN$ that called the routine $\tt D02RAF$, part of the   {\tt NAG} libraries.)

As mentioned in the text, for various reasons, it makes sense to take an additional derivative of the equation. This reduced the non--linearity somewhat, at the expense of increasing the order, which is a small price to pay.  This is because the first derivative results in an overall factor of ${\cal R}$, which can be divided out since it will not vanish for the solutions of interest. A derivative explicitly removes~$\Gamma$ from the equation, however. Now, the only knowledge the system has of the desired choice of $\Gamma$ is through subleading (in the small~$\hbar$ expansion) terms in the boundary conditions, which need to be solved for with all $t_k$ present. This is nicely organized on $x{>}0$ boundary because the presence of the $t_k$ come in one by one at successively higher terms in the $1/x^2$ expansion (see ref.~\cite{Johnson:2020heh}). It is less nice to do analytically on the $x{<}0$ boundary. There, all the $t_k$ contribute at the next order and solving for the order $\hbar\Gamma$ correction requires solving a $k$th order polynomial. By beyond the $k{=}4$ truncation this becomes unpleasant at best.  However, if the system is solved on a large enough region, with a small enough discretization, terms beyond the leading  left boundary condition can be safely ignored, and a good approximation to the the solution found anyway.  (If needed, however, a recursive code for solving for the subleading corrections to the boundary conditions numerically can be employed.)

This works well for $\Gamma{=}\frac12$, while for $\Gamma{=}{-}\frac12$ a different technique was used, because  the numerical approach was less stable due to a more complicated well shape appearing in the interior, which is hard to control in a 13th order differential equation. Ref.~\cite{Carlisle:2005mk} noticed that $\Gamma$ can be changed by an integer using a special ``Backl\"und" transformation, and  actually derived an analytic expression expression showing how to build the new $u(x)$ at $\Gamma{\pm}1$ from the old $u(x)$ (and its derivatives) at $\Gamma$. Here it is:
\be
\label{eq:CJP-backlund}
u_{\Gamma\pm1}=\frac{3({\cal R}^\prime)^2-2{\cal R}{\cal R}^{\prime\prime}\pm8\hbar\Gamma{\cal R}^\prime+4\hbar^2\Gamma^2}{4{\cal  R}^2}\ ,
\ee
(where ${\cal R}{\equiv}{\cal R}(u_\Gamma)$ and  a sign has been switched to match the current conventions). So once $u(x)$ for the case of $\Gamma{=}\frac12$ was  found using {\tt Maple}, the output of {\tt dsolve} contains all the derivatives of $u(x)$ needed to construct $u(x)$ at $\Gamma{=}{-}\frac12$, giving the well structure seen in figure~\ref{fig:truncation-examples-B}.

In fact, a similar story held for the case of $\Gamma{=}0$ used in section~\ref{sec:JT-gravity-regular}. There is also a well structure in the interior, which is hard to solve for  numerically when at high order and with  boundary conditions far from the structure itself. Experience from studying positive integer $\Gamma$ suggested that $\Gamma{=}1$ would be smoother to solve for and that was indeed the case. From there, transformation~(\ref{eq:CJP-backlund}) was used to construct the desired $\Gamma{=}0$ solution.

\subsection{Suggestions for solving the spectrum}
\label{app:notes-and-tips-2}

As mentioned in section~\ref{sec:spectral-density}, in order to solve for the solutions to the eigenvalue problem, $\{E,\psi(E,x)\}$. a matrix  Numerov method~\cite{doi:10.1119/1.4748813} was used, as it was in refs.\cite{Johnson:2019eik,Johnson:2020heh}.  This simply puts the system into a box, and turns the problem into a large matrix diagonalization problem, for a given input potential $u(x)$. This was done using {\tt MATLAB}. A key point is that it is desirable to have  a large number of  eigenvalues in the energy range from zero to the chosen highest energy (determined by the level of the truncation of the string equation). So two choices were made to ensure a good set of solutions. The first was to use a large grid, so a grid of $20000{\times}20000$ was used. The second was to use a large box.  As stated, the spectrum solving method  is essentially putting the system into a box,  and a portion of the output eigenvalues and eigenfunctions will be affected by the edges of the box. Those should be discarded, and  the larger the box, the more useable eigenstates will be available in the reliable energy window.  Since, as already observed in section~\ref{sec:minimal-model-deconstruction}, the solution for $u(x)$ becomes similar to the disc level behaviour   far away enough from the central region, the box can be easily made larger by connecting the solution (solved numerically out to ${-}200{\leq}x{\leq}{+}200$) to a wider region  ({\it e.g.}, ${-}2645{\leq} x {\leq} {+}2645$ for the (2,2) and (0,2) models) where just the exact disc solution $u_0(x)$ is used. (A  smooth (enough) transition between the two solutions was  done at $x{<}{-}100$, corresponding to energies well above the cutoff determined by good matching for the truncation, so this does not affect the physics.)  

\section{Gel'fand--Dikii Polynomials}
\label{app:gelfand-dikii}

In case they are needed, here are some of the higher order Gel'Fand--Dikii polynomials, normalized such that the coefficient of the pure $u^k$ term is unity. In the following equation a prime denotes an $x$--derivative times a factor of~$\hbar$. For high numbers of derivatives, instead, a notation $u^{(n)}$ is used for $n$ primes.  The first five are listed here:
\bea
\label{eq:gelfand-dikii}
{\tilde R}_1[u]&=&u\ ,\\
{\tilde R}_2[u]&=&u^2-\frac{1}{3}u^{''}\ ,\nonumber\\
{\tilde R}_3[u]&=&u^3-\frac12(u^{'})^2-uu^{''}+\frac{1}{10}u^{''''} \ ,\nonumber\\
{\tilde R}_4[u]&=&u^4   -2u(u^{'})^2-2u^2u^{''}     +\frac45 u^{'}u^{'''} +   \frac35 (u^{''})^2\nonumber\\
&&\hskip3.0cm+\frac25 u u^{''''}  -\frac{1}{35}u^{(6)}\ , \nonumber\\
{\tilde R}_5[u]&=&
u^5 
-5u^2(u^{'})^2
-\frac{10}{3}u^3 u^{''}
+\frac{11}{3} u^{''}(u^{'})^2
\nonumber\\
&&+3u (u^{''})^2 
+4uu^{'}u^{'''}
+u^{2} u^{''''}
-\frac {23}{42} (u^{'''})^2\nonumber\\
&&-\frac {19}{21}u^{''''}u^{''}
-\frac{3}{7}u^{(5)}u^\prime 
-\frac{1}{7} u^{(6)} u 
+\frac{1}{126}u^{(8)}
\ .\nonumber
\eea
It will transpire that  
${\tilde R}_6[u]$  and ${\tilde R}_7[u]$ will be needed as well, in order to get the required level of accuracy for the quantities computed  in this paper. They are rather lengthy quantities, so it is not clear if there is much value in listing them here.  Instead, they (and higher order ones) can be easily computed using the recursion relation: 
\be
C_k {\tilde R}_{k+1}^\prime=\frac{\hbar^3}{4} {\tilde R}^{'''}_k - u\hbar {\tilde R}_k^\prime-\frac{\hbar}{2}u^\prime {\tilde R}_k\ ,
\ee 
and the requirement that they vanish when $u$ does. $C_k$ is chosen to normalize the pure $u^k$ term to unity.

More useful therefore are the following three lines of {\tt Maple} code, which can be iterated, with obvious adjustments, to get to arbitrarily high order:

\bea
&>&{\tt R[1] := u(x);}   \\
&>& {\tt DR[2] := (1/4)*h^3*({\tt diff}(R[1], x\$3))} \nonumber \\
&&\hskip 1cm {\tt -u(x)*h*({\tt diff}(R[1], x))} \nonumber \\
&&\hskip 1cm {\tt -(1/2)*h*({\tt diff}(u(x), x))*R[1];  } \nonumber \\
&>&{\tt R[2] := {\tt simplify}(-4*{\tt integrate}(DR[2], x)/(3*h));} \nonumber  
\eea

\medskip
 
 \begin{acknowledgments}
CVJ  thanks Felipe Rosso and for helpful questions and comments,   Douglas Stanford and Edward Witten for  helpful questions, the  US Department of Energy for support under grant  \protect{DE-SC} 0011687, and, especially during the pandemic,  Amelia for her support and patience.    
\end{acknowledgments}

\bibliographystyle{apsrev4-1}
\bibliography{NP_super_JT_gravity}

\end{document}